\documentclass[a4paper,11pt]{article}
\pdfoutput=1 

\usepackage{jheppub} 

\usepackage[T1]{fontenc} 
\usepackage{slashed}
\usepackage{xcolor}
\usepackage{makecell}
\usepackage{rotating}
\usepackage{pdflscape}
\usepackage{float}
\usepackage{multicol}
\usepackage{xspace}
\usepackage{booktabs}
\usepackage{xspace}
\usepackage{pifont}
\usepackage{braket}
\usepackage{bm}
\usepackage{subcaption}
\usepackage[shortlabels]{enumitem}
\usepackage{amssymb, latexsym, amsmath, amsthm}
\usepackage{multirow}
\usepackage{makecell}
\usepackage{mathtools}

\DeclareMathAlphabet\mathbfcal{OMS}{cmsy}{b}{n}

\newcommand{\ome}{(1-\epsilon)}
\newcommand{\e}{\epsilon}

\newcommand{\order}[1]{\ensuremath{\mathcal{O}}\left( #1 \right)\xspace}

\newcommand{\Pqg}{P_{qg}^{(0)}}

\newcommand{\Pgg}{P_{gg}^{(0)}}
\newcommand{\Pqq}{P_{q\bar{q}}^{(0)}}

\newcommand{\PggS}{P_{gg}^{\text{sub},(0)}}

\newcommand{\softorder}{}

\newcommand{\Sb}{S_{b}^{\softorder}}

\newcommand{\PDSdown}{\mathrm{\mathbf{DS}}^\downarrow\xspace}
\newcommand{\PTCdown}{\mathrm{\mathbf{TC}}^\downarrow\xspace}

\newcommand{\PDCdown}{\mathrm{\mathbf{DC}}^\downarrow\xspace}
\newcommand{\PSdown}{\mathrm{\mathbf{S}}^\downarrow\xspace}
\newcommand{\PCdown}{\mathrm{\mathbf{C}}^\downarrow\xspace}

\newcommand{\PPup}{\mathrm{\mathbf{P}}^\uparrow\xspace}
\newcommand{\PPdown}{\mathrm{\mathbf{P}}^\downarrow\xspace}

\newcommand{\NF}{N_\mathrm{F}}

\newcommand{\calSsoft}{\mathcal{S}\kern-0.05em{s\kern-0.05em o\kern-0.05em f\kern-0.05em t}}
\newcommand{\calScol}{\mathcal{S}\kern-0.05em{c\kern-0.05em o\kern-0.05em l}}
\newcommand{\calDsoft}{\mathcal{D}\kern-0.05em{s\kern-0.05em o\kern-0.05em f\kern-0.05em t}}
\newcommand{\calTcol}{\mathcal{T}\kern-0.05em{c\kern-0.05em o\kern-0.05em l}}
\newcommand{\calDcol}{\mathcal{D}\kern-0.05em{c\kern-0.05em o\kern-0.05em l}}

\newcommand{\qb}{\bar{q}}
\newcommand{\Qb}{\bar{Q}}
\newcommand{\Rb}{\bar{R}}

\newcommand{\omy}{(1-y)}
\newcommand{\omz}{(1-z)}

\newcommand{\omxj}{(1-x_j)}

\renewcommand{\xi}{x_{i}}
\newcommand{\xj}{x_{j}}

\newcommand{\ya}{y_{1}}
\newcommand{\yb}{y_{2}}
\newcommand{\za}{z_{1}}
\newcommand{\zb}{z_{2}}
\newcommand{\omya}{\big(1-y_{1}\big)}
\newcommand{\omyb}{\big(1-y_{2}\big)}
\newcommand{\omza}{\big(1-z_{1}\big)}
\newcommand{\omzb}{\big(1-z_{2}\big)}
\newcommand{\omyaomyb}{\big(1-y_{1}\big(1-y_{2}\big)\big)}
\newcommand{\SAB}{s_{IK}}
\newcommand{\SBC}{s_{KM}}

\newcommand{\Se}{S_{\e}}

\newcommand{\dPS}{{\,\,\mathrm{d}PS}}
\newcommand{\nn}{\nonumber \\}

\newcommand{\wt}[1]{\widetilde{#1}}
\newcommand{\wh}[1]{\widehat{#1}}

\renewcommand{\e}{\epsilon}
\renewcommand{\nn}{\nonumber}

\newcommand{\lb}{\lbrace}
\newcommand{\rb}{\rbrace}

\newcommand{\jet}[2]{J^{(#1)}_{#2}\left( \lb p \rb_{#2}\right)}

\newcommand{\X}{{X}^{0}_{5,3}}
\newcommand{\XM}{{X}^{0}_{5,3;M}}
\newcommand{\XL}{{X}^{0}_{5,3;L}}
\newcommand{\XR}{{X}^{0}_{5,3;R}}

\newcommand{\ant}{\X}
\newcommand{\Y}{{X}^{1}_{4,3}}
\newcommand{\YM}{{X}^{1}_{4,3;M}}
\newcommand{\YL}{{X}^{1}_{4,3;L}}
\newcommand{\YR}{{X}^{1}_{4,3;R}}

\newcommand{\A}{{A}^{0}_{5,3}}

\newcommand{\D}{{D}^{0}_{5,3}}

\newcommand{\DL}{{D}^{0}_{5,3;L}}

\newcommand{\F}{{F}^{0}_{5,3}}

\newcommand{\FL}{{F}^{0}_{5,3;L}}
\newcommand{\FR}{{F}^{0}_{5,3;R}}

\newcommand{\At}{\wt{{A}}^{0}_{5,3}}

\newcommand{\AtR}{\wt{{A}}^{0}_{5,3;R}}

\newcommand{\Att}{\wt{\wt{{A}}}^{0}_{5,3}}

\newcommand{\calX}{{\cal X}^{0}_{5,3}}
\newcommand{\calXM}{{\cal X}^{0}_{5,3;M}}
\newcommand{\calXL}{{\cal X}^{0}_{5,3;L}}
\newcommand{\calXR}{{\cal X}^{0}_{5,3;R}}

\newcommand{\calY}{{\cal X}^{1}_{4,3}}
\newcommand{\calYM}{{\cal X}^{1}_{4,3;M}}
\newcommand{\calYL}{{\cal X}^{1}_{4,3;L}}
\newcommand{\calYR}{{\cal X}^{1}_{4,3;R}}

\newcommand{\calAM}{{\cal A}^{0}_{5,3;M}}
\newcommand{\calAL}{{\cal A}^{0}_{5,3;L}}
\newcommand{\calAR}{{\cal A}^{0}_{5,3;R}}

\newcommand{\calBM}{{\cal B}^{0}_{5,3;M}}
\newcommand{\calBL}{{\cal B}^{0}_{5,3;L}}
\newcommand{\calBR}{{\cal B}^{0}_{5,3;R}}

\newcommand{\calDM}{{\cal D}^{0}_{5,3;M}}
\newcommand{\calDL}{{\cal D}^{0}_{5,3;L}}
\newcommand{\calDR}{{\cal D}^{0}_{5,3;R}}

\newcommand{\calFM}{{\cal F}^{0}_{5,3;M}}
\newcommand{\calFL}{{\cal F}^{0}_{5,3;L}}
\newcommand{\calFR}{{\cal F}^{0}_{5,3;R}}

\newcommand{\calKM}{{\cal K}^{0}_{5,3;M}}
\newcommand{\calKL}{{\cal K}^{0}_{5,3;L}}
\newcommand{\calKR}{{\cal K}^{0}_{5,3;R}}

\newcommand{\calAtM}{\widetilde{{\cal A}}^{0}_{5,3;M}}
\newcommand{\calAtL}{\widetilde{{\cal A}}^{0}_{5,3;L}}
\newcommand{\calAtR}{\widetilde{{\cal A}}^{0}_{5,3;R}}

\newcommand{\calBtM}{\widetilde{{\cal B}}^{0}_{5,3;M}}
\newcommand{\calBtL}{\widetilde{{\cal B}}^{0}_{5,3;L}}
\newcommand{\calBtR}{\widetilde{{\cal B}}^{0}_{5,3;R}}

\newcommand{\calAttM}{\widetilde{\widetilde{{\cal A}}}^{0}_{5,3;M}}
\newcommand{\calAttL}{\widetilde{\widetilde{{\cal A}}}^{0}_{5,3;L}}
\newcommand{\calAttR}{\widetilde{\widetilde{{\cal A}}}^{0}_{5,3;R}}

\newcommand{\calGaM}{{\cal G}^{0 (a)}_{5,3;M}}
\newcommand{\calGaL}{{\cal G}^{0 (a)}_{5,3;L}}
\newcommand{\calGaR}{{\cal G}^{0 (a)}_{5,3;R}}

\newcommand{\calGbM}{{\cal G}^{0 (b)}_{5,3;M}}
\newcommand{\calGbL}{{\cal G}^{0 (b)}_{5,3;L}}
\newcommand{\calGbR}{{\cal G}^{0 (b)}_{5,3;R}}

\newcommand{\calHaM}{{\cal H}^{0 (a)}_{5,3;M}}
\newcommand{\calHaL}{{\cal H}^{0 (a)}_{5,3;L}}
\newcommand{\calHaR}{{\cal H}^{0 (a)}_{5,3;R}}

\newcommand{\calHbM}{{\cal H}^{0 (b)}_{5,3;M}}
\newcommand{\calHbL}{{\cal H}^{0 (b)}_{5,3;L}}
\newcommand{\calHbR}{{\cal H}^{0 (b)}_{5,3;R}}

\newcommand{\calEaM}{{\cal E}^{0 (a)}_{5,3;M}}
\newcommand{\calEaL}{{\cal E}^{0 (a)}_{5,3;L}}
\newcommand{\calEaR}{{\cal E}^{0 (a)}_{5,3;R}}

\newcommand{\calEbM}{{\cal E}^{0 (b)}_{5,3;M}}
\newcommand{\calEbL}{{\cal E}^{0 (b)}_{5,3;L}}
\newcommand{\calEbR}{{\cal E}^{0 (b)}_{5,3;R}}

\newcommand{\calEcM}{{\cal E}^{0 (c)}_{5,3;M}}
\newcommand{\calEcL}{{\cal E}^{0 (c)}_{5,3;L}}
\newcommand{\calEcR}{{\cal E}^{0 (c)}_{5,3;R}}

\newcommand{\calEdM}{{\cal E}^{0 (d)}_{5,3;M}}
\newcommand{\calEdL}{{\cal E}^{0 (d)}_{5,3;L}}
\newcommand{\calEdR}{{\cal E}^{0 (d)}_{5,3;R}}

\newcommand{\mapM}{\phi_M}
\newcommand{\mapL}{\phi_L}
\newcommand{\mapR}{\phi_R}

\newcommand{\mapYR}{\rho_R}

\renewcommand{\u}[1]{\underline{#1}}
\renewcommand{\e}{\epsilon}

\definecolor{colA}{RGB}{2,122,176}
\definecolor{colB}{RGB}{209,156,47}
\definecolor{colC}{RGB}{73, 153, 124}
\definecolor{colD}{RGB}{30, 190, 205}
\definecolor{colE}{RGB}{174, 57, 24}

\newcommand{\textA}[1]{{\color{colA}{\text{#1}}}}
\newcommand{\textB}[1]{{\color{colB}{\text{#1}}}}
\newcommand{\textC}[1]{{\color{colC}{\text{#1}}}}
\newcommand{\textD}[1]{{\color{colD}{\text{#1}}}}
\newcommand{\textE}[1]{{\color{colE}{\text{#1}}}}

\title{Generalised Antenna Functions for Higher-Order Calculations}

\author[1]{Elliot Fox,}
\author[1]{Nigel Glover,}
\author[1]{Matteo Marcoli}

\affiliation[1]{Institute for Particle Physics Phenomenology, Department of Physics, Durham University, South Road, Durham, DH1 3LE, UK}

\emailAdd{elliot.fox@durham.ac.uk}
\emailAdd{e.w.n.glover@durham.ac.uk}
\emailAdd{matteo.marcoli@durham.ac.uk}

\preprint{IPPP/24/63}

\abstract{In this paper we discuss the definition, the construction and the implementation of \textit{generalised antenna functions} for final-state radiation up to Next-to-Next-to-Leading Order (NNLO) in QCD. Generalised antenna functions encapsulate the singular behaviour of unresolved emissions when these occur within multiple hard radiators and not just two of them, as for traditional antenna functions. The construction of such objects is possible thanks to the recently proposed algorithm for building \textit{idealised antenna functions} from a target set of infrared limits. Generalised antenna functions bring major simplifications in the assemblage of subtraction terms in the context of the antenna scheme at NNLO and beyond, as well as a substantial computational speedup of higher-order calculations. We discuss in detail the improvements on the formal and practical side for the computation of the NNLO correction to three-jet production at electron-positron colliders, providing a thorough numerical validation of the newly proposed scheme. For this calculation one can expect almost an order of magnitude speedup with respect to the original implementation.}

\begin{document}
\maketitle
\flushbottom

\section{Introduction}\label{sec:introduction}

The High Luminosity Large Hadron Collider (HL-LHC) provides an unparalleled platform to investigate fundamental processes involving Higgs bosons, electroweak bosons, top quarks, and hadronic jets with exceptional precision. By achieving high accuracy in experimental measurements, the HL-LHC will examine in detail the fundamental interactions of particles at short distances. This precision is crucial, especially when new particle discoveries are absent, as even the smallest deviations from Standard Model predictions could reveal new physics. Precision phenomenology, therefore, plays a vital role in the ongoing quest to expand our understanding of the universe, where any minor discrepancy between theoretical predictions and experimental data could guide us toward physics beyond the Standard Model. Continued advancements in fixed-order calculations, parton distribution functions, parton showers and modelling of non-perturbative effects are necessary to reach the level of accuracy demanded by the LHC experiments.

Perturbation theory plays a central role in improving theoretical predictions to match the precision of current or future collider experiments. To achieve percent-level accuracy, theoretical calculations must extend to at least Next-to-Next-to-Leading Order (NNLO) in the strong-coupling expansion. However, these higher-order calculations are complex due to the interplay between real and virtual corrections across multiple phase spaces. Infrared divergences, arising from unresolved real radiation (such as soft or collinear emissions) must be carefully canceled by singularities in virtual matrix elements. Schemes for achieving this cancellation generally fall into two classes - subtraction and slicing.  
Slicing schemes typically restrict the allowed phase space of the real-radiation using a slicing parameter.   The approximate form of the matrix elements below the parameter is known, and its (divergent) contribution is integrated over the unresolved phase space and combined with the virtual matrix elements.   However, since the real matrix elements and their approximations do not fully match outside the singular limit, there remains a residual dependence on the slicing parameter that must be carefully evaluated.  In contrast, subtraction schemes are free from systematic effects: they subtract the singular terms from the real radiation and add them back exactly to the lower-multiplicity virtual contribution, after integration over the unresolved phase space. Such infrared cancellation schemes offer a well-established solution to handle the intricacies of higher-order perturbative calculations for an arbitrary scattering process. On the other hand, subtraction techniques generally require complete control over each individual unresolved configuration and hence demand significant efforts for their development and numerical implementation. 

For Next-to-Leading Order (NLO) calculations, the infrared cancellation is considered solved. Fully general schemes like Catani-Seymour dipole subtraction~\cite{Catani:1996vz} and FKS subtraction~\cite{Frixione:1995ms} were developed in the mid-1990's. Together with automated one-loop matrix-element generators~\cite{madgraph:2011uj,Cascioli:2011va}, these schemes are used to compute fully-differential predictions for generic processes, with limitations coming only from the difficulty of generating high-multiplicity matrix elements. Schemes for matching NLO calculations with parton showers, such as MC@NLO~\cite{Frixione:2002ik} and POWHEG~\cite{Nason:2004rx,Frixione:2007vw} have been developed which systematically combine NLO calculations with all-order parton-shower resummation. These innovations laid the foundations for the state-of-the-art multi-purpose event generators~\cite{powheg:2010xd,madgraph:2011uj,Bellm:2019zci,Sherpa:2019gpd,Bierlich:2022pfr}, see Ref.~\cite{Campbell:2022qmc} for a review. Nevertheless, for some observables NLO predictions are still not precise enough and even higher order corrections, NNLO or even beyond (N$^3$LO), are needed to approach the desired target of percent-level precision. 

Several methods have been proposed to compute NNLO corrections~\cite{Gehrmann-DeRidder:2005btv,Boughezal:2011jf,Currie:2013vh,DelDuca:2016ily,Catani:2007vq,Czakon:2010td,Czakon:2014oma,Gaunt:2015pea,Cacciari:2015jma,Caola:2017dug,Magnea:2018hab,Herzog:2018ily,TorresBobadilla:2020ekr,Bertolotti:2022aih,Devoto:2023rpv}, leading to NNLO-accurate predictions for essentially all $2\to 1$ and $2\to 2$ processes at hadron colliders. From $2020$ onwards, NNLO calculations for $2\to 3$ processes have started to appear~\cite{Chawdhry:2019bji,Kallweit:2020gcp,Chawdhry:2021hkp,Czakon:2021mjy,Chen:2022ktf,Hartanto:2022qhh,Alvarez:2023fhi,Badger:2023mgf,Catani:2022mfv,Buonocore:2022pqq,Buonocore:2023ljm,Mazzitelli:2024ura}, thanks to the calculation of two-loop five-point amplitudes~\cite{Chicherin:2017dob,Gehrmann:2018yef,Chicherin:2020oor,Abreu:2020cwb,Abreu:2021oya,Abreu:2021asb,Badger:2021ega,Badger:2021imn,Abreu:2023rco,Agarwal:2023suw,Abreu:2023bdp,DeLaurentis:2023izi,DeLaurentis:2023nss}. These results represent the current state-of-the-art for NNLO QCD corrections for LHC processes.

The NNLO results currently available have been typically achieved on a case-by-case basis, often requiring significant effort to adapt existing formalisms to new processes. Despite the recent advances mentioned above, two-loop matrix elements continue to pose major challenges, frequently requiring custom integral reduction techniques and the evaluation of new master integrals. Additionally, the cancellation of infrared divergences across higher-multiplicity final states becomes increasingly complex. Existing NNLO methods do not scale easily to processes with higher multiplicities, either because of intrinsic limitations or the complexity of the numerical implementation. Nevertheless, promising recent progress has been made towards formulating more general schemes. In particular, the sector-improved residue subtraction technique~\cite{Czakon:2010td,Czakon:2014oma} has been employed, within the STRIPPER numerical framework, to perform a series of cutting-edge NNLO calculations for high-multiplicity processes up to three-jet production at hadron colliders~\cite{Czakon:2021mjy,Alvarez:2023fhi}. In parallel, efforts focused on explicitly demonstrating the cancellation of infrared singularities for arbitrary scattering processes have been made in the context of the local analytic sector subtraction~\cite{Bertolotti:2022aih,Magnea:2024jqg} and nested soft-collinear subtraction~\cite{Devoto:2023rpv} methods.  Another major hurdle in making NNLO computations more widely accessible is the significant computational cost involved in generating cross-section predictions, further complicating their broader implementation. It is therefore desirable that, in the development of novel infrared cancellation methods as well as in the improvement of well-established ones, efforts are made to render NNLO calculations as computationally efficient as possible.

Among the various methods employed for fully-differential NNLO QCD calculations, the antenna subtraction scheme has been particularly successful. Initially developed for electron-positron annihilation with massless partons~\cite{Gehrmann-DeRidder:2005btv}, this method enabled the calculation of NNLO corrections for three-jet production and related event-shape observables at LEP energies~\cite{Gehrmann-DeRidder:2007foh}. It was based on colour-decomposed antenna functions constructed directly from matrix elements, thereby capturing the necessary infrared behaviour. Each antenna function captures soft and collinear singularities at the same time. Over time, the scheme was extended to address initial-state radiation, making it applicable to processes involving hadrons in the initial state~\cite{Daleo:2006xa,Daleo:2009yj,Boughezal:2010mc,Pires:2010jv,Currie:2013vh}. It has since been applied to numerous LHC processes using the \textsc{NNLOjet} parton-level Monte Carlo framework. Furthermore, the antenna subtraction technique has been adapted to handle the production of heavy coloured particles~\cite{Gehrmann-DeRidder:2009lyc,Abelof:2011jv,Bernreuther:2011jt,Abelof:2011ap,Abelof:2012bga,Abelof:2012rv,Bernreuther:2013uma,Dekkers:2014hna} as well as fragmentation processes including photons~\cite{Gehrmann:2022cih} and identified hadrons~\cite{Gehrmann:2022pzd,Bonino:2024adk} in the final state. In~\cite{Jakubcik:2022zdi,Chen:2023fba,Chen:2023egx} the analytic integration of N$^3$LO antenna functions for final-state radiation have been performed.

Recently there have been several improvements to the original antenna subtraction scheme. The colourful antenna subtraction method~\cite{Chen:2022ktf,Gehrmann:2023dxm} exploits the predictability of the singularity structure of virtual amplitudes in colour space to straightforwardly construct the virtual subtraction terms in a completely general way. The advantage consists in automatically retaining the structure of colour correlations among external QCD particles, making the construction of subtraction terms straightforward even beyond leading-colour. The one-to-one correspondence between integrated and unintegrated antenna functions is then exploited to infer subtraction terms for real emission corrections. In a parallel development, the recently-proposed designer antenna scheme~\cite{Braun-White:2023sgd,Braun-White:2023zwd,Fox:2023bma} offers an algorithm to build antenna functions for any number of real emissions directly from a specified list of unresolved limits.  This eliminates the need to decompose the antenna functions into sub-antenna functions and it reduces the size of the subtraction terms by avoiding the introduction of spurious limits that are inevitably present in the matrix-element-based antenna functions. This paper makes a further important step towards the simplification and ultimate automation of the antenna subtraction scheme for final-state radiation by further reducing the complexity of the subtraction terms, and paves the way for the merging of the two directions described above. Since the CPU requirements of NNLO calculations are considerable, any new scheme needs to be carefully validated: for this purpose, as an example, we consider perhaps the simplest non-trivial process of $e^+e^-\to jjj$ at NNLO.

The structure of the paper is as follows.  In Section~\ref{sec:preliminaries} we review the fundamentals of the antenna subtraction scheme, outlining the challenges we aim to address in this work and introducing the principles of the designer antenna method. A key focus of this Section is the dependence on the choice of momentum mapping when connecting phase spaces of different multiplicities.
Section~\ref{sec:X53} then introduces the generalised real-radiation antenna functions at NNLO, explaining their definition and construction. In Section~\ref{sec:RRsub} we detail the implementation of these new antenna functions in the double-real subtraction terms, using leading-colour three-jet production at electron-positron colliders as an illustrative example. In Section~\ref{sec:RVsub} we explain how the real-virtual subtraction terms are modified within the new framework. At this level, we will introduce a new class of generalised real-radiation one-loop antenna functions. In the context of a subtraction scheme, these real-radiation antenna functions must be integrated over the unresolved phase space. Section~\ref{sec:integration} demonstrates how to achieve this in full analytical fashion for the novel antenna functions. In Section~\ref{sec:VVsub} we assemble the double-virtual subtraction terms and assess the cancellation of infrared singularities. Section~\ref{sec:checks} presents numerical validation of the entire setup.  Finally our conclusions are summarised in Section~\ref{sec:conclusions}.
The appendices provide further details:
Appendix~\ref{app:limX53} lists the unresolved limits of the new real-radiation antenna functions, Appendix~\ref{app:integrated} gives the explicit analytic expressions of the integrated antenna functions and Appendix~\ref{app:sub} gives the full set of subtraction terms required for $e^+e^-\to jjj$ at NNLO.

\section{Preliminaries}\label{sec:preliminaries}

In this Section we briefly cover the necessary background for the remainder of the paper. 

\subsection{Recap of the antenna subtraction method}\label{sec:recap_antenna}

The antenna subtraction scheme for NNLO calculations is presented in detail in~\cite{Gehrmann-DeRidder:2005btv,Currie:2013vh}. In this Section we do not aim at a thorough summary of the method, but rather we want to highlight the underlying structure of the subtraction terms and motivate the generalisations described in the remainder of this paper. 

Antenna functions capture the infrared behaviour of partons becoming unresolved between a pair of hard radiators. The colour connections between the emitted partons and the hard radiators determine the possible structures and infrared patterns that need to be addressed within the antenna subtraction scheme. 

\subsubsection*{Single unresolved}

\begin{figure}[h]
    \centering
    \includegraphics[width=0.5\linewidth]{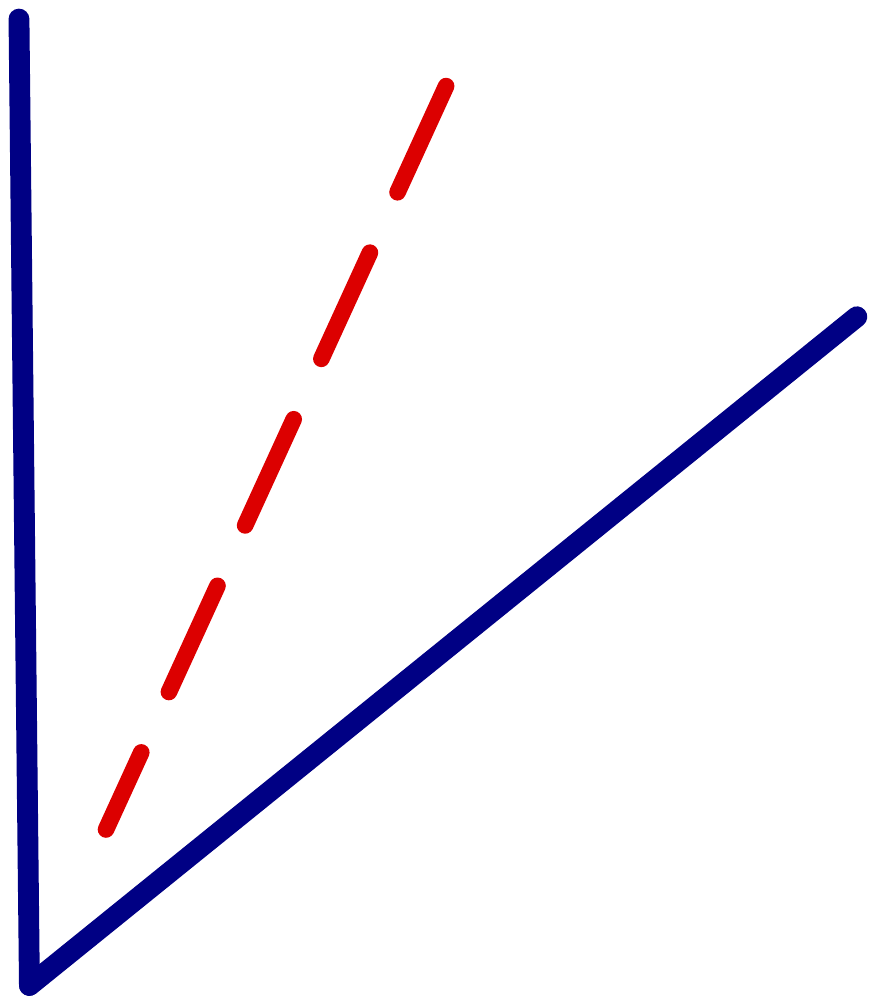}
    \caption{A single-unresolved emission (dashed line) emitted between two hard radiators (continuous lines).}
    \label{fig:X30fig}
\end{figure}

At NLO, only one parton can become unresolved, therefore the only emission topology to be considered is the one depicted in Figure~\ref{fig:X30fig}. Here continuous lines represent hard radiators, while the dashed line indicates the single-unresolved emission. 

This configuration is naturally described by a \textit{tree-level} \textit{three-parton} (two hard radiators and a single emission) antenna function $X_3^0$. In general,  $X_n^{\ell}$ denotes an $\ell$-loop $n$-particle antenna function.
Considering a $(n+1)$-parton tree-level matrix element $M_{n+1}^0(\ldots,i,j,k,\ldots)$ as a real correction to an underlying $n$-particle Born configuration, in the limits where parton $j$ becomes unresolved between $i$ and $k$, we can reproduce the singular behaviour with:
\begin{equation}
    X_{3}^{0}(i^{h},j,k^{h})M^{0}_{n}(\ldots,(\widetilde{ij}),(\widetilde{jk}),\ldots)J^{(n)}_{n}(\lb p \rb_{n}),
\end{equation}
where the superscript $h$ indicates a specific choice of hard radiator. $M^{0}_{n}$ represents the Born-level matrix element, and $(\wt{ij})$, $(\wt{jk})$ denote composite momenta obtained by applying a momentum-conserving on-shell map $(p_i,p_j,p_k)\to(p_{\wt{ij}},p_{\wt{jk}})$ which preserves the correct behaviour in the unresolved limits. This map defines the $n$-particle phase space from the $(n+1)$-particle one. Traditionally, in the context of antenna subtraction, the momentum mapping (antenna mapping) described in~\cite{Kosower:2002su} is used. We anticipate that for the generalised antenna functions introduced below we will make use of other choices of momentum mapping for reasons we will outline later in the paper. Finally, the jet function $J^{(n_p)}_{n_j}$ reconstructs $n_j$ resolved jets from $n_p$ final-state partons with momenta $\lb p \rb_{n_p}$ and implements the definition of the fiducial phase space.

\subsubsection*{Colour connection}
Before analysing the double-unresolved emission case, it is helpful to introduce the idea of colour-connection.  In any colour-string, each gluon is colour-connected to two other partons (either quarks or gluons) and each quark/antiquark is colour-connected to a single parton. Photons are colour-connected to a quark and an antiquark. At NNLO we encounter different configurations depending on whether the two unresolved partons are
\begin{itemize}
    \item colour-unconnected from each other and have no hard radiators in common 
    \item colour-connected to each other
    \item almost colour-connected, which we define to be the case where the unresolved partons are colour-unconnected, but one or more of the hard radiators is colour-connected to both unresolved partons.
\end{itemize}
These different emission topologies need to be addressed separately.

\subsubsection*{Double unresolved: colour unconnected}\label{sec:cu}

\begin{figure}[h]
    \centering
    \includegraphics[width=0.5\linewidth]{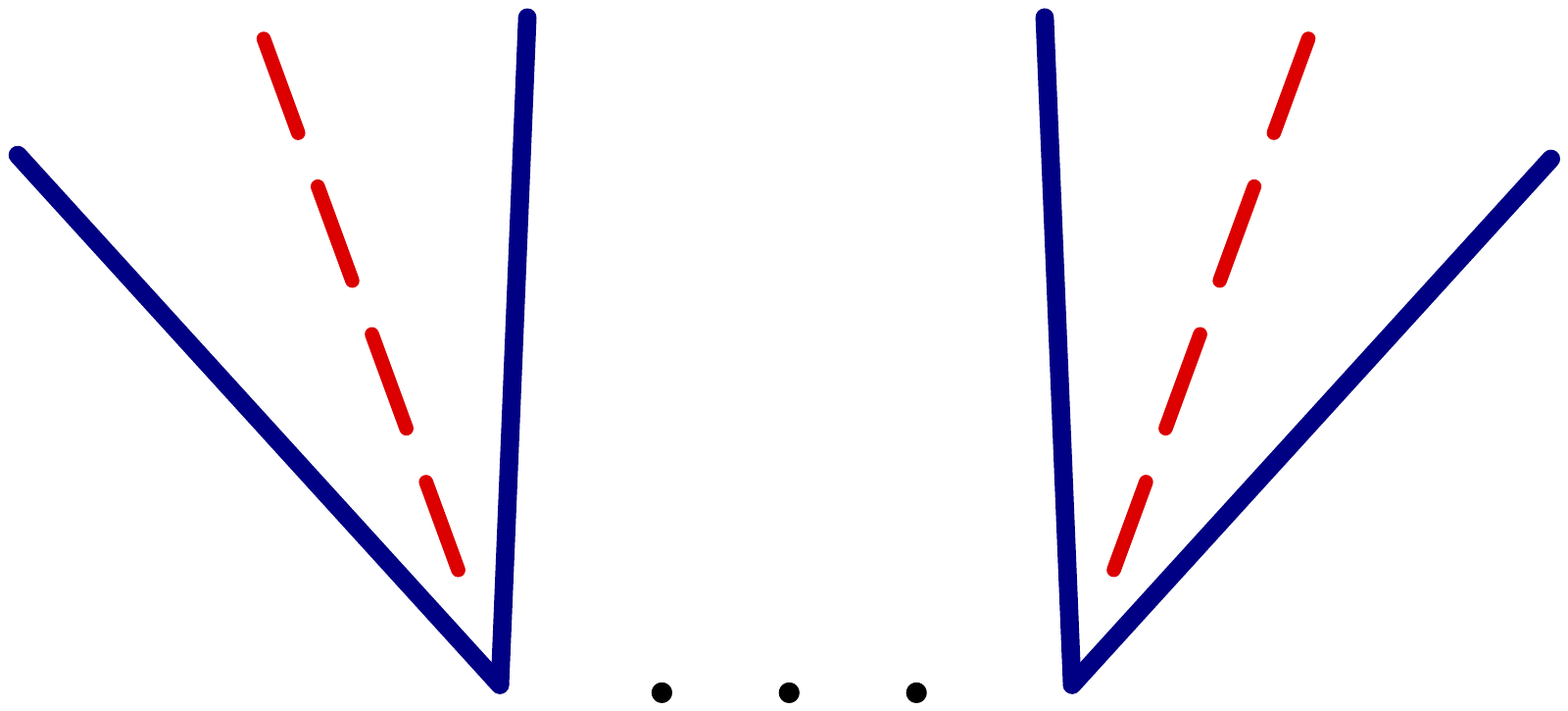}
    \caption{Two colour-unconnected unresolved partons (dashed lines) emitted between two different pairs of hard radiators (continuous lines).}
    \label{fig:X30X30fig}
\end{figure}

We start by considering the particularly simple configuration where the two emissions are well-separated in the colour-string and do not share any common hard radiator. This configuration is labelled \textit{colour-unconnected} and is depicted in Figure~\ref{fig:X30X30fig}. 
In this case, the double-unresolved contribution is purely iterated, hence the singular behaviour of a double-real matrix element $M_{n+2}^0(\ldots,i_1,j_1,k_1,\ldots,i_2,j_2,k_2,\ldots)$ with $j_1$ and $j_2$ becoming unresolved between the neighbouring partons can be fully described by the product of two disconnected $X_{3}^{0}$ antenna functions:
\begin{eqnarray}
   &&\hspace{-0.5cm}X_{3}^{0}(i_1^{h},j_1,k_1^{h})X_{3}^{0}(i_2^{h},j_2,k_2^{h})M^{0}_{n}(\ldots,(\wt{i_1j_1}),(\wt{j_1k_1}),\ldots,(\wt{i_2j_2}),(\wt{j_2k_2}),\ldots)\jet{n}{n}.
\end{eqnarray}
We note the combined action of the two antenna functions produces a map from the $(n+2)$-particle phase space to the $(n+1)$-particle phase space, and from this to the $n$-particle one, whose momenta enter the Born matrix element and the jet function.

\subsubsection*{Double unresolved: colour-connected}\label{sec:cc}

\begin{figure}[h]
    \centering
    \includegraphics[width=0.5\linewidth]{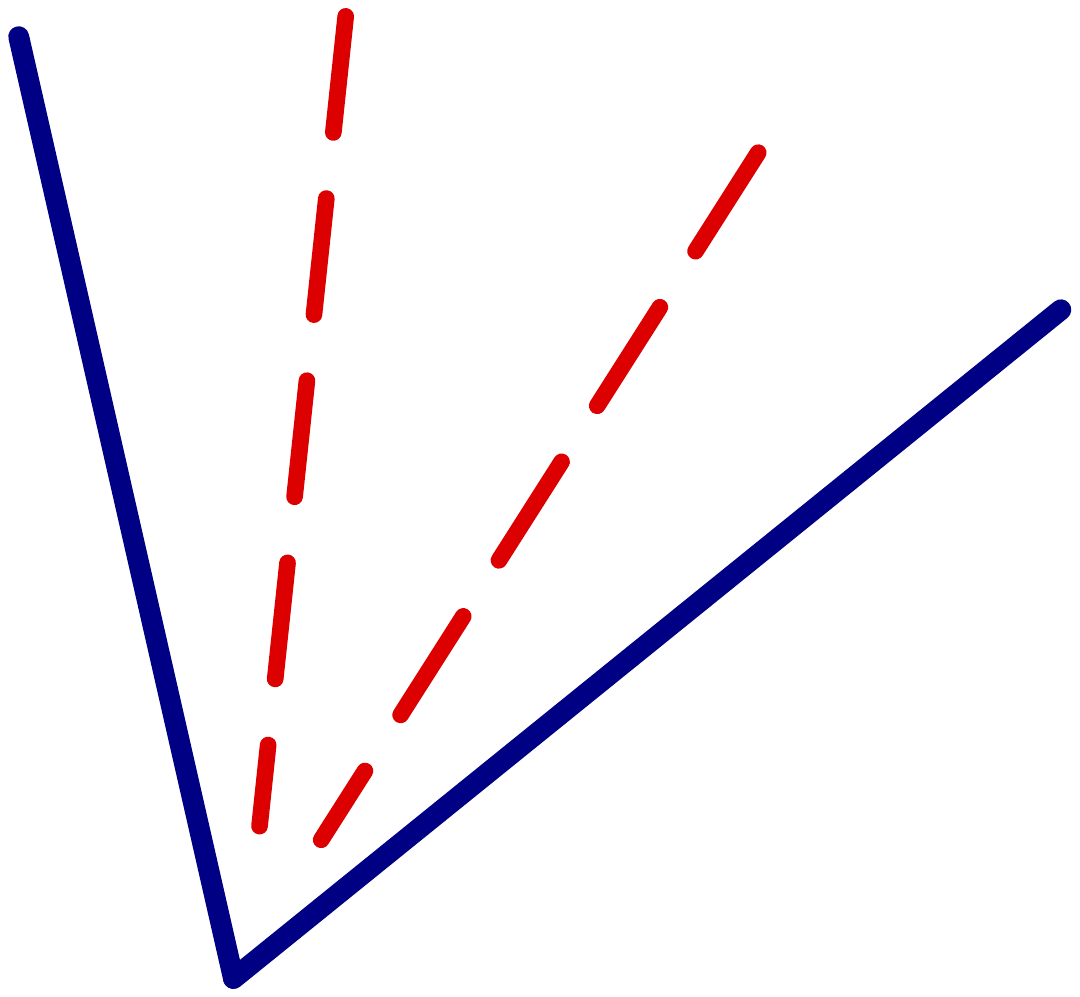}
    \caption{Two colour-connected unresolved partons (dashed lines) emitted between two common hard radiators (continuous lines).}
    \label{fig:X40fig}
\end{figure}

The next configuration we have to consider at NNLO is given by two unresolved partons that are \textit{colour-connected} to each other and one of the hard radiators. This case is depicted in Figure~\ref{fig:X40fig}. 
The singular behavior associated to such a configuration is not simply iterated, and hence cannot be captured by any combination of $X_3^0$ antenna functions. Instead, \textit{tree-level four-parton} (two hard radiators and two unresolved partons) antenna functions $X_4^0$ are used. These remove the singular behaviour of the double-real matrix element $M_{n+2}^0(\ldots,i,j,k,l,\ldots)$ in configurations where $j$ and $k$ are unresolved according to:
\begin{equation}
    X_{4}^{0}(i^{h},j,k,l^{h})M^{0}_{n}(\ldots,(\wt{ijk}),(\wt{jkl}),\ldots)J^{(n)}_{n}(\lb p \rb_{n}),
\end{equation}
where $(\wt{ijk})$, $(\wt{jkl})$ indicate composite momenta obtained by applying a momentum-conserving on-shell map $(p_i,p_j,p_k,p_l)\to(p_{\wt{ijk}},p_{\wt{jkl}})$, which preserves the correct behaviour in the unresolved limits. This map defines the $n$-particle phase space from the $(n+2)$-particle one. While providing the correct double-unresolved behaviour, $X_4^0$ antenna functions inevitably also contribute to spurious limits in single-unresolved regions. In this regard, the $X_4^0$ antenna functions are constructed in such a way that these limits can be subtracted with iterated $X_{3}^{0}X_{3}^{0}$ structures~\cite{Gehrmann-DeRidder:2005btv,Currie:2013vh,Braun-White:2023sgd}. The complete set of terms to address the colour-connected case is then given by
\begin{eqnarray}\label{X40minusX30X30}
&&\phantom{-} X_{4}^{0}(i^{h},j,k,l^{h})M^{0}_{n}(\ldots,(\wt{ijk}),(\wt{jkl}),\ldots)J^{(n)}_{n}(\lb p \rb_{n})\nn\\
&&-X_{3}^{0}(i^{h},j,k^{h})X_{3}^{0}((\wt{ij})^{h},(\wt{jk}),l^{h})M^{0}_{n}(\ldots,(\wt{(\wt{ij})(\wt{jk})}),(\wt{(\wt{jk})l}),\ldots)J^{(n)}_{n}(\lb p \rb_{n})\nn\\
&&-X_{3}^{0}(l^{h},k,j^{h})X_{3}^{0}((\wt{lk})^{h},(\wt{kj}),i^{h})M^{0}_{n}(\ldots,(\wt{i(\wt{jk})}),(\wt{(\wt{jk})(\wt{kl})}),\ldots)J^{(n)}_{n}(\lb p \rb_{n}).
\end{eqnarray}
The momentum mappings in each line of~\eqref{X40minusX30X30} produce different reduced $n$-particle momentum sets. However, the mappings are chosen in such a way that they yield the same momenta in the required unresolved limits, to allow for the desired cancellation between the $X_4^0$ antenna functions and the iterated structures in single-unresolved limits~\cite{Kosower:2002su,Gehrmann-DeRidder:2005btv}. 

\subsubsection*{Double unresolved: almost colour-connected}\label{sec:acc}

\begin{figure}[h]
\begin{subfigure}{0.45\linewidth}
    \centering
    \includegraphics[width=0.9\linewidth]{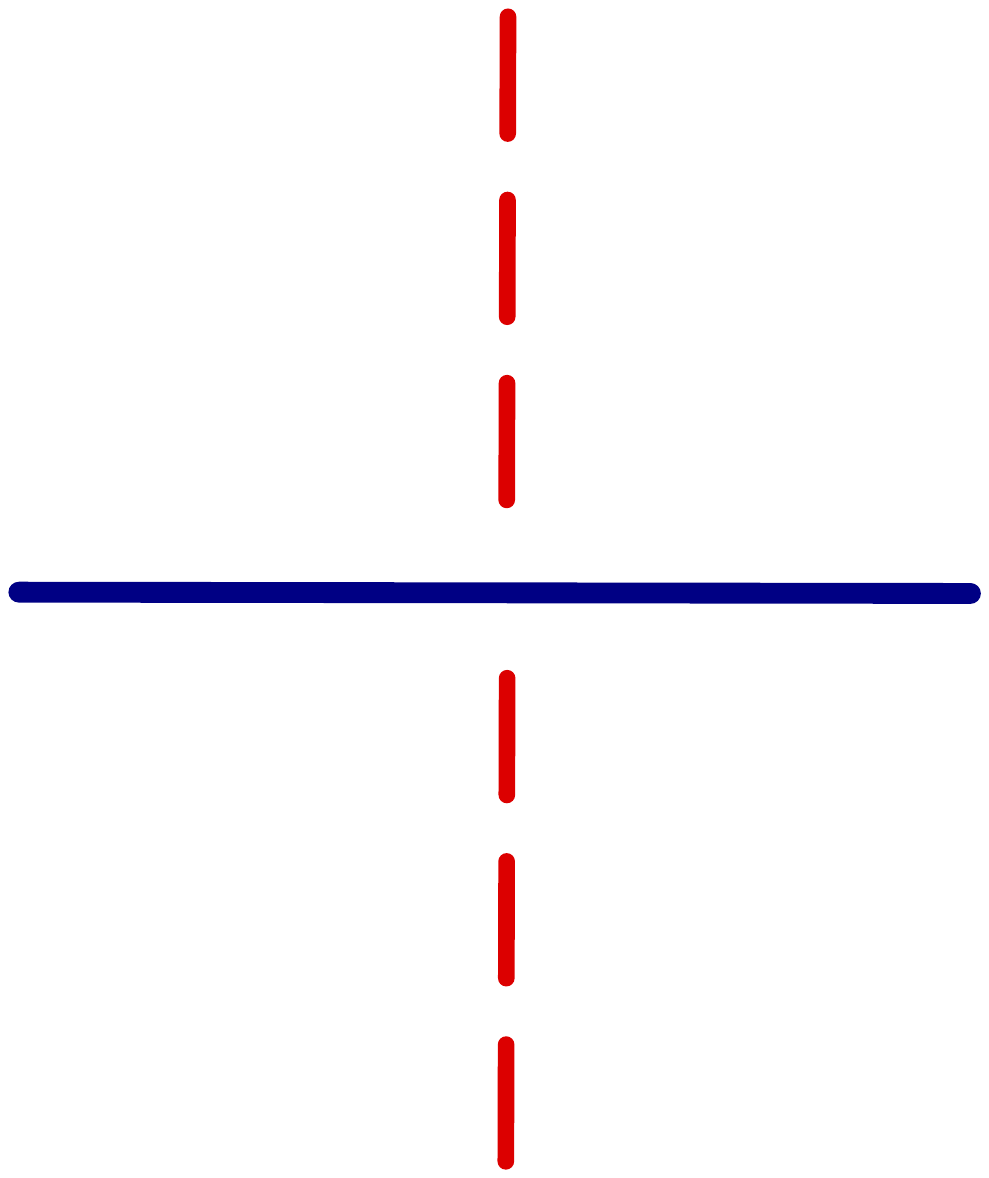}
    \caption{}
    \label{fig:Xt40fig}
\end{subfigure}
\begin{subfigure}{0.45\linewidth}
    \centering
    \includegraphics[width=0.9\linewidth]{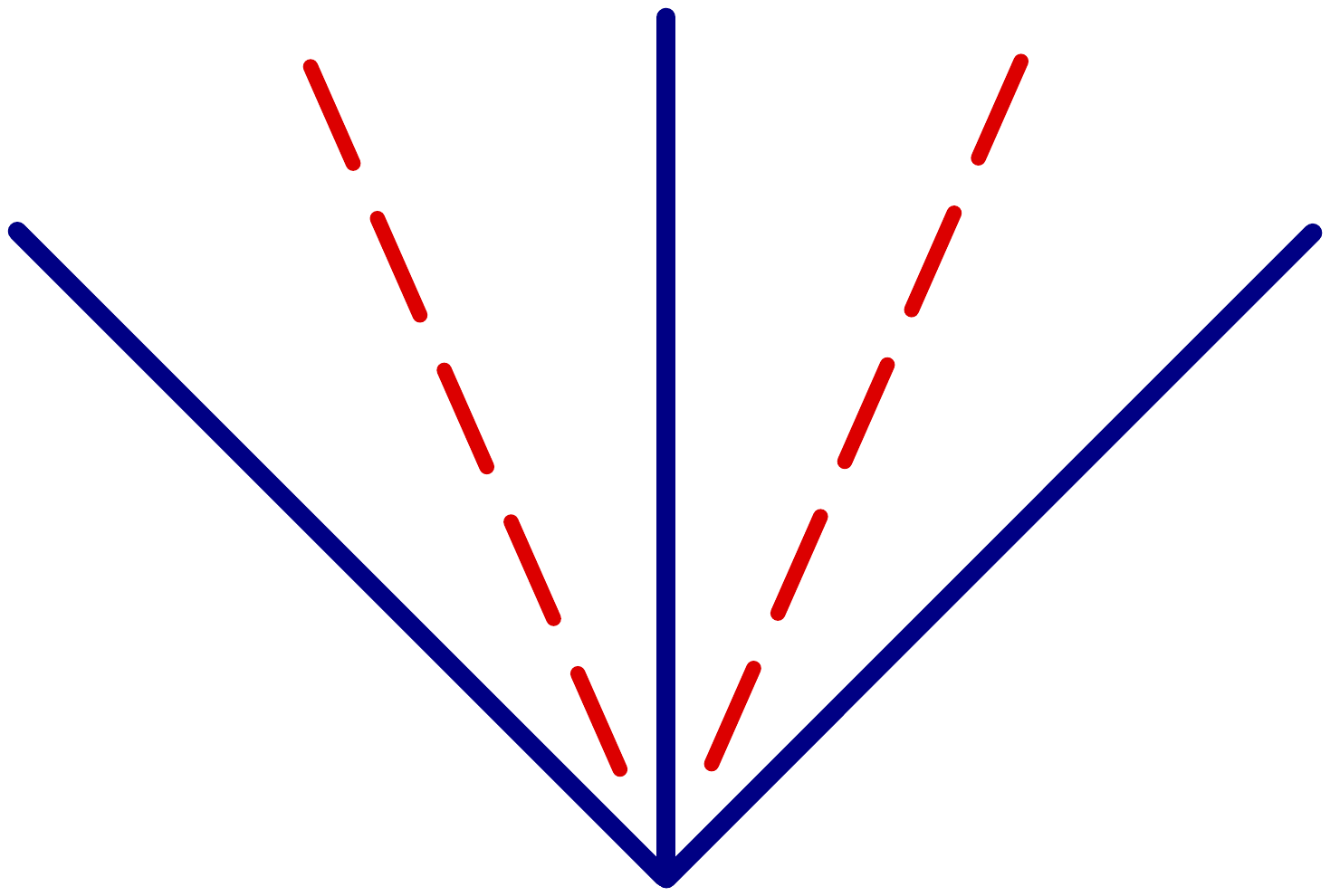}
    \caption{}
    \label{fig:X530fig}
\end{subfigure}
\caption{Two almost colour-connected unresolved partons (dashed lines) sharing both (a) or one (b) hard radiator.}
\label{fig:fullaccfig}
\end{figure}

One final emission topology has to be considered in order to build a subtraction scheme able to deal with arbitrary multiplicity at NNLO. This is the so-called \textit{almost colour-connected} configuration, depicted in Figure~\ref{fig:fullaccfig}. 
Here the two unresolved partons are not colour-connected, but also not entirely disconnected since they share both hard radiators (Figure~\ref{fig:Xt40fig}) or only one of them (Figure~\ref{fig:X530fig}). These two cases are treated differently in the context of antenna subtraction. 

First let us consider the four particle configuration shown in Figure~\ref{fig:Xt40fig}, which we generically label as $\wt{X}_4^{0}$.  These antenna functions are implemented in subtraction terms as:
\begin{eqnarray}\label{Xt40minusX30X30}
&&\phantom{-} \wt{X}_{4}^{0}(i^{h},j,k^{h},l)M^{0}_{n}(\ldots,(\wt{lij}),(\wt{jkl}),\ldots)J^{(n)}_{n}(\lb p \rb_{n})\nn\\
&&-X_{3}^{0}(i^{h},j,k^{h})X_{3}^{0}((\wt{ij})^{h},l,(\wt{jk})^{h})M^{0}_{n}(\ldots,(\wt{(\wt{ij})l}),(\wt{(\wt{jk})l}),\ldots)J^{(n)}_{n}(\lb p \rb_{n})\nn\\
&&-X_{3}^{0}(k^{h},l,i^{h})X_{3}^{0}((\wt{lk})^{h},j,(\wt{il})^{h})M^{0}_{n}(\ldots,(\wt{(\wt{il})j}),(\wt{(\wt{kl})j}),\ldots)J^{(n)}_{n}(\lb p \rb_{n}).
\end{eqnarray}
Note that there are two types of $\wt{X}_4^{0}$.  One type typically addresses subleading-colour contributions to matrix elements which also have two colour-unconnected emissions between the same hard radiators.  The prime example of this is the $\wt{A}_4^{0}$ which describes the subleading-colour emission of two gluons radiated within a quark-antiquark pair. This type of antenna is needed to describe subleading-colour configurations even for higher multiplicity processes.  
The other type of $\wt{X}_4^{0}$ antenna appears with the same colour factor as the corresponding $X_4^{0}$.   An example being the $F_4^{0}$ and $\wt{F}_4^{0}$ antenna functions.  When there are only four gluons in the process, the 
$\wt{F}_4^{0}$ is necessary to capture the almost colour-connected limits.   However, as soon as the multiplicity increases, these $\wt{X}_4^{0}$ would be replaced by the new constructs that we are introducing below. This point is further discussed in Section~\ref{subsec:Xt40}.

From now on we focus on Figure~\ref{fig:X530fig} and assume that at least five partons are present at the double-real level. In the traditional antenna subtraction scheme, dealing with this emission topology is significantly less straightforward than the cases discussed above. As one can guess, the reason for this lies in the fact that the unresolved radiation is distributed among three hard radiators, a situation which is not easily addressed by the two-hard-radiator antenna functions considered so far in the literature. It is still possible to achieve local infrared subtraction for these configurations in the context of the traditional antenna subtraction method, however this requires complex strings of interdependent terms, which typically end up being the largest contribution to the expressions of the double-real subtraction terms. In addition to iterated $X_3^0X_3^0$ contributions, the almost colour-connected subtraction terms rely on suitable combinations of eikonal factors, called \textit{large angle soft terms}~\cite{Weinzierl:2009nz,Gehrmann-DeRidder:2007foh}. The complexity of the subtraction at the double-real level also impacts the integrated structures in the real-virtual subtraction terms~\cite{Gehrmann-DeRidder:2011jwo}. The conventional treatment of the almost colour-connected case is described in detail in~\cite{Gehrmann-DeRidder:2007foh,Currie:2013vh,Gehrmann:2023dxm}.

In this paper we revisit the infrared structure of almost colour-connected singularities at NNLO and aim at a description as concise as the one in the colour-connected case. In particular, considering the parton string $(i^{h},j,k^{h},l,m^{h})$, the goal is to capture the full unresolved behaviour of the configuration in Figure~\ref{fig:X530fig} with something like:
\begin{eqnarray}\label{X53minusX30X30}
&&\phantom{-} X_{5,3}^{0}(i^{h},j,k^{h},l,m^{h})M^{0}_{n}(\ldots,(\wt{ijk}),(\wt{ijklm}),(\wt{klm}),\ldots)J^{(n)}_{n}(\lb p \rb_{n})\nn\\
&&-X_{3}^{0}(i^{h},j,k^{h})X_{3}^{0}((\wt{jk})^{h},l,m^{h})M_{n}^{0}(\ldots,(\wt{ij}),(\wt{(\wt{jk})l}),(\wt{lm}),\ldots)^{(n)}_{n}(\lb p \rb_{n})\nn\\
&&-X_{3}^{0}(m^{h},l,k^{h})X_{3}^{0}((\wt{lk})^{h},j,i^{h})M_{n}^{0}(\ldots,(\wt{ij}),(\wt{j(\wt{kl})}),(\wt{lm}),\ldots)^{(n)}_{n}(\lb p \rb_{n}),
\end{eqnarray}
where $X_{5,3}^{0}(i^{h},j,k^{h},l,m^{h})$ denotes a \textit{tree-level five-parton three-hard-radiator} antenna function. The notation we will adopt for these generalised antenna functions in the remainder of the paper is $X_{n,n_h}^{\ell}$, where $n$ is the total number of partons in the antenna function, $n_h$ is the number of hard radiators, and $\ell$ is the number of loops. If the $n_h$ index is omitted, in analogy with the notation used so far, we assume $n_h=2$. The immediate advantages of this program at the double-real level are:
\begin{itemize}
    \item a more algorithmic construction of generic subtraction terms;
    \item a significant reduction in the size of the expressions of the subtraction terms;
    \item the elimination of large angle soft terms;
    \item the possibility of constructing local subtraction terms which work for a single colour-ordered matrix element, rather than for the full sum.
\end{itemize}
The last point above will also allow us to significantly reduce the computational time required for an NNLO calculation, as we will comment on later. Finally, the simplifications at the double-real level propagate into the real-virtual subtraction terms, as discussed in Section~\ref{sec:RVsub}.

\subsubsection*{Single unresolved at one loop}\label{sec:RVintro}

The removal of single-unresolved divergences at one-loop can also be organised in terms of emission topologies. Even if at the real-virtual level there is only a single emission, in the context of antenna subtraction it matters how the unresolved parton relates to the loop correction. $X_3^0$ antenna-functions are only capable to capture the singular behaviour due to tree-level sub-graphs of a full virtual matrix elements and so new structures are needed, which we illustrate in the following.

\begin{figure}[h]
    \centering
    \includegraphics[width=0.5\linewidth]{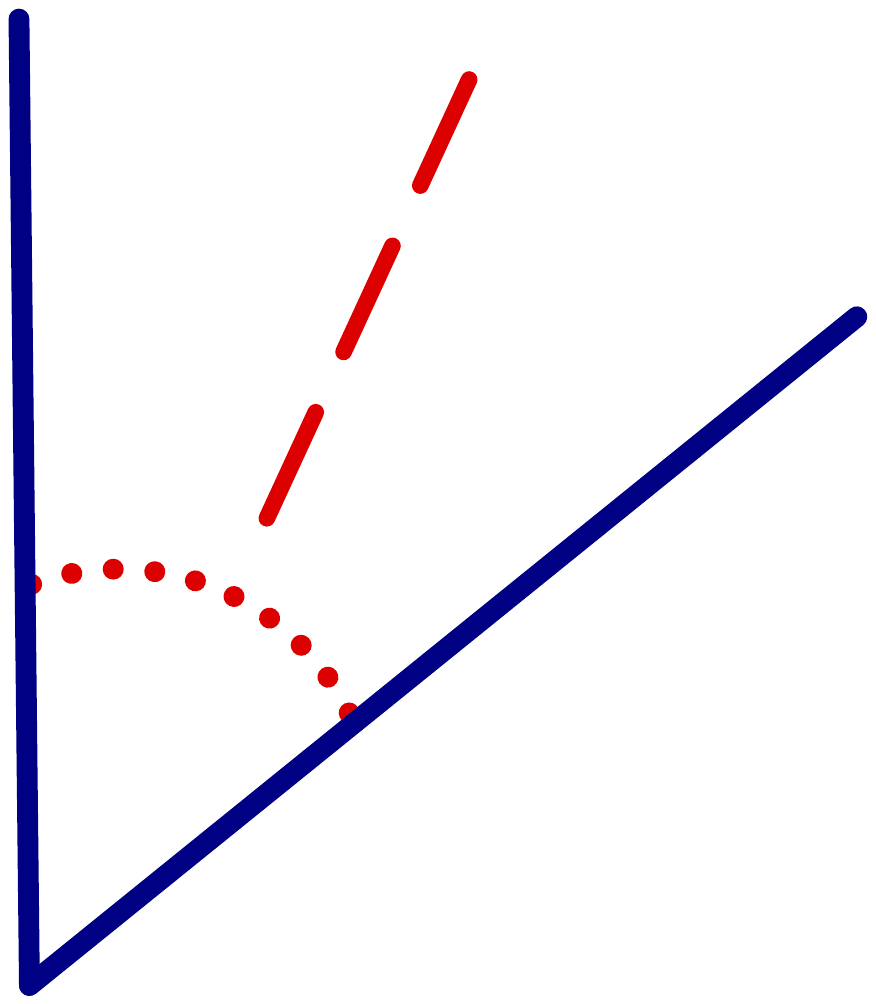}
    \caption{Colour-connected virtual (dotted line) and real (dashed line) partons emitted between the same pair of hard radiators (continuous lines).}
    \label{fig:X31fig}
\end{figure}

The first case we consider is given by an unresolved emission directly connected to a loop. In particular, the soft and collinear singularities are generated by loop graphs where a virtual particle couples to one or more of the hard radiators and the unresolved real emission. We represent this topology in Figure~\ref{fig:X31fig},  with a dotted red line representing a virtual particle exchanged between the two hard radiators.
Such configuration is treated with \textit{three-parton one-loop} antenna functions $X_3^1$~\cite{Gehrmann-DeRidder:2005btv}, which are implemented within the real-virtual subtraction term as:
\begin{equation}
    X_{3}^{1}(i^{h},j,k^{h})M^{0}_{n}(.,(\wt{ij}),(\wt{jk}),.)J^{(n)}_{n}(\lb p \rb_{n}).
\end{equation}

\begin{figure}[h]
\begin{subfigure}{0.45\linewidth}
    \centering
    \includegraphics[width=0.9\linewidth]{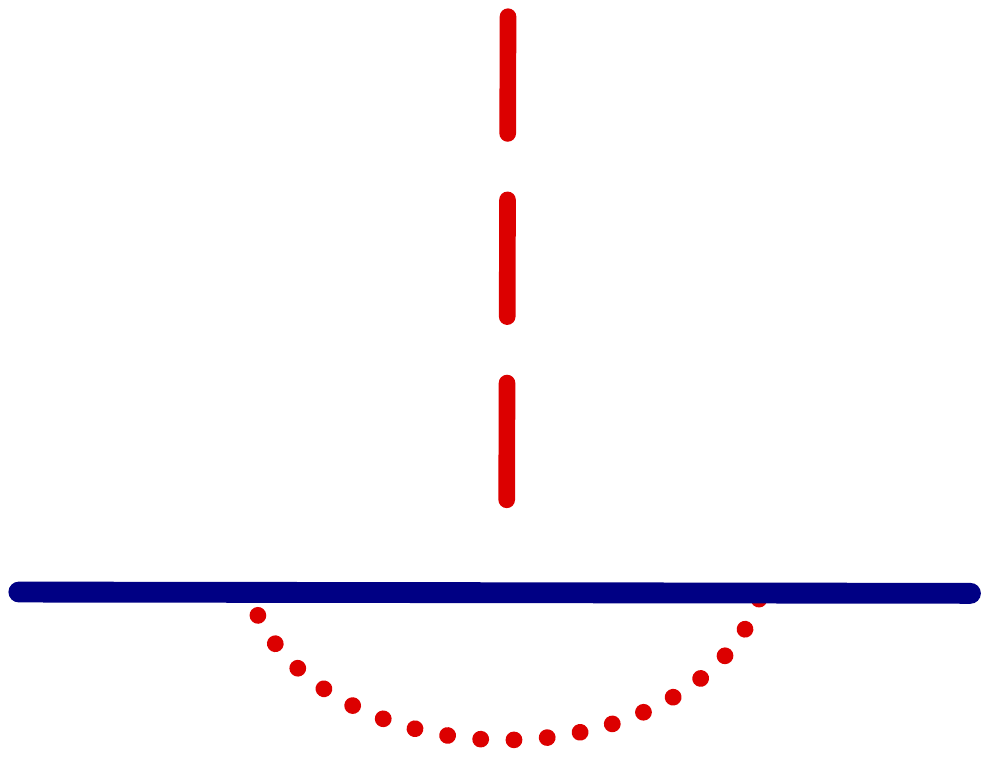}
    \caption{}
    \label{fig:Xt31fig}
\end{subfigure}
\begin{subfigure}{0.45\linewidth}
    \centering
    \includegraphics[width=0.9\linewidth]{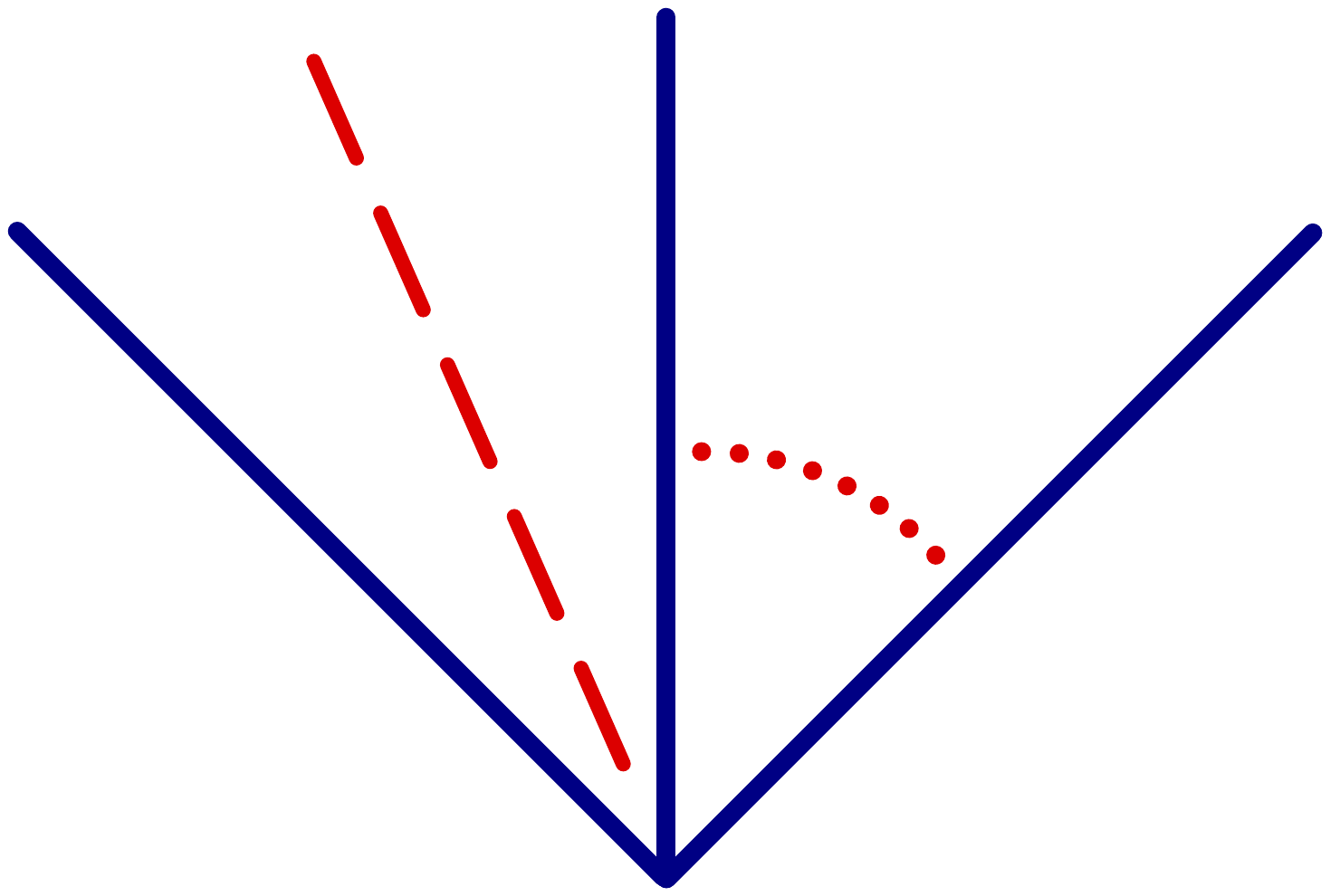}
    \caption{}
    \label{fig:X431fig}
\end{subfigure}
\caption{Colour-unconnected virtual (dotted line) and real (dashed line) partons sharing both (a) or only one (b) hard radiator (continuous line).}
\label{fig:fullX431fig}
\end{figure}

The other possibility at NNLO is given by colour-unconnected virtual and real emissions with one or more common hard radiators, represented in Figure~\ref{fig:fullX431fig}. 
As for the double-real emission discussed above, we distinguish the cases in which the virtual and real particles share both or only one hard radiator, depicted in Figure~\ref{fig:Xt31fig} and Figure~\ref{fig:X431fig} respectively. The former configuration is practically analogous to the colour-connected case in Figure~\ref{fig:X31fig}, typically appears in subleading-colour contributions to matrix elements, and is therefore addressed with subtraction terms like:
\begin{equation}
    \wt{X}_{3}^{1}(i^{h},j,k^{h})M^{0}_{n}(\ldots,(\wt{ij}),(\wt{jk}),\ldots)J^{(n)}_{n}(\lb p \rb_{n}).
\end{equation}
The configuration in Figure~\ref{fig:X431fig}, instead, naturally involves three hard radiators, hence it is not straightforwardly described by conventional antenna functions. Indeed, Figure~\ref{fig:X431fig} can be loosely seen as the real-virtual counterpart of Figure~\ref{fig:X530fig}. To properly deal with this case, in Section~\ref{sec:RVsub} we will introduce \textit{four-parton three-hard-radiator one-loop} antenna functions $X_{4,3}^1$. They will be implemented as:
\begin{equation}
    X_{4,3}^{1}(i^{h},j,k^{h}, a^{h})M^{0}_{n}(\ldots,(\wt{ij}),(\wt{jk}),(\wt{ka}),\ldots)J^{(n)}_{n}(\lb p \rb_{n}).
\end{equation}
As we will discuss in detail, the contribution given by these new antenna functions is a byproduct of the implementation of $X_{5,3}^0$ antenna functions at the double-real level. $X_{4,3}^{1}$ antenna functions absorb the integrated counterpart of the large-angle eikonal factors, which are traditionally needed in almost colour-connected configurations~\cite{Gehrmann-DeRidder:2007foh}.

\vspace{1cm}

The remainder of this paper is dedicated to the construction of generalised antenna functions and to their implementation in a consistent subtraction scheme for NNLO calculation. The possibility of building antenna functions with more than two hard radiators is a direct consequence of the designer antenna method, which is briefly outlined below.

\subsection{Recap of designer antenna principles}\label{sec:recap_designer}

In the original formulation of the antenna subtraction scheme, the antenna functions used to construct subtraction terms were extracted from matrix elements~\cite{Gehrmann-DeRidder:2004ttg,Gehrmann-DeRidder:2005svg,Gehrmann-DeRidder:2005alt}. Recently, work has been done to instead construct antenna functions directly from the desired unresolved limits in a new formalism known as the \textit{designer antenna method}~\cite{Braun-White:2023sgd,Braun-White:2023zwd,Fox:2023bma}. The antenna functions obtained from this approach, labelled \textit{idealised}, offer several advantages with respect to their traditional counterparts, which are summarised in the design principles:
\begin{enumerate}[I.]
\item each antenna function has exactly two hard particles (``radiators'') which cannot become unresolved;
\item each antenna function captures all (multi-)soft limits of its unresolved particles;
\item where appropriate, (multi-)collinear and mixed soft and collinear limits are decomposed over ``neighbouring'' antenna functions;
\item antenna functions do not contain any spurious (unphysical) limits;
\item antenna functions only contain singular factors corresponding to physical propagators;
\item where appropriate, antenna functions obey physical symmetry relations (such as line reversal).
\end{enumerate}
The iterative algorithm for constructing real-radiation antenna functions according to the principles above is described in detail in~\cite{Braun-White:2023sgd} and relies on:
\begin{itemize}
    \item a list of “target functions” $L_i$, which describe the behaviour of the colour-ordered matrix element squared in a given unresolved limit;
    \item a set of “down-projectors” $\PPdown_i$ which map the invariants of the full phase space into the subspace relevant for the limit $L_i$;
    \item a set of “up-projectors” $\PPup_i$ which restore the full antenna phase space. 
\end{itemize}
With such ingredients the construction of an antenna function $X$ is given by:
\begin{eqnarray}
    X_{1} &=& \textbf{P}_{1}^{\uparrow}L_{1}\nn\\
    X_{2} &=& X_{1}+\textbf{P}_{2}^{\uparrow}\left(L_2-\textbf{P}_{2}^{\downarrow}X_1\right)\nn\\
    \vdots\nn\\
    X_{N} &=& X_{N-1}+\textbf{P}_{N}^{\uparrow}\left(L_N-\textbf{P}_{N}^{\downarrow}X_{N-1}\right)\nn
\end{eqnarray}
The generalisation of the algorithm to include $\ell$-loop antenna functions and address virtual corrections is described in~\cite{Braun-White:2023zwd}.

\subsection{Mapping dependence}

In this Section we elaborate on the choice of the momentum map which relates different-multiplicity phase spaces in the context of local subtraction. In any subtraction scheme, one has to guarantee that the local subtraction terms in the numerical evaluation of the real radiation contributions precisely cancel against the analytically integrated terms that are added back to the virtual contributions. We are particularly interested in how the choice of momentum maps in the unintegrated subtraction terms influences their analytic integration. 

The traditional antenna subtraction method with two hard radiators relies on the class of maps described in~\cite{Kosower:2002su} for the numerical implementation of the real radiation subtraction terms. The main advantage of the antenna mapping is the smooth interpolation between all infrared configurations. However, it is implemented via complicated relations among momenta and invariants, which clearly make it not suitable for analytical integration. 
In fact, as we formalize below, for the case of two hard radiators the integration over the unresolved phase space is independent of the choice of mapping, and one can therefore choose a simpler mapping for the analytical integration of the antenna subtraction terms. However, the case of  multiple-hard-radiator antenna functions requires a more careful treatment. 

We begin by considering an unintegrated subtraction term,
\begin{align}
\label{eq:unintegratedterm}
    X(\{p\}) \, M(\{\wt{p}\},\{\wt{q}\}) \,
    J^{(n_{\wt{p}}+n_q)}_{n_j}(\{\wt{p}\},\{\wt{q}\}) \, dPS_{n_p+n_q}({p},{q})
\end{align}
where the $n_p$ momenta $\{p\}$ and the $n_q$ momenta $\{q\}$ are the original momenta appearing in the phase space $dPS_{n_p+n_q}$. The mapping acts on the momenta appearing in the antenna $X$,
\begin{align}
    \{p\} \to \{\wt{p}\}
\end{align}
(where there are $n_{\wt{p}}$ hard radiators in the set $\{\wt{p}\}$) in a way that
preserves momentum conservation and on-shellness
but leaves each momentum in the set $\{q\}$ unaffected,
\begin{align}
    q \to \wt{q} \equiv q.
\end{align}
The jet function $J$ selects $n_j$ jets from the final-state partons. 
The matrix elements $M$ and the associated jet function $J$ depend on the mapped momentum set.

In general, we aim to select a mapping such that the phase space factorises,
\begin{align}
    dPS_{n_p+n_q}({p},{q}) \to dPS_X({p}/\{\wt{p}\}) \,
    dPS_{n_{\wt{p}}+n_q}
    (\{\wt{p}\},\{\wt{q}\})
\end{align}
and that the integration over the antenna phase space $dPS_X$ can be performed,
\begin{align}
    \int X(\{p\}) \,  dPS_X({p}/\{\wt{p}\}) 
    = {\cal X} (\{\wt{p}\}),
\end{align}
so that the integrated subtraction term is given by, 
\begin{align}
\label{eq:integratedterm}
    {\cal X} (\{\wt{p}\}) \,
    M(\{\wt{p}\},\{\wt{q}\}) \, 
    J^{(n_{\wt{p}}+n_q)}_{n_j}(\{\wt{p}\},\{\wt{q}\}) \,
    dPS_{n_{\wt{p}}+n_q}(\{\wt{p}\},\{\wt{q}\}).
\end{align}
Two mappings are \textit{equivalent} if ${\cal X} (\{\wt{p}\})$ is the same for both mappings. 

\subsubsection*{Antenna functions with two hard radiators}

When there are two hard radiators, the generic mapping has the form,
\begin{align}
    \{p_{1},\ldots,p_{n_p}\} \to \{\wt{p}_{I},\wt{p}_{J}\}.
\end{align}
The only available scale is the invariant mass of the antenna, \begin{align}
    s_{1\ldots n_p} \equiv s_{IJ}
\end{align}
which, because of momentum conservation, is the same for all possible mappings.
On dimensional grounds the result of the integration of the antenna over the antenna phase space must satisfy,
\begin{align}
\label{eq:calX2}
    {\cal X} (\{\wt{p}_{I},\wt{p}_{J}\}) = c (\e) \, (s_{IJ})^{d}
\end{align}
where $d$ is a constant fixed by the dimensionality of the antenna and the antenna phase space, and $c(\e)$ is a constant.
It is straightforward to check that different $n_p\to 2$ mappings (that preserve momentum conservation and on-shellness) always produces the same values of $d$ and $c(\e)$.
For this reason, one has the freedom to choose different mappings for the analytical integration, where simplicity is preferred, and the numerical implementation, for which regularity across the phase space is better suited. 
In the two-hard-radiator case, the numerically integrated real-emission subtraction terms are then guaranteed to match and cancel the analytically integrated ones added back in virtual corrections. 

\subsubsection*{Antenna functions with three hard radiators}

For antenna functions with three hard radiators, the mapping has the form
\begin{align}
    \{p_{1},\ldots,p_{n_p}\} \to \{\wt{p}_{I},\wt{p}_{J},\wt{p}_{K}\}.
\end{align}
In this case, there are multiple scales available,
\begin{align}
    s_{IJ}, \qquad s_{IK}, \qquad s_{JK}, \qquad s_{IJK},
\end{align}
as well as composite scales, e.g.,
\begin{align}
    s_{IJ}+s_{JK}, \qquad s_{IJ}+s_{IK}, \qquad s_{JK}+s_{IK}
\end{align}
so that, while the overall dimensionality of ${\cal X}$ is fixed (and mapping independent), the dependence on the individual scales is not. In general,
\begin{align}
\label{eq:calX3}
    {\cal X} (\{\wt{p}_{I},\wt{p}_{J},\wt{p}_{K}\}) = \sum_i c_i \, (s_{IJ})^{\alpha_i} (s_{JK})^{\beta_i}(s_{JK})^{\gamma_i} (s_{IJK})^{\delta_i} + \ldots
\end{align}
where $c_i$ is a scaleless function of ratios of scales and $d = \alpha_i+\beta_i+\gamma_i+\delta_i$. Here $+\ldots$ represents possible contributions involving composite scales. Different mappings can lead to different values of $\{c_i,\alpha_i,\beta_i,\gamma_i,\delta_i\}$, and 
therefore it is clear that 
${\cal X} (\{\wt{p}_{I},\wt{p}_{J},\wt{p}_{K}\})$
is not necessarily the same for all mappings. For the cases we investigated, each mapping gave a different dependence on the scales.  It may be possible to find classes of mappings that lead to the same value of ${\cal X} (\{\wt{p}_{I},\wt{p}_{J},\wt{p}_{K}\})$.  However, to be guaranteed to avoid any mismatch between real and virtual subtraction terms, one must use exactly the same mapping for the numerical implementation as for the analytic integration. 

It follows from the observations above that the momentum mapping for the new antenna functions will be more constrained than the traditional one. For clarity, in the remainder of this paper, we will denote with $\wt{ij}$ or $\wt{ijk}$ momenta reconstructed applying a single- or double-unresolved antenna mapping~\cite{Kosower:2002su}. Where alternative choices are required, we will introduce dedicated notation.

\section{Tree-level five-particle three-hard-radiator antenna functions}\label{sec:X53}

The main objective of this Section is the construction of five-particle antenna functions $\X(i^h_a,j_b,k_c^h,l_d,m_e^h)$, aimed at describing the singular behaviour of the configuration in Figure~\ref{fig:X530fig}. The particle types are denoted by $a$, $b$, $c$, $d$ and $e$, and carry four-momenta $p_i$, $p_j$, $p_k$, $p_l$ and $p_m$ respectively. Particles $a$, $c$ and $e$ are the three hard radiators and the antenna functions must have the correct limits when particles $b$ and $d$ become unresolved. 

\newcommand{\qt}{\wt{q}}
\newcommand{\qbt}{\wt{\qb}}
\newcommand{\Qbt}{\wt{\Qb}}
\newcommand{\Qt}{\wt{Q}}
\newcommand{\gt}{\wt{g}}

\subsection{Unresolved limits}

The starting point for the application of the designer antenna algorithm~\cite{Braun-White:2023sgd} consists in a list of infrared limits a given antenna functions needs to reproduce. For the $X_{5,3}^0(i^h_a,j_b,k^h_c,l_d,m^h_e)$ antenna functions, these limits occur when momenta $j$, $l$ are either soft or collinear with the hard radiator momenta $i$, $k$, $m$. In particular, we have: double-soft $j$, $l$, triple-collinear $j||k^h||l$, double-collinear $i^h||j$, $k^h||l$, $i^h||j$ and $l||m^h$, $j||k^h$ and $l||m^h$, single-soft $j$ and $l$, and single-collinear $i^h||j$, $j||k^h$, $k^h||l$  and $l||m^h$.  In addition, there are the four soft-collinear limits, when $j$ is soft with either $k^h||l$ or $l||m^h$ and when $l$ is soft with either $i^h||j$ or $j||k^h$.  In the designer antenna approach~\cite{Braun-White:2023sgd} these soft-collinear limits are automatically satisfied.  The list of eleven limits for $X_{5,3}^0(i^h_a,j_b,k^h_c,l_d,m^h_e)$ in the notation of~\cite{Braun-White:2023sgd} is then given by:
\begin{align}
\label{eq:X53limits}
    L_{1}(i^{h},j,k^{h},l,m^{h})  =& S_{b}(i^{h},j,k^{h}) S_{d}(k^{h},l,m^{h})   &&\quad \text{double-soft $j$, $l$}\nn\\
    L_{2}(i^{h},j,k^{h},l,m^{h})  =& P_{ab}(i^{h},j)P_{cd}(k^{h},l)              &&\quad \text{double-collinear $i\parallel j$, $k\parallel l$} \nn\\
    L_{3}(i^{h},j,k^{h},l,m^{h})  =& P_{ab}(i^{h},j)P_{ed}(m^{h},l)              &&\quad \text{double-collinear $i\parallel j$, $l\parallel m$} \nn\\
    L_{4}(i^{h},j,k^{h},l,m^{h})  =& P_{cb}(k^{h},j)P_{ed}(m^{h},l)              &&\quad \text{double-collinear $b\parallel c$, $l\parallel m$} \nn\\
    L_{5}(i^{h},j,k^{h},l,m^{h})  =& S_b(i^{h},j,k^{h})X_{3}^{0}(k^{h},l,m^{h})  &&\quad \text{single-soft $j$} \nn\\
    L_{6}(i^{h},j,k^{h},l,m^{h})  =& S_d(k^{h},l,m^{h})X_{3}^{0}(i^{h},j,k^{h})  &&\quad \text{single-soft $l$} \nn\\
    L_{7}(i^{h},j,k^{h},l,m^{h})  =& P_{ab}(i^{h},j)X_{3}^{0}(k^{h},l,m^{h})     &&\quad \text{single-collinear $i\parallel j$} \nn\\
    L_{8}(i^{h},j,k^{h},l,m^{h})  =& P_{ed}(m^{h},l)X_{3}^{0}(i^{h},j,k^{h})     &&\quad \text{single-collinear $l\parallel m$} \nn\\
    L_{9}(i^{h},j,k^{h},l,m^{h})  =& P_{bcd}(j,k^{h},l)                          &&\quad \text{triple-collinear $j\parallel k \parallel l$} \nn\\
    L_{10}(i^{h},j,k^{h},l,m^{h}) =& P_{bc}(k^{h},j)X_{3}^{0}((j+k)^{h},l,m^{h}) &&\quad \text{single-collinear $j\parallel k$} \nn\\
    L_{11}(i^{h},j,k^{h},l,m^{h}) =& P_{cd}(k^{h},l)X_{3}^{0}(i^{h},j,(k+l)^{h}) &&\quad \text{single-collinear $k\parallel l$}.
\end{align}
The tree-level soft factor $\Sb$ is given by the eikonal for particle $b$ radiated between two hard radiators:
\begin{align}
    S_g(i^h,j_g,k^h) &= S_{\gamma}(i^h,j_{\gamma},k^h) =   \frac{2s_{ik}}{s_{ij}s_{jk}},  \\
    S_q(i^h,j_q,k^h) &= S_{\qb}(i^h,j_{\bar{q}},k^h) =0,
\end{align}
where we use Lorentz-invariant momentum structures
\begin{equation}
	s_{i,\ldots,n} \equiv (p_{i}+...+p_{n})^2.
\end{equation}
The splitting functions $P_{ab}(i^h,j)$ are {\em not singular} in the limit where the hard radiator $a$ becomes soft and are related to the usual spin-averaged splitting functions, cf.~\cite{Altarelli:1977zs,Dokshitzer:1977sg}, by, 
\begin{align} 
\label{eqn:Pqg}
\Pqg(i^h,j) &= \frac{1}{s_{ij}} \Pqg(\xj) \\
\Pqg(i,j^h) &= 0,\\
\label{eqn:Pqq}
\Pqq(i^h,j) &= \frac{1}{s_{ij}} \Pqq(\xj),\\
\Pqq(i,j^h) &= \frac{1}{s_{ij}} \Pqq\omxj,\\
\label{eqn:Pgg}
\Pgg(i^h,j) &= \frac{1}{s_{ij}} \PggS(\xj)\hfill\\
\Pgg(i,j^h) &= \frac{1}{s_{ij}}\PggS\omxj
\end{align}
with
\begin{eqnarray}
\Pqg(\xj) &=& \left(\frac{2\omxj}{\xj} + \ome \xj \right)\\
\Pqq(\xj) &=& \left( 1 -\frac{2\omxj\xj}{\ome} \right) = \Pqq\omxj\\
\label{eq:PggS}
\PggS(\xj)&=& \left( \frac{2\omxj}{\xj} + \xj \omxj \right) 
\end{eqnarray}
and
\begin{equation}
\PggS(\xj) + \PggS\omxj \equiv \Pgg(\xj).
\end{equation}
Here, the momentum fraction $\xj$ is defined with reference to one of the other particles in the antenna, e.g. 
\begin{equation}
\xj = \dfrac{p_j\cdot p_k}{(p_i+p_j)\cdot p_k} = \dfrac{s_{jk}}{(s_{ik}+s_{jk})}. 
\end{equation}
Finally, $P_{bcd}$ denotes a triple-collinear splitting function~\cite{campbell,Catani:1998nv,Braun-White:2022rtg}. For explicit expressions see, for example~\cite{Braun-White:2022rtg}.
Of course, not all limits are required or will be present for all particle types.  For example, if $b$ (or $d$) is a quark there is no singular behaviour when its momentum become soft, hence no soft limit associated to it.

In single-unresolved limits ($L_5$--$L_8$, $L_{10}$, $L_{11}$) we require that $X_{5,3}^0(i^h_a,j_b,k^h_c,l_d,m^h_e)$ factorises onto three-parton tree-level antenna functions. We recall that an individual $X_3^0(i_a,j_b,k_c)$ antenna function describes the $b$ soft limit and the $a^h\parallel b$ or $b\parallel c^h$ collinear limits.  In the designer antenna approach, these are denoted as:
\begin{equation}
\begin{split}
    L_1(i^h,j,k^h) &= \Sb(i^h,j,k^h) \, , \\
    L_2(i^h,j;k) &= P_{ab}(i^h,j) \, , \\
    L_3(k^h,j;i) &= P_{cb}(k^h,j) \, .
\end{split}
\label{eq:NLOL}
\end{equation}

We note that the product of two single-unresolved antenna functions,
\begin{equation}
\label{eq:X3X3}
    X_{3}^{0}(i^h_a,j_b,k^h_c) X_{3}^{0}(k^h_c,l_d,m^h_e)
\end{equation}
correctly describes singular limits when $b$ and $d$ are unresolved and emitted between pairs of hard radiators $(a,c)$ and $(c,e)$. Namely, Eq.~\eqref{eq:X3X3} captures a significant subset of the limits that are expected for the $\X$. However, it does not get every limit correct. More precisely, every limit involving $c$ being collinear with any other parton is only partially described, because $c$ is shared between both antenna functions. In other words, Eq.~\eqref{eq:X3X3} gives limits $L_1$--$L_8$ in Eq.~\eqref{eq:X53limits} exactly, while only partially describes limits $L_9$--$L_{11}$, given that they involve collinear configurations with the shared hard radiator $c$. It is nevertheless advantageous to consider Eq.~\eqref{eq:X3X3} as the starting point for the construction of the new antenna functions.

We define a {\em middle} component of $\X$ (that we denote as $\XM$) that fully satisfies limits $L_1$--$L_8$
and gives no purely collinear (i.e. not soft) contribution to limits $L_9$--$L_{11}$ by removing the collinear limits with the shared hard radiator $c$:
\begin{align}
\label{eq:XM}
    \XM 
    &\equiv X_{3}^{0}(i^h_a,j_b,k^h_c) X_{3}^{0}(k^h_c,l_d,m^h_e) 
    \nonumber \\ 
    & \hspace{1cm} -
    \PCdown_{jk} ((1-\PSdown_{j})X_{3}^{0}(i^h_a,j_b,k^h_c))
    \PCdown_{kl} ((1-\PSdown_{l})X_{3}^{0}(k^h_c,l_d,m^h_e)).
\end{align} 
In Eq.~\eqref{eq:XM}, the combination 
$\PCdown_{jk} ((1-\PSdown_{j})X_{3}^{0}(i^h_a,j_b,k^h_c))$ selects the non-soft $j$ part of the antenna function when $j||k$. In general each $X_3^0$ antenna contains three singular limits, leading to nine terms in the product. After removing the collinear (but not soft) component, eight terms remain in $\XM$.

We note that $\XM$ naturally involves all five particles. However, each term in $\XM$ is composed of invariants made from the momenta in one antenna multiplied by invariants made from the momenta in the other.  No term involves an invariant that spans momenta from one three-parton antenna function to the other. This feature will turn out to be very useful in constructing suitable momentum mappings and in performing the analytic integration of the $\XM$ antenna function.

The remaining limits only affect a subset of the five momenta.  The triple-collinear $j\parallel k^{h}\parallel l$ limit $L_9$ directly involves the three collinear momenta, plus a spectator to determine the momentum fractions, which can be either $i$ or $m$. 
The single-collinear limit $L_{10}$ involves four momenta $j,\ldots ,m$ (which all appear in the argument of the $X_{3}^{0}$), provided that the spectator for the single-collinear splitting is chosen to be $m$.  Similarly, the single-collinear limit $L_{11}$ involves four momenta $i,\ldots ,l$ when $i$ is chosen as spectator to define the momentum fractions of the splitting. 

We therefore break the $\X$ into three parts:
\begin{align}
    \X(i^h,j,k^h,l,m^h) &= \XM(i^h,j,k^h,l,m^h) \nn\\
                        &+ \XL(i^h,j,k^h,l,m^h) \nn\\
                        &+ \XR(i^h,j,k^h,l,m^h),
\end{align}
where $\XM$ satisfies limits $L_1$--$L_8$, while $\XL$ (\textit{left} component) and $\XR$ (\textit{right} component) are built from limits $L_9$--$L_{11}$. $\XL$ depends on momenta $i,j,k,l$ and not on momentum $m$, while $\XR$ depends on momenta $j,k,l,m$ and not on momentum $i$. To assemble $\XL$ and $\XR$, we follow closely the procedure detailed in Ref.~\cite{Braun-White:2023sgd} for the construction of four-parton antenna functions. We systematically ensure that combination $\X$ satisfies Eq.~\eqref{eq:X53limits} by injecting $\XM$ as the starting point of the designer algorithm and then working through limits $L_9$--$L_{11}$ sequentially to avoid any double-counting of singular behaviour. 
Limits $L_9$--$L_{11}$ introduce denominators $s_{jk}$, $s_{kl}$ and $s_{jkl}$. Terms that have denominators $s_{jk}$ are assigned to $\XR$ (using $m$ as the reference momentum for defining the collinear momentum fractions), while those that have denominators $s_{kl}$ are assigned to $\XL$ (using $i$ as the reference momentum). The $\XL$ antenna function depends functionally on the first four momenta, while the $\XR$ antenna function depends functionally on the final four momenta.
$\XL$ therefore describes part of $L_9$ and $L_{10}$ while $\XR$ describes the remainder of $L_9$ and $L_{11}$.

\subsection{Momentum mappings}
\label{sec:Xmappings}
Within a local subtraction term, antenna functions induce a momentum mapping from the real-emission phase space to the reduced phase space obtained by redistributing the momenta of the additional emissions among the ones of the hard particles. In this Section we define the appropriate double-unresolved mapping to be used with the new antenna functions. Each of the three components introduced above ($\XM$, $\XL$ and $\XR$) is associated with a different 
$\{i,j,k,l,m\}  \to \{I,K,M \}$ mapping.
Each mapping must: 
\begin{itemize}
    \item[(a)] preserve both momentum conservation and on-shellness conditions;
    \item[(b)] behave correctly in the limits relevant to that part of the $\X$ antenna;
    \item[(c)] be analytically integrable over the phase space of the unresolved emissions.
\end{itemize} 
To be precise, point $(c)$ is not strictly necessary to achieve local subtraction, but it is mandatory to have analytical control over the integrated form of the subtraction terms and avoid potentially demanding point-by-point numerical integrations. 

\subsubsection{Momentum mapping for $\XM$: $\mapM$}
For the $\XM$ part of $\X$, we use the mapping, $\mapM$,
\begin{alignat}{2}
\label{eq:X53Mmapping}
                       &p_I  =&   p_i+p_j - \frac{s_{ij}}{s_{ik}+s_{jk}} p_k, \nonumber \\
\mapM: \hspace{1cm}    &p_K  =&  \left(1+ \frac{s_{ij}}{s_{ik}+s_{jk}} +\frac{s_{lm}}{s_{lk}+s_{mk}} \right)p_k, \\ 
                       &p_M  =& p_l+p_m - \frac{s_{lm}}{s_{lk}+s_{mk}} p_k, \nonumber 
\end{alignat}
which clearly satisfies both momentum conservation,
\begin{align}
    p_I + p_K + p_M = p_i+p_j+p_k+p_l+p_m,
\end{align}
and on-shellness,
\begin{align}   p_I^2=p_K^2=p_M^2=0.
\end{align}  
This mapping amounts to the iterated use of a dipole mapping~\cite{Catani:1996vz}, first on the triplet $(i,j,k)$ with momentum $k$ rescaled, and then on the triplet $(k,l,m)$, again with momentum $k$ rescaled. The final mapping is independent of the order of the individual dipole mappings. To make our notation for this mapping explicit, we will use the following convention:
\begin{align}
    I &\equiv [ij\u{k}],\qquad \qquad
    K \equiv [ij\u{k}lm],\qquad \qquad
    M \equiv [\u{k}lm],   
\end{align}
where the momentum that is rescaled is underscored.

 $\XM$ describes $L_1$--$L_8$ of Eq.~\eqref{eq:X53limits}, therefore, the $\phi_M$ mapping must collapse appropriately in these limits. Indeed, we find that
\allowdisplaybreaks 
\begin{align*}
\label{eq:mapMlimits}
\{I, K, M\}& \xrightarrow[]{\text{$j$,$l$ soft}} \{i,k,m\},\\
\{I, K, M\}& \xrightarrow[]{\text{$i||j$ and $k||l$}} \{i+j,k+l,m\},\\
\{I,K,M\}& \xrightarrow[]{\text{$i||j$ and $l||m$}} \{i+j,k,l+m\},\\
\{I,K,M\}& \xrightarrow[]{\text{$j||k$ and $l||m$}} \{i,j+k,l+m\},\\
\{I,K,M\}& \xrightarrow[]{\text{$j$ soft}} \{i,[\u{k}l],[lm]\},\\
\{I,K,M\}& \xrightarrow[]{\text{$l$ soft}} \{[ij],[j\u{k}],m\},\\
\{I,K,M\}& \xrightarrow[]{\text{$i||j$}} \{i+j,[\u{k}l],[lm]\},\\
\{I,K,M\}& \xrightarrow[]{\text{$l||m$}} \{[ij],[j\u{k}],l+m\}.
\end{align*}
Here $\{[ij],[j\u{k}]\}$ denotes the result of a dipole mapping where $k$ is the scaled momentum,
\begin{align}
    p_{[ij]} &= p_i + p_j - \frac{s_{ij}}{s_{ik}+s_{jk}} p_k, \nonumber \\
    p_{[j\u{k}]} &= \phantom{p_i+p_j-}  \frac{s_{ijk}}{s_{ik}+s_{jk}} p_k.
\end{align}
Similarly, $\{[\u{k}l],[lm]\}$ is obtained by a dipole mapping where $k$ is the scaled momentum.
In addition, the limits where either $j$ or $l$ are soft and one other pair is collinear are also correctly described,
\begin{align*}
\{I,K,M\}& \xrightarrow[]{\text{$j$ soft $k||l$}} \{i,k+l,m\},\\
\{I,K,M\}& \xrightarrow[]{\text{$j$ soft $l||m$}} \{i,k,l+m\},\\
\{I,K,M\}& \xrightarrow[]{\text{$l$ soft $i||j$}} \{i+j,k,m\},\\
\{I,K,M\}& \xrightarrow[]{\text{$l$ soft, $j||k$}} \{i,j+k,m\}.
\end{align*}

\subsubsection{Momentum mapping for $\XL$: $\mapL$}

For the $\XL$ part of $\X$, we use the tripole mapping~\cite{Gehrmann-DeRidder:2003pne}, $\mapL$,
\begin{alignat}{2}
\label{eq:X53Lmapping}
    & p_I =&  p_{i}+p_{j}+p_{l}-\frac{s_{ijl}}{s_{ik}+s_{jk}+s_{kl}}p_{k} \nonumber \\
\mapL: \hspace{1cm}    &    p_K  =&  \frac{s_{ijkl}}{s_{ik}+s_{jk}+s_{kl}}p_{k} \\
   & p_M  = & p_m, \nonumber
\end{alignat}
which clearly satisfies both momentum conservation,
\begin{align}
    p_I + p_K + p_M = p_i+p_j+p_k+p_l+p_m,
\end{align}
and on-shellness,
\begin{align}   p_I^2=p_K^2=p_M^2=0.
\end{align}  
To make our notation for this mapping explicit, we will use the convention:
\begin{align}
    I &\equiv [ijl],\qquad \qquad
    K \equiv [j\u{k}l],\qquad \qquad
    M \equiv m,   
\end{align}
where the momentum that is rescaled is underscored.

$\XL$ describes the triple-collinear limit $L_{9}$ as well as the single-collinear $k||l$ limit $L_{11}$.  In this case, we find that
for the $\mapL$ mapping, 
\begin{align*}
\{I,K,M\}& \xrightarrow[]{\text{$j||k||l$}} \{i,j+k+l,m\}\\
\{I,K,M\}& \xrightarrow[]{\text{$k||l$}} \{[ij],[j\u{(k+l)}],m\}.
\end{align*}
Here $\{[ij],[j\u{(k+l)}]\}$ denotes the result of a dipole mapping where the collinear momentum $(k+l)$ satisfies $(k+l)^2 = 0$ and is rescaled:
\begin{align}
    p_{[ij]} &= p_i + p_j - \frac{s_{ij}}{s_{ik}+s_{il}+s_{jk}+s_{jl}} (p_k+p_l), \nonumber \\
    p_{[j\u{(k+l)}]} &= \phantom{p_i+p_j-} \frac{s_{ij}+s_{ik}+s_{il}+s_{jk}+s_{jl}}{s_{ik}+s_{il}+s_{jk}+s_{jl}} (p_k+p_l).
\end{align}

\subsubsection{Momentum mapping for $\XR$: $\mapR$}

The mapping $\mapR$ is obtained by line-reversal $\{ i \leftrightarrow m, j \leftrightarrow l \}$ from $\mapL$.  By symmetry, the $\mapR$ mapping correctly describes the $j||k||l$ and $j||k$ limits.

\subsection{Summary}\label{sec:summary_X530}
Our notation for the $\X$ antenna functions is summarised in Table.~\ref{tab:5to3mappings}.
Putting together the different components, we can write down the complete implementation of the $\X$ antenna function in relation to its respective reduced matrix element within a double-unresolved subtraction term:
\begin{align}
\label{eq:Xsplit}
\X(i,j,k,l,m) &M_n^{0} (\ldots,\{ijk\},\{ijklm\},\{klm\},\ldots)
    \equiv \nn \\
 ~\qquad   &\XL(i,j,k,l,m) M_n^{0} (\ldots,[ijl],[j\u{k}l],m,\ldots
    )\nn \\
+    &\XM(i,j,k,l,m) M_n^{0} (\ldots,[ij\u{k}],[ij\u{k}lm],[\u{k}lm],\ldots)\nn \\
+    &\XR(i,j,k,l,m) M_n^{0} (\ldots,i,[j\u{k}l],[jlm],\ldots) . 
\end{align}
\begin{table}[h]
\centering
\begin{tabular}{ccccc}
& & I & K & M \\ 
\multirow{4}{*}{$\X(i,j,k,l,m)$}
&generic mapping & $\{ijk\}$ & $\{ijklm\}$ & $\{klm\}$ \\
&$\XM$ mapping & $[ij\u{k}]$ & $[ij\u{k}lm]$ & $[\u{k}lm]$ \\
&$\XL$ mapping & $[ijl]$ & $[j\u{k}l]$ & $m$ \\
&$\XR$ mapping & $i$ & $[j\u{k}l]$ & $[jlm]$ \\
\end{tabular}
\caption{Notation for the mappings for the $\X$ antenna functions with three hard radiators. Details of the $\XM$ ($X^0_{5,3;L(R)}$) mappings are given in Eq.~\eqref{eq:X53Mmapping}  \eqref{eq:X53Lmapping} respectively. }
\label{tab:5to3mappings}
\end{table}
In Table~\ref{tab:X530} we list all the $\X$ antenna functions required for an NNLO calculation up to line reversal, charge conjugation and quark-flavour swaps. We organize them according to the partonic species of the \textit{reconstructed hard partons}, namely the constituents of the underlying resolved three-particle configuration in infrared limits. We distinguish gluon-gluon-gluon, quark-gluon-gluon, quark-antiquark-gluon and quark-antiquark-quark antenna functions. Within these classes, a specific $\X$ has a particular partonic content and a particular order of reconstructed hard partons. The names of the new antenna functions are chosen to follow the conventions for traditional antenna functions~\cite{antennafull} as closely as possible. Where needed, we introduced dedicated labels ($a$, $b$, \ldots) to distinguish $\X$ antenna functions which differ by a rearrangement of their partonic content. Finally, we indicate what is the pair of $X_3^0$ antenna functions needed to assemble the $\XM$ component, according to~\eqref{eq:XM}. We notice there are no $\X$ antenna functions with two pairs of identical-flavour quarks. This is due to the fact that the infrared limits associated to a string like $(i_a^h,j_q,k^h_{\qb},l_q,m^h_{\qb})$, where $a$ indicates a generic parton, are almost entirely covered by the $(i_a^h,j_q,k^h_{\qb},l_Q,m^h_{\Qb})$ configuration. The only additional limit is represented by the $j_q\parallel k^h_{\qb} \parallel l_q$ triple-collinear limit, which can easily be accommodated by a $C_4^0$ antenna function~\cite{Gehrmann-DeRidder:2005btv}. Hence, there is no need to introduce dedicated $\X$ antenna functions.

We give the explicit expressions of all the $\X$ antenna functions in ancillary files. In Appendix~\ref{app:limX53} we illustrate the infrared limits of each antenna function in Table~\ref{tab:X530}.

\begingroup
\renewcommand{\arraystretch}{1.5}
\begin{table}[h]
    \centering
\begin{tabular}{c|ccc}
    \textbf{Type} & \thead{\textbf{Name and} \\ \textbf{parton content}} & \thead{\textbf{Reconstructed} \\ \textbf{hard partons}} & $X_3^0 \otimes X_3^0$\\
    \hline
    quark-antiquark-gluon
    & \begin{tabular}{@{}c@{}} $\A(i^h_{q},j_{g},k^h_{g},l_{g},m^h_{\qb})$ \\ $B_{5,3}^{0}(i^h_{\qb},j_{g},k^h_{Q},l_{\Qb},m^h_{q})$ \\ $\At(i^h_{\qb},j_{\gamma},k^h_{q},l_{g},m^h_{g})$ \\ $\wt{B}^0_{5,3}(i^h_{\qb},j_{\gamma},k^h_{q},l_{\Qb},m^h_{Q})$\end{tabular}
    & \begin{tabular}{@{}c@{}} $q g \bar{q}$ \\ $q g \bar{q}$ \\ $\bar{q} q g$ \\ $\bar{q} q g$ \end{tabular} 
    & \begin{tabular}{@{}c@{}} $D\otimes \overline{D}$ \\ $A\otimes \overline{E}$ \\ $A \otimes D$ \\ $A \otimes E$ \end{tabular} \\ 
    \hline
    quark-antiquark-quark
    & \begin{tabular}{@{}c@{}} $\Att(i^h_{q},j_{\gamma},k^h_{\qb},l_{g},m^h_{Q})$ \end{tabular}
    & \begin{tabular}{@{}c@{}} $q \qb Q$ \end{tabular} 
    & \begin{tabular}{@{}c@{}} $A \otimes A$ \end{tabular} \\
    \hline
    quark-gluon-gluon
    & \begin{tabular}{@{}c@{}} $\D(i^h_{q},j_{g},k^h_{g},l_{g},m^h_{g})$ \\ $E_{5,3}^{0(a)}(i^h_{q},j_{\Qb},k^h_{Q},l_{g},m^h_{g})$ \\ $E_{5,3}^{0(b)}(i^h_{q},j_{g},k^h_{\Qb},l_{Q},m^h_{g})$ \\ $E_{5,3}^{0(c)}(i^h_{q},j_{g},k^h_{g},l_{\Qb},m^h_{Q})$ \\ $E_{5,3}^{0,(d)}(i^h_{Q},j_{\Qb},k^h_{q},l_{g},m^h_{g})$ \\ $K^0_{5,3}(i^h_{q},j_{\Qb},k^h_{Q},l_{\bar{R}},m^h_{R})$  \end{tabular}
    & \begin{tabular}{@{}c@{}} $q g g$ \\ $q g g$ \\ $q g g$ \\ $q g g$ \\ $g q g$ \\ $q g g$ \end{tabular} 
    & \begin{tabular}{@{}c@{}} $D\otimes F$ \\ $E\otimes D$ \\ $A\otimes \overline{G}$ \\ $D\otimes G $ \\ $\overline{E} \otimes D$ \\ $E\otimes E$ \end{tabular} \\
    \hline
    gluon-gluon-gluon 
    & \begin{tabular}{@{}c@{}} $\F(i^h_g,j_g,k^h_g,l_g,m^h_g)$ \\ $G_{5,3}^{0 (a)}(i^h_{\qb},j_{q},k^h_{g},l_{g},m^h_{g})$  \\ $G_{5,3}^{0 (b)}(i^h_{g},j_{\qb},k^h_{q},l_{g},m^h_{g})$ \\ $H_{5,3}^{0 (a)}(i^h_{\Qb},j_{Q},k^h_{g},l_{\qb},m^h_{q})$ \\ $H_{5,3}^{0 (b)}(i^h_{g},j_{\Qb},k^h_{Q},l_{\qb},m^h_{q})$ \end{tabular}
    & \begin{tabular}{@{}c@{}} $g g g$ \\ $g g g$ \\ $g g g$ \\ $g g g$ \\ $g g g$ \end{tabular} 
    & \begin{tabular}{@{}c@{}} $F\otimes F$ \\ $\overline{G} \otimes F$ \\ $G\otimes D$ \\ $\overline{G}\otimes G$ \\ $G\otimes E$ \end{tabular} \\
\end{tabular}
    \caption{List of $\X$ antenna functions, organised by the partonic species of the reconstructed hard partons of the underlying resolved three-particle configuration. The symbols $q$ ($\qb$), $Q$ ($\Qb$), $R$, ($\bar{R}$) represent different flavours of quarks (antiquarks). We use the symbol $\gamma$ to denote a photon or an abelian gluon. The $X_3^0 \otimes X_3^0$ column refers to the assemblage of $\XM$ in~\eqref{eq:XM}, with the notation $\overline{X}_3^0(i_a,j_b,k_c)=X_3^0(k_c,j_b,i_a)$. Note that $\overline{A} = A$, $\overline{F} = F$.}
    \label{tab:X530}
\end{table}
\endgroup
We note that because $E_3^0(i,j,k)$ and $G_3^0(i,j,k)$ only contain the limit where $j$ and $k$ are collinear, with parton $i$ being a mere spectator whose nature is irrelevant,  we can identify the following relations between $\X$ antenna functions:
\begin{align}
    B_{5,3;i}^{0}(i,j,k,l,m) &\equiv E_{5,3;i}^{0 (b)}(i,j,k,l,m),\\
    K_{5,3;i}^{0}(i,j,k,l,m) &\equiv H_{5,3;i}^{0 (b)}(i,j,k,l,m), \\
    E_{5,3;i}^{0(a)}(i,j,k,l,m) &\equiv G_{5,3;i}^{0 (b)}(i,j,k,l,m),     
\end{align}
for $i = M,L,R$.
Furthermore, the line reversal symmetry of the antenna induces additional relations between the three components of the same $\X$ antenna functions,
\begin{align}
 A_{5,3;M}^{0}(i,j,k,l,m) &\equiv A_{5,3;M}^{0}(m,l,k,j,i),\\ 
 A_{5,3;L}^{0}(i,j,k,l,m) &\equiv A_{5,3;R}^{0}(m,l,k,j,i),\\
 \wt{\wt{A}}_{5,3;M}^{0}(i,j,k,l,m) &\equiv \wt{\wt{A}}_{5,3;M}^{0}(m,l,k,j,i),\\
 \wt{\wt{A}}_{5,3;L}^{0}(i,j,k,l,m) &\equiv \wt{\wt{A}}_{5,3;R}^{0}(m,l,k,j,i),\\
 F_{5,3;M}^{0}(i,j,k,l,m) &\equiv F_{5,3;M}^{0}(m,l,k,j,i),\\
F_{5,3;L}^{0}(i,j,k,l,m) &\equiv F_{5,3;R}^{0}(m,l,k,j,i),
\end{align}
while the particle content implies further relations between the $L$ and $R$ parts of some different $\X$ antenna functions,
\begin{align}
 A_{5,3;L}^{0}(i,j,k,l,m) &\equiv D_{5,3;L}^{0}(i,j,k,l,m),\\
 \wt{A}_{5,3;L}^{0}(i,j,k,l,m) &\equiv \wt{\wt{A}}_{5,3;L}^{0}(i,j,k,l,m),\\
 F_{5,3;R}^{0}(i,j,k,l,m) &\equiv D_{5,3;R}^{0}(i,j,k,l,m),
 \\
 E_{5,3;L}^{0(a)}(i,j,k,l,m) &\equiv E_{5,3;R}^{0(b)}(m,l,k,j,i),\\
 G_{5,3;L}^{0(a)}(i,j,k,l,m) &\equiv E_{5,3;R}^{0(c)}(m,l,k,j,i),\\
 E_{5,3;L}^{0(d)}(i,j,k,l,m) &\equiv \wt{B}_{5,3;R}^{0}(m,l,k,j,i). 
\end{align}

\subsection{Link to $\wt{X}_4^0$ antenna functions}
\label{subsec:Xt40}

As discussed in the introduction, almost colour-connected limits are also present for processes with only four particles at the double-real level. The $\wt{X}_4^0$ antenna functions describing these limits were constructed directly from the relevant infrared limits in Ref.~\cite{Braun-White:2023sgd}.  However, it is interesting to see whether these special cases can also be constructed using the same approach as for the $\X$ antenna functions.

The list of limits for the
$\wt{X}_4^0(i^h_a,j_b,l_d,k^h_c)$ antenna is given by~\cite{Braun-White:2023sgd}
\begingroup
\allowdisplaybreaks
\begin{align}
\label{eq:listX40t}
    \wt{L}_1(i^h,j,l,k^h) &= S_{b}(i^h,j,k^h)S_{d}(i^h,l,k^h) \, , \nn\\
    \wt{L}_2(i^h,j,l,k^h) &= P_{ab}(i^h,j) P_{cd}(k^h,l) \, , \nn\\
    \wt{L}_3(i^h,j,l,k^h) &= P_{ad}(i^h,k) P_{cb}(k^h,j) \, , \nn\\
    \wt{L}_4(i^h,j,l,k^h) &= S_{b}(i^h,j,k^h) \, X_{3}^{0}(i^h,l,k^h) \, , \nn\\
    \wt{L}_5(i^h,j,l,k^h) &= S_{d}(i^h,l,k^h) \, X_{3}^{0}(i^h,j,k^h) \, , \nn\\
    \wt{L}_6(i^h,j,l;k^h) &= P_{abd}(i^h,j,l) \, , \nn\\
    \wt{L}_7(i^h,j,l,k^h) &= P_{ab}(i^h,j) X_{3}^{0}((i+j)^h,l,k^h) \, , \nn\\
    \wt{L}_8(i^h,j,l,k^h) &= P_{ad}(i^h,l) X_{3}^{0}((i+l)^h,j,k^h) \, , \nn\\
    \wt{L}_9(i^h,j,l,k^h) &= P_{cdb}(k^h,l,j) \, , \nn\\
    \wt{L}_{10}(i^h,j,l,k^h) &= P_{cd}(k^h,k) X_{3}^{0}(i^h,j,(l+k)^h) \, , \nn\\
    \wt{L}_{11}(i^h,j,l,k^h) &= P_{cb}(k^h,j) X_{3}^{0}(i^h,l,(k+j)^h) \, .
\end{align}
\endgroup
We see that one can do a similar construction as for the $\X$ antenna, by introducing 
\begin{align}
\label{eq:XMX40t}
    \wt{X}_{4,M}^{0} (i_a^h,j_b,l_d,k_c^h)
    &\equiv X_{3}^{0}(i^h_a,j_b,k^h_c) X_{3}^{0}(k^h_c,l_d,i^h_a) 
    \nonumber \\ 
    & \hspace{1cm} -
    \PCdown_{jk} ((1-\PSdown_{j})X_{3}^{0}(i^h_a,j_b,k^h_c))
    \PCdown_{kl} ((1-\PSdown_{l})X_{3}^{0}(k^h_c,l_d,i^h_a)) \nonumber \\
    & \hspace{1cm} -
    \PCdown_{ij} ((1-\PSdown_{j})X_{3}^{0}(i^h_a,j_b,k^h_c))
    \PCdown_{il} ((1-\PSdown_{l})X_{3}^{0}(k^h_c,l_d,i^h_a))     
\end{align}
which fully satisfies the double-soft, double-collinear and soft-collinear limits $\wt{L}_1$--$\wt{L}_5$. The remaining limits
$\wt{L}_6$--$\wt{L}_{11}$ are the single- and triple-collinear limits that are precisely captured by the $\XL$ and $\XR$ antenna functions. When the particle types are appropriate, we can therefore build an $\wt{X}_{4}^{0}$ function by combining $\wt{X}_{4,M}^{0}$, $\XL$ or $\XR$ functions as, for example,
\begin{align}
\label{eq:F40tconstruct}
  \wt{F}_{4}^{0} (i,j,l,k) &= \wt{F}_{4,M}^{0} (i,j,l,k) \nn\\
  &+ \FL(i,j,k,l,A)  + \FR(A,l,i,j,k)\nn \\
  &+ \FL(i,l,k,j,A)  + \FR(A,j,i,l,k),\\
\label{eq:D40tconstruct}
  \wt{D}_{4}^{0} (i,j,l,k) &= \wt{D}_{4,M}^{0} (i,j,l,k) \nn\\
  &+ \DL(i,j,k,l,A)  + \AtR(A,l,i,j,k)\nn \\
  &+ \DL(i,l,k,j,A)  + \AtR(A,j,i,l,k),
\end{align}
where, because $\XL$ ($\XR$) does not depend on the fifth particle momentum, $A$ is a dummy momentum. 
We have explicitly checked that Eqs.~\eqref{eq:F40tconstruct} and ~\eqref{eq:D40tconstruct} correctly generate the proper infrared singular structure (and the correct infrared poles when integrated) for the $\wt{F}_4^0$ and $\wt{D}_4^0$ antenna functions respectively.

\section{Double-real subtraction terms}\label{sec:RRsub}
As an example of the use of the generalised antenna functions, we consider the process $e^+e^-\to jjj$ at leading colour.
We follow closely the notation in Ref.~\cite{Gehrmann-DeRidder:2007foh} to demonstrate the similarities and most importantly the differences with the original antenna subtraction.

The real-radiation correction to the cross section ${\rm d}\sigma_{NNLO,N^2}^{RR}$ is obtained by summing the contributions of the colour-ordered matrix elements for the six permutations of gluons (labelled here as $i$, $j$ and $k$). The quarks carry momentum $p_1$ and $p_2$ and the gluons carry momentum $p_i$, $p_j$ and $p_k$ in the set $\{p_3,p_4,p_5\}$.
\begin{align}\label{RRLC}
{\rm d}\sigma_{NNLO,N^2}^{RR} &= N_5 N^2 d\Phi_5(\lbrace p\rbrace_{5};q) \frac{1}{3!} \sum_{(i,j,k) \in P(3,4,5)} M_5^{0}(1,i,j,k,2)\,J_3^{(5)}(\lbrace p\rbrace_{5}).
\end{align}
Here we use $M_n^{0}(1,\ldots,2)$ to denote the squared tree-level leading-colour colour-ordered amplitude $|A_n^{0}(1,\ldots,2)|^2$ for the annihilation of an electron and a positrion to two quarks and three gluons, while $N_5$ is an overall normalisation and $d\Phi_5$ the five particle phase space. In general we denote:
\begin{equation}
\label{eq:Nndef}
N_n = 4 \pi  \alpha\,
\sum_q e_q^2 \left(4\pi\alpha_s\right)^{(n-2)}
\left(N^2-1\right)\,.
\end{equation}

The double-real subtraction term using the generalised $\X$ antenna functions is given by,   
\begin{align}\label{eq:RRsubLC}
{\rm d}\sigma_{NNLO,N^2}^{S} &= N_5 N^2 d\Phi_5(\lbrace p\rbrace_{5};q) \frac{1}{3!} \sum_{(i,j,k) \in P(3,4,5)} \Biggl \{\nn \\
\textA{1}\quad &+D_3^{0}(1,i,j)\,M_4^{0}((\wt{1i}),(\wt{ij}),k,2)\,J_3^{(4)}(\lbrace p\rbrace_{4}) \nn\\
\textA{2}\quad &+F_3^{0}(i,j,k)\,M_4^{0}(1,(\wt{ij}),(\wt{kj}),2)\,J_3^{(4)}(\lbrace p\rbrace_{4}) \nn\\
\textA{3}\quad &+D_3^{0}(2,k,j)\,M_4^{0}(1,i,(\wt{jk}),(\wt{k2}))\,J_3^{(4)}(\lbrace p\rbrace_{4}) \nn\\
---&-----------------------\nn\\
\textB{4}\quad &+\,D_4^0(1,i,j,k)\,M_3^{0}((\wt{1ij}),(\wt{kji}),2)\,J_3^{(3)}(\lbrace p\rbrace_{3}) \nn\\
\textB{5}\quad &-\,D_3^{0}(1,i,j)\,D_3^{0}((\wt{1i}),(\wt{ji}),k)\,M_3^{0}((\wt{(\wt{1i})(\wt{ij})}),(\wt{k\wt{(ji})}),2)\,J_3^{(3)}(\lbrace p\rbrace_{3}) \nn\\
\textB{6}\quad &-\,F_3^{0}(i,j,k)\,D_3^{0}(1,(\wt{ij}),(\wt{kj}))\,M_3^{0}((\wt{1\wt{(ij})}),(\wt{(\wt{kj})(\wt{ji})}),2)\,J_3^{(3)}(\lbrace p\rbrace_{3}) \nn\\
&\nn\\
\textB{7}\quad &+\,D_4^0(2,k,j,i)\,M_3^{0}(1,(\wt{ijk}),(\wt{2kj}))\,J_3^{(3)}(\lbrace p\rbrace_{3}) \nn\\
\textB{8}\quad &-\,F_3^{0}(i,j,k)\,D_3^{0}(2,(\wt{jk}),(\wt{ij}))\,M_3^{0}(1,(\wt{(\wt{ij})(\wt{jk})}),(\wt{2\wt{(kj})}))\,J_3^{(3)}(\lbrace p\rbrace_{3}) \nn\\
\textB{9}\quad &-\,D_3^{0}(2,k,j)\,D_3^{0}((\wt{2k}),(\wt{jk}),i)\,M_3^{0}(1,(\wt{i\wt{(jk})}),\wt{((\wt{2k}), (\wt{jk}))})\,J_3^{(3)}(\lbrace p\rbrace_{3}) \nn\\
---&-----------------------\nn\\
\textC{10}\quad &+\,A_{5,3}^0(1,i,j,k;2)\,M_3^{0}((\{1ij\}),(\{1ijk2\}),(\{jk2\})))\,J_3^{(3)}(\lbrace p\rbrace_{3}) \nn\\
\textC{11}\quad &-\,D_3^{0}(1,i,j)\,D_3^{0}(2,k,(\wt{ij}))\,M_3^{0}((\wt{1i}),[\u{(\wt{ji})}k],[k2])\,J_3^{(3)}(\lbrace p\rbrace_{3}) \nn\\
\textC{12}\quad &-\,D_3^{0}(2,k,j)\,D_3^{0}(1,i,(\wt{kj}))\,M_3^{0}([1i],[\u{(\wt{kj})}i],(\wt{2k}))\,J_3^{(3)}(\lbrace p\rbrace_{3}) \Biggr \rbrace ,
\end{align}
where the single-unresolved $X_{3}^{0}$ and double-unresolved $X_{4}^{0}$ are as given in Ref.~\cite{Braun-White:2023sgd}.
This is to be compared and contrasted with the subtraction term given by combining Eqs. (6.3) and (3.31) of Ref.~\cite{Gehrmann-DeRidder:2007foh}. First of all, we sum over the six colour orderings, rather than the three pairs of line-reversed colour orderings, because the new subtraction term is able to locally remove the divergent behaviour associated to each individual colour ordering. We introduce coloured line labels to facilitate the discussion below.

The first three terms (\textA{1-3}) in Eq.~\eqref{eq:RRsubLC} represent the single-unresolved limits and correspond to the first six lines of Eq.~(6.3) in Ref.~\cite{Gehrmann-DeRidder:2007foh}. 
Terms \textB{4-9} in Eq.\eqref{eq:RRsubLC} describe the double-unresolved colour-connected singularities, as anticipated in~\eqref{X40minusX30X30}. The $D_4^0$ antenna functions (lines \textB{4} and \textB{7}) contain both single and double-unresolved singularities.   The single-unresolved contributions are removed by lines \textB{5,6} and \textB{8,9} respectively.  These terms correspond roughly to the blocks with $D_{4,a}^{0}$ and $D_{4,b}^{0}$ in Eq.~(6.3) in Ref.~\cite{Gehrmann-DeRidder:2007foh}. The main difference is that the $D_{4}^{0}$ is designed to contain only the required infrared limits, while 
$D_{4,a}^{0}$ and $D_{4,b}^{0}$ were extracted by partitioning $D_{4}^{0}$ based on matrix elements.

The major change comes in the double-unresolved almost colour-connected contribution in lines \textC{10-12} in Eq.\eqref{eq:RRsubLC}.  These terms replace both the remainder of Eq. (6.3) and the large angle soft contribution in Eq. (3.3) of Ref.~\cite{Gehrmann-DeRidder:2007foh}.  For the first nine terms, all the mappings (whether $3\to 2$ or $4\to 2$) are antenna-like (denoted by $(\wt{ij})$ etc).
We note that in double-unresolved limits both antenna- and dipole-like mappings collapse correctly, but in single-unresolved limits we must be consistent between terms that should cancel. As noted in Section~\ref{sec:Xmappings} and specifically Eq.~\eqref{eq:Xsplit}, line \textC{10} is divided into three parts, each with different (dipole-like) mappings,
\begin{align}
&\textC{10} &&\,A_{5,3}^0(1,i,j,k,2)\,A_3^{0}(\{1ij\},\{1ijk2\},\{jk2\})\,J_3^{(3)}(\lbrace p\rbrace_{3})  \nn\\
&\textC{10L}  &\equiv &\,A_{5,3;L}^0(1,i,j,k,2)\,A_3^{0}([1i\u{j}],[i\u{j}k],2)\,J_3^{(3)}(\lbrace p\rbrace_{3}) \nn\\
&\textC{10M}  &+&\,A_{5,3;M}^0(1,i,j,k,2)\,A_3^{0}([1i\u{j}],[1i\u{j}k2],[\u{j}k2])\,J_3^{(3)}(\lbrace p\rbrace_{3}) \nn\\
&\textC{10R} &+&\,A_{5,3;R}^0(1,i,j,k,2)\,A_3^{0}(1,[i\u{j}k],[\u{j}k2])\,J_3^{(3)}(\lbrace p\rbrace_{3}). 
\end{align}

The mappings in terms \textC{11} and \textC{12} must correctly cancel the single-unresolved contributions in term \textC{10}.  This means that the mapping used in the second $X_3^0$ antenna function must also be dipole-like. 
To determine the mappings needed, we consider the soft $i$ limit. In this limit, term \textA{1} gives the correct unresolved behaviour, while there are additional contributions from terms \textA{3},\textB{4,5} and \textC{10,11,12}. By design, the contributions from terms \textB{4} and \textB{5} cancel, and therefore we need pairwise cancellation between  terms \textA{3} and \textC{12}, and  between terms \textC{10} and \textC{11}. 
For the \textA{3} and \textC{12} cancellation, we find
\begin{eqnarray}
    \textA{3}\quad&+D_3^{0}(2,k,j)S_{1i(\wt{jk})}M_3^{0}(1,(\wt{jk}),(\wt{k2}))\\
    \textC{12}\quad&-D_3^{0}(2,k,j)S_{1i(\wt{jk})}M_3^{0}(1,(\wt{jk}),(\wt{k2}))
\end{eqnarray}
where $S_{abc}$ is the soft eikonal function.
These contributions do indeed cancel if the mapping for the first $X_3^0$ antenna in term \textC{12} is antenna-like. 
For the \textC{10} and \textC{11} cancellation, we find
\begin{eqnarray}
    \textC{10}\quad&+S_{1ij}D_3^{0}(2,k,j)M_3^{0}(1,[\u{j}k],[k2])\\
    \textC{11}\quad&-S_{1ij}D_3^{0}(2,k,j)M_3^{0}(1,[\u{j}k],[k2])
\end{eqnarray}
which again does cancel, so long as the mapping for the second $X_3^0$ antenna in term \textC{12} is a dipole mapping with $j$ rescaled. Similar considerations in the $k$ soft limit mean that the mapping for the first $X_3^0$ antenna in both terms \textC{11} and \textC{12} is an antenna mapping, while the mapping for the second $X_3^0$ antenna must be a dipole mapping with momentum $j$ rescaled.  
One can check that this solution works in single-collinear regions too.

With the appropriate choice of mapping, the single-unresolved limits are addressed by the (NLO-like) terms \textA{1}-\textA{3}. For example, the limit where particle $i$ is unresolved is fully described by term~\textA{1}. All the other terms with divergences when $i$ becomes unresolved must then cancel between each other.

On the other hand, the double-unresolved behaviour of the matrix element is fully captured by \textB{4},\textB{7} and \textC{10}, with the $X_3^0X_3^0$ terms cancelling the additional divergences of \textA{1}-\textA{3}. Hence, the correct behaviour across the whole phase space is guaranteed.

Overall, the subtraction term in~\eqref{eq:RRsubLC} represents a dramatic simplification compared to Ref.~\cite{Gehrmann-DeRidder:2007foh}. Similar simplifications occur for every colour factor that previously required large angle soft terms, in particular the $N^2$, $N^0$ and $\NF N^{-1}$ contributions in Ref.~\cite{Gehrmann-DeRidder:2007foh}. QED-like contributions, such as the $N^{-2}$ factor in~Ref.~\cite{Gehrmann-DeRidder:2007foh}, do not have such structures and therefore only benefit from the use of idealized antenna functions.

To confirm that~\eqref{eq:RRsubLC} indeed does locally reproduce the single- and double-unresolved behaviour of the matrix element, we performed pointwise numerical tests similar to those described in~\cite{NigelGlover:2010kwr,Gehrmann:2023dxm}. We note that to account for azimuthal contributions in the collinear limits of the real matrix elements, we employ azimuthal averaging as described in Ref.~\cite{Pires:2010jv}. As expected, we find local cancellation for a single squared colour-ordering. To compare our implementation to the original subtraction terms for $e^+ e^- \to jjj$~\cite{Gehrmann-DeRidder:2007foh,Gehrmann:2017xfb}, a sum over colour orderings is required. We observe no degradation in the subtraction performances relying on the new setup, which demonstrates that there are no numerical drawbacks in combining dipole mappings with antenna mappings. We performed analogous tests for all the other double-real subtraction terms in Appendix~\ref{app:RRsub} and drew the same conclusions.

\section{Real-virtual subtraction term}\label{sec:RVsub} 

We now consider the corresponding real-virtual subtraction term. 
The real-virtual correction to the cross section ${\rm d}\sigma_{NNLO,N^2}^{RV}$ is obtained by summing the contributions of the colour ordered matrix elements for the two permutations of gluons (labelled here as $i$ and $j$ carrying momentum $p_3$ and $p_4$ respectively):
\begin{align}\label{RVLC}
{\rm d}\sigma_{NNLO,N^2}^{RV} &= N_4 N^2 \left(\frac{\alpha_s}{2\pi}\right) d\Phi_4(\lbrace p\rbrace_{4};q) \frac{1}{2!} \sum_{(i,j) \in P(3,4)} M_4^{1}(1,i,j,2)\,J_3^{(4)}(\lbrace p\rbrace_{4})
\end{align}
Here we use $M_n^{1}(1,\ldots,2)$ to denote the interference between the one-loop and tree-level leading-colour colour-ordered amplitudes: $2\text{Re}\left[A_n^0(1,\ldots,2)(A^{1}_n(1,\ldots,2))^{\dagger}\right]$, with $N_4$ being an overall normalisation as in Eq.~\eqref{eq:Nndef} and $d\Phi_4$ the four particle phase space.

The real-virtual subtraction term is given by,
\begin{align}
\label{eq:RVsubLC}
{\rm d}\sigma_{NNLO,N^2}^{T} &= N_4 N^2 \left(\frac{\alpha_s}{2\pi}\right) d\Phi_4(\lbrace p\rbrace_{4};q) \frac{1}{2!} \sum_{(i,j) \in P(3,4)} \Biggl \{ \nn \\
\textA{1}\quad &-\bigg[ 
 +\mathcal{D}_3^0(s_{1i})
 +\mathcal{F}_3^0(s_{ij})
 +\mathcal{D}_3^0(s_{j2})\bigg]\,M_4^{0}(1,i,j,2)\,J_3^{(4)}(\lbrace p\rbrace_{4})\nn\\
---&-----------------------\nn\\
\textB{2}\quad &+\bigg[
 +\mathcal{D}_3^0(s_{1i})
 +\mathcal{F}_3^0(s_{ij})
\bigg]  D_3^{0}(1,i,j)\,M_3^{0}((\wt{1i}),(\wt{ij}),2)\,J_3^{(3)}(\lbrace p\rbrace_{3})  \nn\\
\textB{3}\quad &+\bigg[
 +\mathcal{F}_3^0(s_{ij})
 +\mathcal{D}_3^0(s_{j2})
\bigg]  D_3^{0}(2,j,i)\,M_3^{0}(1,(\wt{ij}),(\wt{j2}))\,J_3^{(3)}(\lbrace p\rbrace_{3})  \nn\\
---&-----------------------\nn\\
\textC{4}\quad  &+\mathcal{D}_3^0(s_{j2})D_3^{0}(1,i,j)\,M_3^{0}([1i],[i\underline{j}],2)\,J_3^{(3)}(\lbrace p\rbrace_{3})  \nn\\
\textC{5}\quad  &+\mathcal{D}_3^0(s_{1i})D_3^{0}(2,j,i)\,M_3^{0}(1,[\underline{i}j],[j2])\,J_3^{(3)}(\lbrace p\rbrace_{3})  \nn\\
---&-----------------------\nn\\
\textD{6}\quad  &+\,D_{3}^{1}(1,i,j)\,M_3^{0}((\wt{1i}),(\wt{ij}),2)\,J_3^{(3)}(\lbrace p\rbrace_{3})\nn\\
\textD{7}\quad  &+\,D_{3}^{1}(2,j,i)\,M_3^{0}(1,(\wt{ij}),(\wt{j2}))\,J_3^{(3)}(\lbrace p\rbrace_{3})\nn\\
\textD{8}\quad  &+\,D_3^{0}(1,i,j)\,M_3^{1}((\wt{1i}),(\wt{ij}),2)\,J_3^{(3)}(\lbrace p\rbrace_{3})\nn\\
\textD{9}\quad  &+\,D_3^{0}(2,j,i)\,M_3^{1}(1,(\wt{ij}),(\wt{j2}))\,J_3^{(3)}(\lbrace p\rbrace_{3})\nn\\
---&-----------------------\nn\\
\textE{10}\quad  &+\,D_{4,3}^{1}(1,i,j,2)\,M_3^{0}(\{1i\},\{ij\},\{2\})\,J_3^{(3)}(\lbrace p\rbrace_{3})\nn\\
\textE{11}\quad  &+\,D_{4,3}^{1}(2,j,i,1)\,M_3^{0}(\{1\},\{ij\},\{j2\})\,J_3^{(3)}(\lbrace p\rbrace_{3}) \Bigg \rbrace
\end{align}
where the final-final integrated antenna functions are denoted by $\mathcal{X}_3^0$ and are as given in~\cite{Braun-White:2023sgd}, and the single-unresolved one-loop antenna function $X_{3}^{1}$ is as given in Ref.~\cite{Braun-White:2023zwd}.
This is to be compared and contrasted with the subtraction term given by combining Eqs. (6.5) and (3.35) of Ref.~\cite{Gehrmann-DeRidder:2007foh}. The real-virtual subtraction term has a twofold purpose: removing from the matrix element the implicit divergent behaviour in single-unresolved limits and the explicit $\e$-poles exactly across the phase space. In the following we show how equation~\eqref{eq:RVsubLC} achieves this. The real-virtual subtraction term is structured as follows:  
\begin{enumerate}
    \item[(a)] Terms \textA{1}, \textB{2}, \textB{3}, \textC{4} and \textC{5} are fixed by the integration of terms \textA{1,2,3}, \textB{5,6} and \textB{8,9}, and \textC{11,12} of  Eq.~\eqref{eq:RRsubLC}.  
Specifically,
\begin{align}\label{correspondence}
    Eq.~\eqref{eq:RRsubLC} &\to Eq.~\eqref{eq:RVsubLC} \nn\\
    \textA{1,2,3} &\to \textA{1} \nn \\
    \textB{5,6} &\to \textB{2} \nn \\
    \textB{8,9} &\to \textB{3} \nn \\
    \textC{12} &\to \textC{4} \nn \\
    \textC{11} &\to \textC{5}.
\end{align}
Note that the mappings appearing in terms \textC{4,5} precisely match those in terms \textC{11,12} of Eq.~\eqref{eq:RRsubLC}. There is no freedom to choose different mappings for terms \textC{4,5} because, although the integration from RR to RV is mapping independent, the numerical integration of terms \textC{4,5} over the rest of the phase space is mapping dependent.  We must use the same mapping as the second mapping in the corresponding RR term, namely the dipole mapping with $j$ rescaled for term \textC{4} and $i$ rescaled for \textC{5}.

\item[(b)] Terms \textB{2}, \textB{3}, \textC{4} and \textC{5} completely cancel the divergent behaviour in single-unresolved limits of term \textA{1}. This is explained by the fact that the RR terms to the left of the arrows in~\eqref{correspondence} cancel in double-unresolved configurations. After analytical integration over a single emission one obtains the RV terms to the right of the arrows, which have to cancel in the residual single-unresolved limits.

\item[(c)] Terms \textD{6-9} subtract the single-unresolved behaviour of the real-virtual matrix element~\cite{Currie:2013vh}.  

\item[(d)] Term \textA{1} exactly reproduces the $\e$-pole structure of the real-virtual matrix element, but in general, terms \textB{2}-\textD{9} are not poles free. Therefore, we add terms \textE{10,11} precisely to subtract the leftover $\e$-poles, and we design them in such a way they do not produce any divergence in unresolved limits. We discuss below how this is achieved. 

\end{enumerate}

\subsection{One-loop four-particle three-hard-radiator antenna functions}\label{sec:X431} 

We introduce a generalised antenna function $X_{4,3}^{1}(i_{a}^h, j_{b}, k_{c}^{h},l_{d}^h)$ for each possible $X_3^0(i_a^h,j_b,k_c^h)$ antenna function appearing in the real-virtual subtraction term. Since each $X_3^0$ can appear with up to three different types of mapping (antenna or dipole with either of the hard radiators being rescaled), we identify three parts for each $\Y$ function:
\begin{align}
&\textE{10} &&\,\Y(1,i,j,2)\,M_3^{0}(\{1i\},\{ij\},\{2\})\,J_3^{(3)}(\lbrace p\rbrace_{3})  \nn\\
&\textE{10L}  &\equiv &\,\YL(1,i,j,2)\,
M_3^{0}([\u{1}i],[ij],2)\,J_3^{(3)}(\lbrace p\rbrace_{3}) \nn\\
&\textE{10M}  &+&\,\YM(1,i,j,2)\,M_3^{0}((\wt{1i}),(\wt{ij}),2)\,J_3^{(3)}(\lbrace p\rbrace_{3}) \nn\\
&\textE{10R} &+&\,\YR(1,i,j,2)\,M_3^{0}([1i],[i\u{j}],2)\,J_3^{(3)}(\lbrace p\rbrace_{3}). 
\end{align}
The mappings for the three cases are summarised in Table~\ref{tab:4to3mappings}.
\begin{table}[h]
\centering
\begin{tabular}{ccccc}
& & I & K & L \\ 
\multirow{4}{*}{$\Y(i,j,k,l)$}
&generic mapping & $\{ijk\}$ & $\{jkl\}$ & $\{l\}$ \\
&$\YM$ mapping & $\wt{ij}$ & $\wt{jk}$ & $l$ \\
&$\YL$ mapping & $[\u{i}j]$ & $[jk]$ & $l$ \\
&$\YR$ mapping & $[ij]$ & $[j\u{k}]$ & $l$ \\
\end{tabular}
\caption{Notation for the mappings for the $\Y$ antenna functions with three hard radiators.}
\label{tab:4to3mappings}
\end{table}
The explicit expression for the $\Y$ antenna functions are completely determined by the leftover $\e$-poles that they need to remove. 

For the specific leading-colour subtraction term considered in this example, we first recall that the poles of the one-loop reduced matrix element can be expressed by means of integrated NLO antenna functions as:
\begin{equation}
    Poles(M_3^1((\wt{1i}),(\wt{ij}),2))=(-{\cal D}_3^0(s_{1ij})-{\cal D}_3^0(s_{(\wt{ij})2}))M_3^0((\wt{1i}),(\wt{ij}),2)
\end{equation}
With this we obtain:
\begin{align}
    D_{4,3;L}^{1}(1,i,j,2) &\equiv 0,\nn \\
    D_{4,3;M}^{1}(1,i,j,2) &= -Poles\bigg[\left({\cal D}_3^0(s_{1i})+{\cal F}_3^0(s_{ij})-{\cal D}_3^0(s_{1ij})-{\cal D}_3^0(s_{(\wt{ij})2})\right)D_{3}^{0}(1,i,j)\nn\\
    &\qquad\qquad +D_{3}^{1}(1,i,j)\bigg] \nn \\ 
    &=\left(\frac{1}{\e^2}+\frac{1}{\e}\left(\log\left(\frac{s_{1ij}\mu^2}{(s_{1j}+s_{ij})s_{\wt{ij}2}}\right)+\frac{5}{3}\right)\right)
    D_{3}^{0}(1,i,j) \nn \\
    D_{4,3;R}^{1}(1,i,j,2) &= \nn -Poles\left({\cal D}_3^0(s_{j2})D_{3}^{0}(1,i,j)\right),\\
    &=\left(-\frac{1}{\e^2}+\frac{1}{\e}\left(\log\left(\frac{s_{j2}}{\mu^2}\right)-\frac{5}{3}\right)\right)
    D_{3}^{0}(1,i,j),
\end{align}
where $\mu$ is the renormalisation scale. The left ($L$) component vanishes because, for this specific subtraction term, there is no occurrence of a dipole mapping where momentum $1$ is rescaled.
We see that term \textE{10R} exactly cancels the poles of term \textC{4}, while term \textE{10M} removes those of terms \textB{2} and \textD{6,8}.

Away from the unresolved limits,  the $\YL$, $\YM$ and $\YR$ contributions multiply reduced matrix elements with different mappings.  However,  in the unresolved limits, the antenna and dipole mappings coincide. This means that $\YL$, $\YM$ and $\YR$ can be combined together into an overall non-divergent contribution. For example, in the $i$ soft limit: 
\begin{align}
    D_{4,3;L}^{1}(1,i,j,2) &\equiv  0,\nn \\
    D_{4,3;M}^{1}(1,i,j,2) &\to  \left(\frac{1}{\e^2}+\frac{1}{\e}\left(\log\left(\frac{\mu^2}{s_{j2}}\right)+\frac{5}{3}\right)\right)
    S_{1ij}, \nn \\
    D_{4,3;R}^{1}(1,i,j,2) &\to  \left(-\frac{1}{\e^2}+\frac{1}{\e}\left(\log\left(\frac{s_{j2}}{\mu^2}\right)-\frac{5}{3}\right)\right)
    S_{1ij},
\end{align}
with $\YM$ and $\YR$ clearly cancelling each other. This is also true for both the $i\parallel j$ and $j\parallel k$ single-collinear limits. We remark that this holds in general \textit{by construction}: terms \textA{1}-\textD{9}, despite not being $\e$-finite, do correctly subtract the divergent behaviour of the matrix element in single-unresolved limits, both for the finite part \textit{and} the explicit poles. Therefore, the poles mismatch encapsulated in terms \textE{10} and \textE{11} has to vanish in singular configurations. 

In Table~\ref{tab:X431} we list all the $\Y$ antenna functions. As explained above, each $\Y$ collects residual explicit singularities from terms coming with specific $X_3^0$ and $X_3^1$ antenna functions. Since standard three-parton antenna functions can be classified as quark-antiquark, quark-gluon and gluon-gluon ones according to the partonic species of the two reconstructed hard radiators, we can straightforwardly adopt the same separation for the $\Y$ too. For each $\Y$, we indicate the corresponding $X_3^0$ and $X_3^1$. 
\begingroup
\renewcommand{\arraystretch}{1.5}
\begin{table}[t]
\centering
\begin{tabular}{c|ccc}
    \textbf{Type} & \thead{\textbf{Name and} \\ \textbf{parton content}} & $X_3^0$ & $X_3^1$\\
    \hline
    quark-antiquark
    & \begin{tabular}{@{}c@{}} $A_{4,3}^{1}(i^h_q,j_g,k^h_{\qb},b^h)$ \\ $\wh{A}_{4,3}^{1}(i^h_q,j_g,k^h_{\qb},b^h)$ \\ $\wt{A}_{4,3}^{1}(i^h_q,j_g,k^h_{\qb},b^h)$ \end{tabular}
    & \begin{tabular}{@{}c@{}} $A_3^0$ \\ $A_3^0$ \\ $A_3^0$ \end{tabular}
    & \begin{tabular}{@{}c@{}} $A_3^1$ \\ $\wh{A}_3^1$ \\$\wt{A}_3^1$ \end{tabular} \\ 
    \hline
    quark-gluon
    & \begin{tabular}{@{}c@{}} $D_{4,3}^{1}(i^h_q,j_g,k^h_g,b^h)$ \\ $\wh{D}_{4,3,g}^{1}(i^h_q,j_g,k^h_g,b^h)$ \\ $\wt{D}_{4,3,g}^{1}(i^h_q,j_g,k^h_g,b^h)$ \\ $E_{4,3}^{1}(i_q^h,j_{\bar{Q}},k_Q^h,b^h)$ \\ $\wh{E}_{4,3}^{1}(i_q^h,j_{\bar{Q}},k_Q^h,b^h)$ \\ $\wt{E}_{4,3}^{1}(i_q^h,j_{\bar{Q}},k_Q^h,b^h)$ \end{tabular}
    & \begin{tabular}{@{}c@{}} $D_3^0$ \\ $D_3^0$ \\ $D_3^0$ \\ $E_3^0$ \\ $E_3^0$ \\ $E_3^0$ \end{tabular}
    & \begin{tabular}{@{}c@{}} $D_3^1$ \\ $\wh{D}_3^1$ \\ $\wt{D}_3^1$ \\ $E_3^1$ \\ $\wh{E}_3^1$ \\ $\wt{E}_3^1$ \end{tabular} \\
    \hline
    gluon-gluon 
    & \begin{tabular}{@{}c@{}} $F_{4,3}^{1}(i^h_g,j_g,k^h_g,b^h)$ \\ $\wh{F}_{4,3}^{1}(i^h_g,j_g,k^h_g,b^h)$ \\ $G_{4,3}^{1}(i_g^h,j_{\bar{Q}},k_Q^h,b^h)$ \\ $\wh{G}_{4,3}^{1}(i_g^h,j_{\bar{Q}},k_Q^h,b^h)$ \\ $\wt{G}_{4,3}^{1}(i_g^h,j_{\bar{Q}},k_Q^h,b^h)$ \end{tabular}
    & \begin{tabular}{@{}c@{}} $F_3^0$ \\ $F_3^0$ \\ $G_3^0$ \\ $G_3^0$ \\ $G_3^0$ \end{tabular}
    & \begin{tabular}{@{}c@{}} $F_3^1$ \\$\wh{F}_3^1$ \\ $G_3^1$ \\ $\wh{G}_3^1$ \\ $\wt{G}_3^1$ \end{tabular} \\ 
\end{tabular}
\caption{List of $X_{4,3}^{1}$ antenna functions, organised by the partonic species of the two hard radiators in the underlying three-parton antenna functions, identified by removing the generic spectator parton $b$. The corresponding three-parton tree-level and one-loop antenna functions are indicated in the $X_3^0$ and $X_3^1$ column respectively.}
\label{tab:X431}
\end{table}
\endgroup
We provide expressions for all the unintegrated $\Y$ antenna functions in Appendix~\ref{app:unintegratedX431}. 

\vspace{1cm}

We performed pointwise tests~\cite{Gehrmann:2023dxm} to confirm that the subtraction term in~\eqref{eq:RVsubLC} correctly removes both the explicit $\e$-poles and local single-unresolved limits of the real-virtual matrix element. As for the double-real case, we find that the cancellation works for individual colour-ordered matrix elements, and observe the same performances when comparing with original subtraction terms~\cite{Gehrmann-DeRidder:2007foh,Gehrmann:2017xfb}. All the other real-virtual subtraction terms in Appendix~\ref{app:RVsub} passed the tests too. 


\section{Analytical integration}\label{sec:integration}

In this Section we describe how the momentum mappings introduced in the previous Sections for the generalised antenna functions induce a phase-space factorisation which allows us to perform the integration over the degrees of freedom of the unresolved radiation in fully analytical fashion. As usual, we work in dimensional regularisation with the number of dimensions $d=4-2\e$.

\subsection{Integration of $X_{5,3}^0$ antenna functions}

\subsubsection{$\XM$}
The $\mapM$ mapping is defined in Eq.~\eqref{eq:X53Mmapping}.
For this mapping, we can write the five particle phase space in factorised form as,
\begin{align}
\label{eq:PSfactorXM}
    dPS_5(i,j,k,l,m) &= dPS_{\XM}(i,j,k,l,m) dPS_{3}(I,K,M)
\end{align}
where
\begin{align}
\label{eq:dPSXM}
   dPS_{\XM}(i,j,k,l,m) &=
   \frac{1}{\Se^2} \left(\frac{e^{\e\gamma}}{2\Gamma(1-\e)}\right)^2
   d\za\, d\zb\, d\ya\, d\yb\,
   \za^{-\e}
   \omza^{-\e}
   \zb^{-\e}
   \omzb^{-\e}
   \nonumber \\
   & \times 
   \ya^{-\e}
   \omya^{1-2\e}
   \yb^{-\e}
   \omyb^{2-3\e}
   \SAB^{1-\e}
   \SBC^{1-\e}  
\end{align}
with
\begin{align}
    \Se = 8\pi^2 (4\pi)^{-\e} e^{\e\gamma}.
\end{align}
The relations between the invariants of un-mapped momenta and the integration variables $y_1$, $y_2$, $z_1$ and $z_2$ (which all run between $0$ and $1$) are given by
\begin{align}
  s_{ij} &= \ya\omyb\SAB, \nonumber \\
  s_{jk} &=\omya\omyb\omza\SAB, \nonumber \\
  s_{ik} &= \za\omya\omyb\SAB, \nonumber \\
  s_{kl} &=\omya\omyb\omzb\SBC, \nonumber \\
  s_{km} &=\zb\omya\omyb\SBC, \nonumber \\
  s_{lm} &= \yb\SBC, \nonumber \\
  s_{ijk} &=  \omyb\SAB, \nonumber \\
  s_{klm} &= \omyaomyb\SBC .
\end{align}
As a basic check of the formulae above, we confirmed that integrating $dPS_{\XM}$ over $z_{1(2)}$ and $y_{1(2)}$ and then integrating inclusively over the three-particle phase space $\dPS_3(I,K,M)$ yields the full $d$-dimensional five-particle phase space.

Because of the product-like nature of Eq.~\eqref{eq:XM}, every
term in an $\XM$ antenna can be completely factorised and written as
\begin{align}\label{I_M}
    I_{\XM}(a_1,a_2,a_3,a_4,a_5,a_6,a_7,a_8) &= 
s_{ij}^{a_1}
s_{ik}^{a_2}
s_{jk}^{a_3}
s_{ijk}^{a_4}
s_{kl}^{a_5}
s_{km}^{a_6}
s_{lm}^{a_7}
s_{klm}^{a_8}
\end{align}
where the exponents $a_1,\ldots,a_4$ are fixed by the $X_3^0(i,j,k)$ antenna function, while
$a_5,\ldots,a_8$ are fixed by the $X_3^0(k,l,m)$ antenna function.  Because of the dimensionality of each antenna function, we have:
\begin{align*}
a_1+a_2+a_3+a_4 &=-1, \\
a_5+a_6+a_7+a_8 &=-1.
\end{align*}
The integration of a term like $I_{\XM}$ in Eq.~\eqref{I_M} over $dPS_{\XM}$ can be performed in full analytical fashion, retaining exact dependence on the regulator $\e$. The integral has the following closed form:
\begin{align}\label{eq:I_M_res}
  &\int I_{\XM}(a_1,a_2,a_3,a_4,a_5,a_6,a_7,a_8) dPS_{\XM}(i,j,k,l,m) \nonumber \\
  &\qquad = \frac{1}{\Se^2} \left(\frac{e^{\e\gamma}}{2\Gamma(1-\e)}\right)^2 \SAB^{-\e} \SBC^{-\e}
  \nonumber \\
  &\qquad \times I(a_2-\e,a_3-\e,a_6-\e,a_5-\e,a_1-\e,a_2+a_3+a_5+a_6+1-2\e,
  \nonumber \\
  &\qquad \qquad a_7-\e,a_1+a_2+a_3+a_4+a_5+a_6+2-3\e,a_8)
\end{align}
with
\begin{align}\label{I_M_res2}
    &I(b_1,b_2,b_3,b_4,b_5,b_6,b_7,b_8,b_9) \nonumber \\
    &\qquad = 
    \int_0^1 
     \za^{b_1} \omza^{b_2} d\za\,
     \int_0^1 \zb^{b_3} \omzb^{b_4} 
       d\zb\, \nonumber \\
      & \times
     \int_0^1\int_0^1 \ya^{b_5} \omya^{b_6} 
     \yb^{b_7} \omyb^{b_8} \omyaomyb^{b_9} d\ya\, d\yb\,
    \nonumber \\
    &\qquad=
    \frac{\Gamma(1+b_1) \Gamma(1+b_2)}{\Gamma(2+b_1+b_2)}
    \frac{\Gamma(1+b_3) \Gamma(1+b_4)}{\Gamma(2+b_3+b_4)}
    \nonumber \\
    &\qquad \times
    \frac{\Gamma(1+b_5) \Gamma(1+b_6)}{\Gamma(2+b_5+b_6)}
    \frac{\Gamma(1+b_7) \Gamma(1+b_8)}{\Gamma(2+b_7+b_8)}
    \nonumber \\
 &\times     {}_{3}F_{2} ([1+b_5,1+b_8,-b_9],[2+b_5+b_6,2+b_7+b_8],1).
\end{align}
The generalised hypergeometric function ${}_{p}F_{q}([a_1,\ldots,a_p],[b_1,\ldots,b_q],z)$ can be written as:
\begin{equation}\label{hypergeometric}
    {}_{p}F_{q}([a_1,\ldots,a_p],[b_1,\ldots,b_q],z)=\sum_{n=0}^{\infty}\dfrac{(a_1)_n\ldots(a_p)_n}{(b_1)_n\ldots(b_q)_n}\dfrac{z}{n!},
\end{equation}
where $(a)_n$ denotes the Pochhammer symbol:
\begin{equation}
    (a)_n = \frac{\Gamma(a+n)}{\Gamma(a)}.
\end{equation}
The expression above allows one to expand the exact result in Eqs.~\eqref{eq:I_M_res} and~\eqref{I_M_res2} up to any desired power in $\e$. We used {\tt HypExp}\cite{Huber:2005yg,Huber:2007dx} to check all our results.

We note that we could achieve a relatively straightforward analytic integration of a double-unresolved divergent function of five distinct momenta, retaining differential information with respect to independent hard scales. On one hand, this is due to the simplicity of the integrand, which in turn descends from the designer antenna algorithm, which naturally produces simple physical denominators. On the other hand, the phase-space factorisation induced by the iterated dipole mapping in~\eqref{eq:X53Mmapping} allows for a sequential integration of the almost completely factorised integral $I$ in~\eqref{I_M_res2}. The only term spoiling the separation of the $y_1$ and $y_2$ integrals is $(1-y_1(1-y_2))^{b_9}$, associated with the $s_{klm}$ invariant, which is at the origin of the ${}_{3}F_{2}$ hypergeometric function. The specific choice of mapping also ensures that the dependence from the hard scales $s_{IK}$ and $s_{KM}$ of the final result is trivial and completely factorised.

\subsubsection{$\XL$ ($\XR$)}
The $\mapL$ ($\mapR$) mapping is defined in Eq.~\eqref{eq:X53Lmapping}.
For this mapping, we can write the five-particle phase space in factorised 
form as
\begin{align}
\label{eq:PSfactorXL}
dPS_5(i,j,k,l,m) &= dPS_{\XL}(i,j,k,l,m) dPS_{3}(I,K,M)
\end{align}
where we can exploit the fact that one of the five momenta is untouched ($m$=$M$ for $\XL$ or $i$=$I$ for $\XR$) and rely the four-particle phase space parametrisation (tripole parametrisation) described in~\cite{Gehrmann-DeRidder:2003pne}:
\begin{align}
\label{eq:dPSXL}
   dPS_{\XL}(i,j,k,l,m) &=
   \frac{1}{\Se^2} \left(\frac{e^{\e\gamma}}{2\Gamma(1-\e)}\right)^2
   d\za\, d\zb\, d\ya\, d\yb\,
   \za^{-\e}
   \omza^{1-2\e}
   \zb^{-\e}
   \omzb^{-\e}
   \nonumber \\
   & \times 
   \ya^{1-\e}
   \omya^{1-2\e}
   \yb^{-\e}
   \omyb^{-\e}
   \SAB^{2-2\e} ,
\end{align}
with a similar expression for $\XR$ with $\SAB \to \SBC$.
As a basic check, we verified that integrating~\eqref{eq:dPSXL} over $z_{1(2)}$ and $y_{1(2)}$ and then integrating inclusively over the three-particle phase space $\dPS_3(I,K,M)$ yields the full $d$-dimensional five-particle phase space.

The terms in an $\XL$ ($\XR$) antenna functions are products of invariants involving two or more of the four momenta $i,~j,~k,~l$ ($j,~k,~l,~m$). Therefore, such expressions can be all be straightforwardly integrated following the procedure outlined in Ref.~\cite{Gehrmann-DeRidder:2003pne} for the analytical integration of traditional four-particle antenna functions.

As indicated by Eq.~\eqref{eq:masterX2}, after integration over $\dPS_{X^0_{5,3;L(R)}}$, the final result will depend on a single hard scale $\SAB$ (or $\SBC$), which hence appears simply as an overall factor. 

\subsubsection{Summary}

As a shorthand, we use the notation $\calX  (\SAB,\SBC)$ to denote the integrated $\X$ function where, 
\begin{equation}
\label{eq:masterX}
\calX  (\SAB,\SBC) = \calXM (\SAB,\SBC) + \calXL (\SAB) + \calXR (\SBC). 
\end{equation}
We remark that the explicit scale dependence of each term~\eqref{eq:masterX} is different, due to the different phase-space factorisation for the respective mappings. In particular:
\begin{eqnarray}
\label{eq:masterX2}
\calXM (\SAB,\SBC)&=& \left( \frac{\SAB}{\mu^2} \right)^{-\e}\left(\frac{\SBC}{\mu^2} \right)^{-\e}f_M(\e), \\
\calXL (\SAB)&=& \left( \frac{\SAB}{\mu^2} \right)^{-2\e}f_L(\e), \\
\calXR (\SBC)&=& \left( \frac{\SBC}{\mu^2} \right)^{-2\e}f_R(\e),
\end{eqnarray}
where $f_M$, $f_L$ and $f_R$ are constant series in the dimensional regulator $\e$. In Appendix~\ref{app:integratedX530} we provide the analytically integrated form of all the $\X$ antenna functions.

We notice that the deepest infrared poles ($\e^{-4}$ and $\e^{-3}$) of an integrated $\X$ antenna function, whose $\XM$ component has been constructed according to~\eqref{eq:XM} from $X_{3,A}^0X_{3,B}^0$, coincide with the deepest poles of ${\cal X}_{3,A}^0{\cal X}_{3,B}^0$. This happens because both the residual terms in~\eqref{eq:XM}, as well as $\XL$ and $\XR$, do not have soft singularities and hence give only up to $\e^{-2}$ poles after integration. This serves as a basic sanity check of the analytic integration procedure.

\subsection{Integration of $\Y$ antenna functions}

For any mapping where the fourth spectator momentum is unaffected, the $\calY$ function is a single-scale function, is independent of the choice of how the three active momenta are mapped and depends only on their invariant mass. The analytic integration of $\YR$ ($\YL$) can be straightforwardly carried out using the appropriate dipole mapping (see Table~\ref{tab:4to3mappings}).  On the other hand, $\YM$ is easy to integrate with a dipole mapping where either of the two hard momenta is rescaled. For this reason, in the following we explicitly provide formulae for the $\YR$ integration.

For the $\YR$ part of $\Y$, we use the dipole mapping, $\mapYR$,
\begin{alignat}{2}
\label{eq:X43Rmapping}
    & p_I =&  p_{i}+p_{j}-\frac{s_{ij}}{s_{ik}+s_{jk}}p_{k} \nonumber \\
\mapYR: \hspace{1cm}    &    p_K  =&  \frac{s_{ijk}}{s_{ik}+s_{jk}}p_{k} \nn\\
   & p_L  = & p_l.
\end{alignat}
As usual, we will use the notation
\begin{align}
    I &\equiv [ij],\qquad \qquad
    K \equiv [j\u{k}],\qquad \qquad
    L \equiv l,   
\end{align}
where the momentum that is rescaled is underscored.

For the $\mapYR$ mapping  defined in Eq.~\eqref{eq:X43Rmapping} we can write the four particle phase space in factorised form as,
\begin{align}
\label{eq:PSfactorYR}
    dPS_4(i,j,k,l) &= dPS_{\YR}(i,j,k,l) dPS_{3}(I,K,L)
\end{align}
where
\begin{align}
\label{eq:dPSYR}
   dPS_{\YR}(i,j,k,l) &=
   \frac{1}{\Se} \left(\frac{e^{\e\gamma}}{2\Gamma(1-\e)}\right)
   dz\, dy\, 
   z^{-\e}
   \omz^{-\e}
   y^{-\e}
   \omy^{1-2\e}
   \SAB^{1-\e}.
\end{align}
The invariants before mapping are related to the invariant masses after mapping by
\begin{align}
    s_{ij} &= y s_{IK},\nn \\
    s_{ik} &=  (1-z) (1-y) \SAB,\nn \\
    s_{jk} &=  z (1-y) \SAB,\nn \\
    s_{ijk} &=  \SAB,\nn \\
    s_{ik}+s_{jk} &= (1-y) \SAB, \nn \\
    s_{ij}+s_{ik} &= (1-z(1-y)) \SAB, \nn \\
    s_{lk} &= (1-y)   s_{KL}.
\end{align}

The generic form of the $\YR$ integrand is a product of an $X_3^0(i,j,k)$ antenna and a logarithm (see Appendix~\ref{app:unintegratedX431}). The $X_3^0$ depends only on the invariants constructed from the momenta $i$, $j$ and $k$ and has the form,
\begin{align}
    X_3^0(i,j,k) &= \sum s_{ij}^{a_1}
s_{ik}^{a_2}
s_{jk}^{a_3}
s_{ijk}^{a_4}
\end{align}
where the sum runs over the terms in the antenna and on dimensional grounds,
\begin{align}
&a_1+a_2+a_3+a_4 = -1.
\end{align}
The logarithm can in general be expressed as,
\begin{align}
    \log\left(s_{ij}^{b_1}
s_{ik}^{b_2}
s_{jk}^{b_3}
s_{ijk}^{b_4}
[s_{ik}+s_{jk}]^{b_5}
[s_{ij}+s_{ik}]^{b_6}
s_{lk}^{b_7}
s_{KL}^{b_8}\right)
\end{align}
where, to have a dimensionless argument in the logarithm,
\begin{align}
    &b_1+b_2+b_3+b_4 +b_5+b_6+b_7+b_8= 0.
    \label{eq:bcanc}
\end{align}
Moreover, by construction, the dependence on the fourth momentum $l$ has to cancel after analytic integration, which requires:
\begin{align}
    \label{eq:bcanc2}
    &b_7+b_8 = 0.
\end{align}
Note that for the dipole mapping where $k$ is rescaled and the spectator $l$ is unchanged, so that $s_{LK}\propto s_{lk}$, Eq.~\eqref{eq:bcanc2} ensures that the dependence of the integrand on the momentum $l$ (and on the scale $s_{LK}$) drops out. Therefore every term that we need to integrate has the form of the generalised integrand
\begin{align}
&I_{\YR}(a_1,a_2,a_3,a_4,b_1,b_2,b_3,b_4,b_5,b_6,b_7,b_8) \nn \\
&\qquad \equiv s_{ij}^{a_1}
s_{ik}^{a_2}
s_{jk}^{a_3}
s_{ijk}^{a_4} 
 \log\left(s_{ij}^{b_1}
s_{ik}^{b_2}
s_{jk}^{b_3}
s_{ijk}^{b_4}
[s_{ik}+s_{jk}]^{b_5}
[s_{ij}+s_{ik}]^{b_6}
s_{lk}^{b_7}
s_{LK}^{b_8}\right) \nn\\
&\qquad
=
\frac{\partial}{\partial\delta} \left(s_{ij}^{a_1+b_1 \delta}
s_{ik}^{a_2+b_2 \delta}
s_{jk}^{a_3+b_3 \delta}
s_{ijk}^{a_4+b_4 \delta}
[s_{ik}+s_{jk}]^{b_5 \delta}
[s_{ij}+s_{ik}]^{b_6 \delta}
s_{lk}^{b_7 \delta}
s_{LK}^{b_8 \delta}\right) \Bigg|_{\delta=0} \nn \\
\end{align}

Integrating $I_{\YR}$ over $dPS_{\YR}$ we find that

\begin{align}
\label{eq:integratedYR}
&\int I_{\YR} dPS_{\YR} = 
\frac{1}{\Se} \left(\frac{e^{\e\gamma}}{2\Gamma(1-\e)}\right) \SAB^{-\e} \nn\\
&
\times \frac{\partial}{\partial\delta} K(a_3\!+\!b_3\delta\!-\!\e,a_2\!+\!b_2\delta\!-\e,
a_1\!+\!b_1\delta\!-\!\e,1\!+\!a_2\!+\!a_3\!+\!(b_2\!+\!b_3\!+\!b_5\!+\!b_7)\delta\!-\!2\e,b_6\delta) \Bigg|_{\delta=0} 
\end{align}
where
\begin{align}\label{I431}
    &K(c_1,c_2,c_3,c_4,c_5) \nonumber \\
    &\qquad = 
     \int_0^1\int_0^1 \ya^{c_1} \omya^{c_2} 
     \yb^{c_3} \omyb^{c_4} \omyaomyb^{c_5} d\ya\, d\yb\,
    \nonumber \\
    &\qquad =
    \frac{\Gamma(1+c_1) \Gamma(1+c_2)}{\Gamma(2+c_1+c_2)}
    \frac{\Gamma(1+c_3) \Gamma(1+c_4)}{\Gamma(2+c_3+c_4)}
    \nonumber \\
 &\qquad \times     _{3}\!\!F_{2} ([1+c_1,1+c_4,-c_5],[2+c_1+c_2,2+c_3+c_4],1).
\end{align}
Expressions for the dipole mapping for $\YL$ in which $i$ is the scaled momentum can be obtained from Eq.~\eqref{eq:integratedYR} with the exchange 
\begin{align}
    \{a_1 \leftrightarrow a_3, a_5 \leftrightarrow a_6, b_1 \leftrightarrow b_3, b_5 \leftrightarrow b_6\}.
\end{align}

Similarly to the integration of the $\X$ antenna functions, the result in Eq.~\eqref{I431} allows to obtain the expression of the integrated $\Y$ antenna functions up to any desired power in $\e$.

\subsubsection{Summary}
As a shorthand, we use the notation $\calY (s_{IK})$ to denote the integrated $\Y$ function where,
\begin{equation}
\label{eq:masterY}
\calY  (s_{IK}) =  \calYM(s_{IK}) +\calYL(s_{IK}) + \calYR(s_{IK}) ,
\end{equation}
where the scale dependence is the same for each component, so we can combine the integrated results as:
\begin{equation}
    {\cal X}_{4,3}^1 (s_{IK})= \left( \frac{s_{IK}}{\mu^2} \right)^{-\e}g(\e),
\end{equation}
where the $g$ is a constant series in $\e$. In Appendix~\ref{app:integratedX431} we provide the analytically integrated form of all the $\Y$ antenna functions. 

\section{Double-virtual subtraction term}\label{sec:VVsub}

Finally, we consider the corresponding double-virtual subtraction term. 
The double-virtual contribution to the cross section ${\rm d}\sigma_{NNLO,N^2}^{VV}$ is given by
\begin{align}\label{VVLC}
{\rm d}\sigma_{NNLO,N^2}^{VV} &= N_3 N^2 \left(\frac{\alpha_s}{2\pi}\right)^2 d\Phi_3(\lbrace p\rbrace_{3};q)  M_3^{2}(1,i,2)\,J_3^{(3)}(\lbrace p\rbrace_{3}) .
\end{align}
Here we use $M_3^{2}(1,i,2)$ to denote the sum of the leading-colour contributions of the interference of two-loop with tree-level and one-loop squared amplitudes:
\begin{equation}
M_3^{2}(1,i,2)=2\text{Re}\left[A_n^0(1,i,2)(A^{2}_3(1,i,2))^{\dagger}\right]+|A_3^1(1,i,2)|^2.
\end{equation}
$N_3$ is an overall normalisation and $d\Phi_3$ is the three particle phase space.

The double-virtual subtraction term reads
\begin{align}
\label{eq:VVsubLC}
{\rm d}\sigma_{NNLO,N^2}^{U} &= N_3 N^2 \left(\frac{\alpha_s}{2\pi}\right)^2
d\Phi_3(\lbrace p\rbrace_{3};q)  \Biggl \{ \nn \\
 \textB{1}&-[\mathcal{D}_{4}^{0}(s_{1i})+\mathcal{D}_{4}^{0}(s_{2i})]M_{3}^{0}(1,i,2) \,J_3^{(3)}(\lbrace p\rbrace_{3}) \nonumber\\
 \textC{2}&-[\mathcal{A}_{5,3}^0(s_{1i},s_{2i})]M_{3}^{0}(1,i,2) \,J_3^{(3)}(\lbrace p\rbrace_{3}) \nonumber\\
---&-----------------------\nn\\
 \textD{3}&-\left[\mathcal{D}_{3}^{1}(s_{1i})
 +\frac{b_{0}}{\e}\left(\left(\frac{s_{1i}}{\mu_{R}^2}\right)^{-\e}-1\right)\mathcal{D}_{3}^{0}(s_{1i})\right ]M_{3}^{0}(1,i,2) \,J_3^{(3)}(\lbrace p\rbrace_{3}) \nonumber\\
 \textD{4}&-\left[\mathcal{D}_{3}^{1}(s_{2i})
 +\frac{b_{0}}{\e}\left(\left(\frac{s_{2i}}{\mu_{R}^2}\right)^{-\e}-1\right)\mathcal{D}_{3}^{0}(s_{2i})\right ]M_{3}^{0}(1,i,2) \,J_3^{(3)}(\lbrace p\rbrace_{3}) \nonumber\\
 \textD{5}&-[\mathcal{D}_{3}^{0}(s_{1i})+\mathcal{D}_{3}^{0}(s_{2i})]M_{3}^{1}(1,i,2)\,J_3^{(3)}(\lbrace p\rbrace_{3})  \nonumber\\
 \textE{6}&-[\mathcal{D}_{4,3}^{1}(s_{1i})
           +\mathcal{D}_{4,3}^{1}(s_{2i})]M_{3}^{0}(1,i,2) \,J_3^{(3)}(\lbrace p\rbrace_{3}) \Bigg \rbrace,
\end{align}
where $\mathcal{X}_4^{0}$ denotes the integrated four particle antenna functions from \cite{Braun-White:2023sgd}, $\mathcal{X}_3^{1}$ denotes the integrated one-loop three particle antenna functions from \cite{Braun-White:2023zwd}, and 
$\mathcal{X}_{5,3}^{0}$ and $\mathcal{X}_{4,3}^{1}$ represent the integrated generalised antenna functions introduced earlier.
Eq.~\eqref{eq:VVsubLC} is to be compared with the subtraction term given by combining Eqs. (6.6) of Ref.~\cite{Gehrmann-DeRidder:2007foh}. 

We note that terms \textB{1} and \textC{2} are the integrated counterparts of terms \textA{4} and \textA{7} and \textC{10} in Eq.~\eqref{eq:RRsubLC}.   All terms in Eq.~\eqref{eq:RRsubLC} have now been accounted for in either Eq.~\eqref{eq:RVsubLC} or Eq.~\eqref{eq:VVsubLC}.
The remaining terms in Eq.~\eqref{eq:VVsubLC} are the integrated versions of terms \textD{6--9} and \textE{10-11} in Eq.~\eqref{eq:RVsubLC}.  The $\mathcal{X}_3^{1}$ antenna are presented with the renormalisation scale set to the mass of the antenna, while the unintegrated antenna are subtracted at general $\mu$.  The contributions in terms \textC{3} and \textC{4} proportional to $b_0/\e$ restore the full $\mu$ dependence of the integrated antenna function, where $b_0$ is the leading-colour contribution to the one-loop QCD beta-function:
\begin{equation}
    \beta_0 = b_0 N + b_{0,F} N_f = \frac{11}{6}N-\frac{1}{3}N_f,
\end{equation}
with $N_f$ the number of light fermions. 

We verified that the cancellation of infrared singularities between the subtraction term in~\ref{eq:VVsubLC} and the leading-colour two-loop matrix element happens correctly. The same holds for all the double-virtual subtraction terms in Appendix~\ref{app:VVsub}. This is a strong check of the validity of the implementation we discussed so far and the analytical integration of the new antenna functions.

\section{Numerical validation}\label{sec:checks}

In this Section we discuss a series of checks we performed to probe the correctness of the newly defined antenna functions and of their numerical implementation. We remark that the tests discussed in the following not only validate the $X^0_{5,3}$ and $X^1_{4,3}$ antenna functions, which represent the main novelty of this paper, but also the idealised final-final $X^0_{3}$, $X^0_{4}$ and $X_3^1$ antenna functions introduced in~\cite{Braun-White:2023zwd,Braun-White:2023sgd}, which have been fully incorporated in our numerical setup for the first time in the context of this work. In general, the results we obtain stand as a solid validation of the designer antenna scheme for final-state radiation up to NNLO.

Our main target is to replicate a full NNLO calculation for $e^+e^-\to jjj$, comparing the results and the performance we obtain with the long-established implementation available in~\textsc{NNLOjet}~\cite{Gehrmann:2017xfb} based on the traditional antenna subtraction scheme. 

As a preliminary step, we performed checks at the cross section level for simpler processes: $e^+e^-\to jjj$ at NLO, $e^+e^-\to jj$ at NNLO, $\wt{\chi}\to jj$ (neutralino decay to hadrons~\cite{Gehrmann-DeRidder:2005svg}) at NNLO. We observe complete agreement between results obtained with the traditional and idealised antenna functions. Moreover, to assess the performances of different momentum mappings, we repeated the tests replacing antenna mappings with dipole mappings and found again agreement with no degradation in the numerical convergence. Such computations allowed us to validate all the quark-antiquark and quark-gluon $X_3^0$, $X_4^0$ and $X_3^1$ idealised antenna functions, without the additional complications given by the almost colour-connected sector.

For the NNLO computation of $e^+e^-\to jjj$, we consider $\mu=\sqrt{s}=m_Z=91.186$ GeV. The fiducial phase space is defined imposing a lower bound $y_{cut}=0.05$ on the following event-shapes: 
\begin{itemize}
    \item one-minus-thrust $1-T$ with:
    \begin{equation}\label{thrust}
        T = \text{max}_{|\vec{n}|=1}\left(\frac{\sum_{j}|\vec{p_j}\cdot \vec{n}|}{\sum_{k}|\vec{p_{k}}|}\right),
    \end{equation}
    with the sum running over all external particles. We define the thrust axis $\vec{n}_T$ to be the vector which maximises the expression in the parentheses;
    \item $C$-parameter C, defined through the eigenvalues $\lambda_i$ of the linearised momentum tensor:
    \begin{equation}
    \Theta^{\alpha\beta} = \frac{1}{\sum_k |\vec{p_k}|} \, 
    \sum_{k}\frac{p_k^\alpha p_k^\beta}{ |\vec{p_k}|}\,, \qquad 
    (\alpha,\beta = 1,2,3)\;,
    \end{equation}
    as
    \begin{equation}
    \text{C}= 3\, \left( \lambda_1\lambda_2 + \lambda_2\lambda_3 + \lambda_3\lambda_1
    \right) \;;
    \label{eq:c}
    \end{equation}
    \item total jet broadening TJB, defined starting from the hemisphere broadening:
    \begin{equation}
        B_m=\frac{\sum_{i\in H_m}|p_i\times \vec{n}_T|}{2\sum_i |\vec{p_i}|},\quad m=1,2\;,
    \end{equation}
    as
    \begin{equation}
        \text{TJB}=B_1+B_2;
    \end{equation}
    where the hemispheres $(H_1,H_2)$ are separated by the plane through the origin which is normal to the thrust axis $\vec{n}_T$;
    \item wide jet broadening WJB:
    \begin{equation}
        \text{WJB}=\max(B_1,B_2);
    \end{equation}
    \item two-to-three jet transition variable in the Jade algorithm $y_{23}$ 
    computed as the largest value of $y_{cut,J}$ for which an event is identified as a two-jet event by the Jade algorithm;
    \item  heavy jet mass HJM, defined starting from the single-jet mass:
    \begin{equation}
        \rho_m=\frac{\left(\sum_{i\in H_m}p_i\right)^2}{Q^2},\quad m=1,2\,,
    \end{equation}
    where $H_m$ are the two hemispheres separated by the plane orthogonal to $\vec{n_T}$, as
    \begin{equation}
        \text{HJM}=\max(\rho_1,\rho_2);
    \end{equation}
\end{itemize}
For phenomenological studies, one typically chooses a smaller value of $y_{cut}$, however we want to prevent our numerical results from being dominated by the large logarithmic behaviour of the cross section in the two-jet limit $y_{cut}\to 0$. This allows us to precisely probe potential discrepancies between the traditional and new implementations in the fully-resolved three-jet region. Since our purpose here is the validation of the new framework, rather than the delivery of theoretical predictions, we present results for the NNLO coefficient only in the form of ratios to the traditional setup. Different colour factors in the matrix elements require different combinations of antenna functions. Therefore, to better ensure the correctness of the new ingredients, we separate our result by colour factors. We neglect the singlet contribution, which is infrared-finite~\cite{Gehrmann-DeRidder:2007foh}.

\begin{figure}[h]
    \centering
    \includegraphics[width=\linewidth]{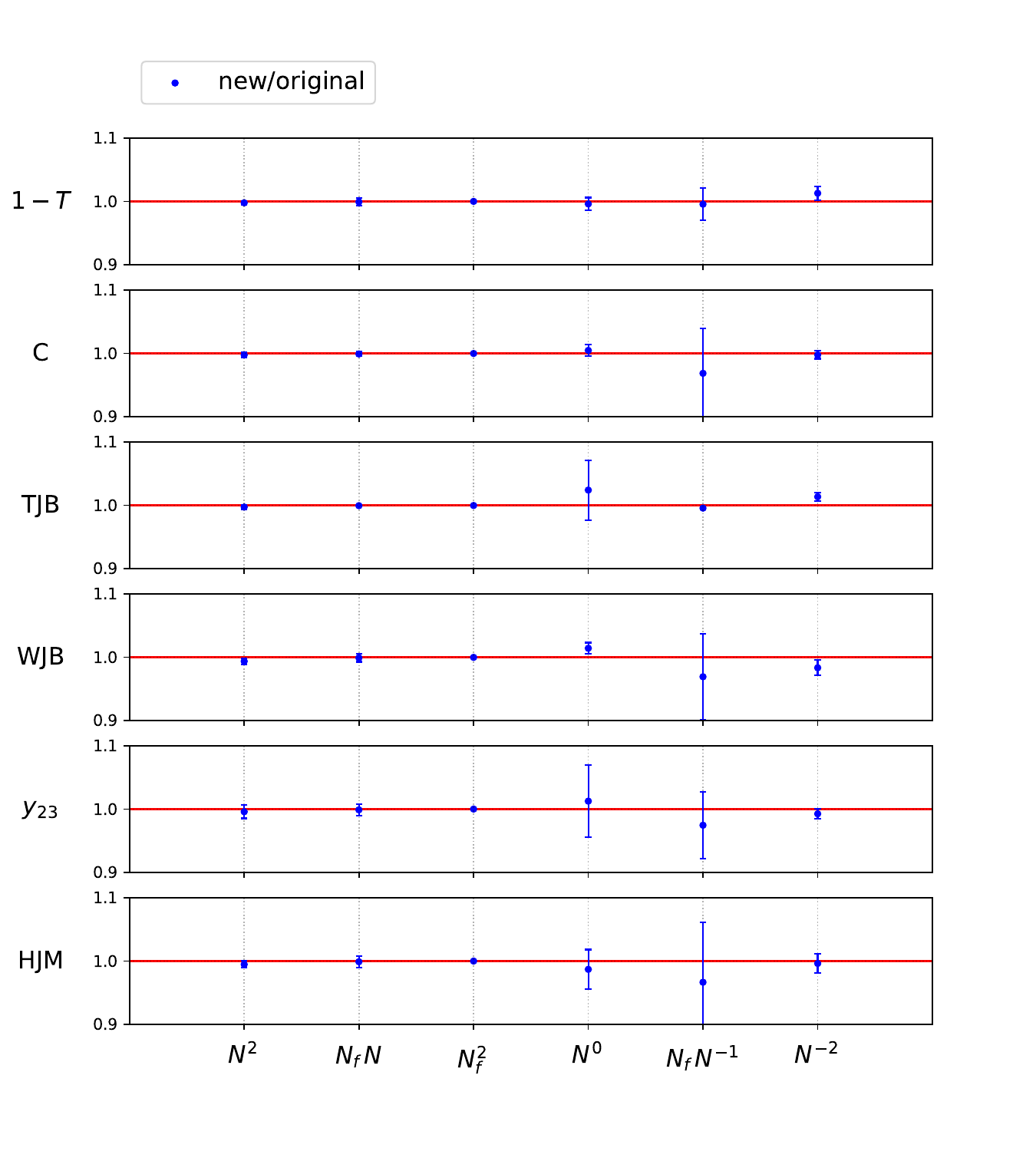}
    \caption{Ratio between the NNLO coefficient for the total cross section of $e^+e^-\to jjj$ calculated with the new and the original implementation of antenna subtraction as the integral under the distribution in the observables indicated on the left. The results are separated by colour factor and the error bars represent Monte Carlo uncertainties.}
    \label{fig:totalXS}
\end{figure}

For each event-shape we compute the NNLO corrections to the total cross section as:
\begin{equation}
    \sigma^{\text{tot.}}_{NNLO} = \int_{y_{cut}}^{\mathcal{O}_{\text{max}}}\text{d}\mathcal{O}\,\dfrac{\text{d}\sigma}{\text{d}\mathcal{O}}\Bigg|_{NNLO},
\end{equation}
where $\mathcal{O}_{\text{max}}$ is the appropriate upper bound. In Figure~\ref{fig:totalXS} we show the comparison between the new and the original implementation separated by colour factor. The error bars represent the Monte Carlo uncertainty. In general, we observe percent-level (or better) agreement on the NNLO correction, which is quite beyond the accuracy typically required for phenomenological applications, but it is relevant in the context of a proof-of-principle validation of the newly proposed framework. 

For some of the event-shapes, the $N^0$ and $N_fN^{-1}$ colour factor exhibit larger error bars. This is due to the fact that the Monte Carlo integration struggles to adapt to these subleading-colour contributions, due to the oscillatory behaviour of the integrand. In a full calculation, these colour factor have a small numerical impact on the NNLO coefficient: $\lesssim 10\%$ for $N^0$ and $\lesssim 1\%$ for $N_fN^{-1}$, therefore one typically does not have to resolve them with the same accuracy as the leading-colour $N^2$ or $N_fN$ contributions (colour sampling). Here we pushed the numerical integration to a point where we can confidently claim very good agreement between the two implementations, without exaggerating the computational cost of this validation. For the remaining colour factors we reach percent or even sub-percent accuracy on the ratio, showing excellent compatibility with the original implementation. Overall, the comparison in Figure~\ref{fig:totalXS} stands as a very solid confirmation of the correctness of the new approach.

\begin{figure}[h]
    \centering
    \includegraphics[width=\linewidth]{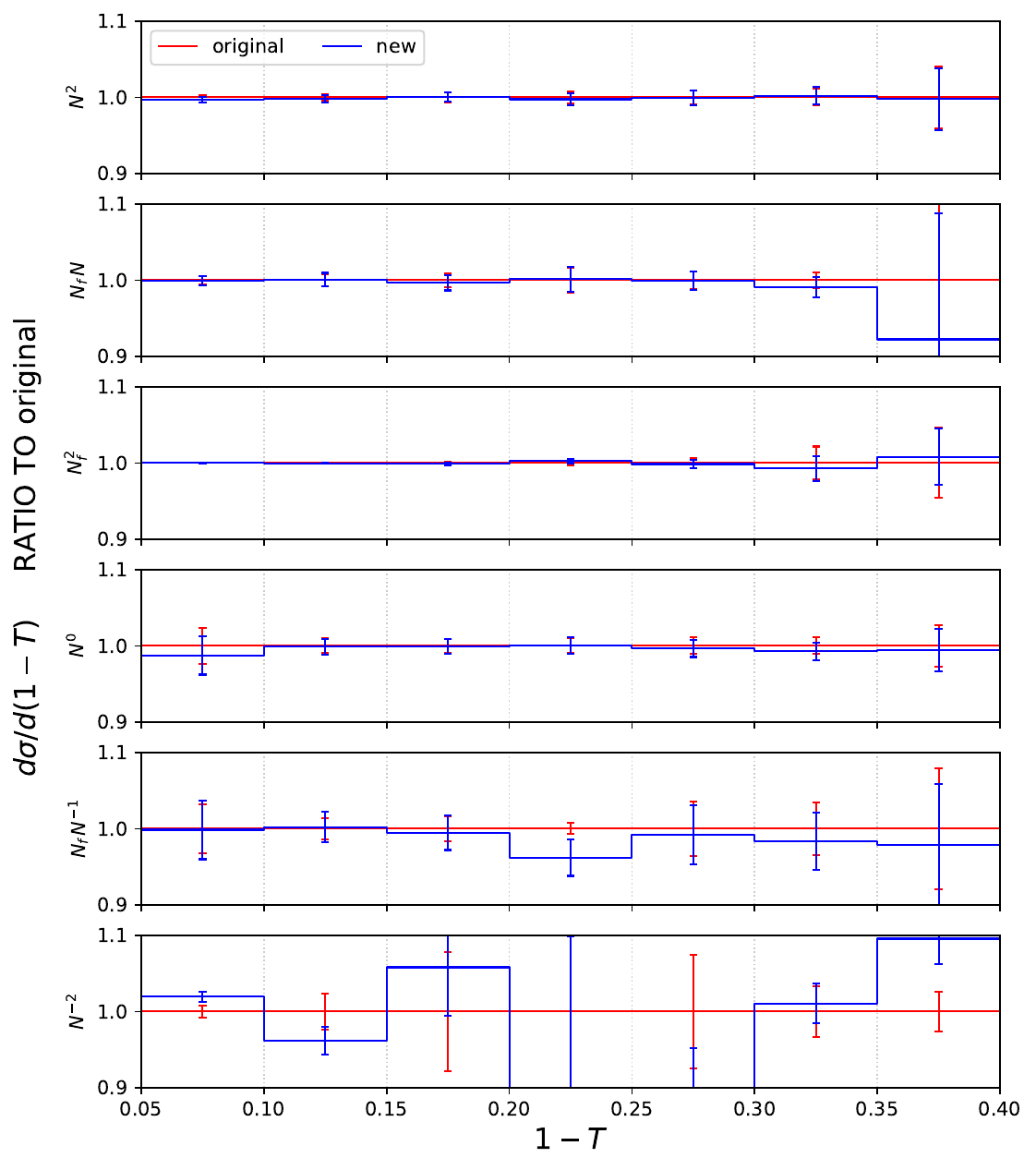}
    \caption{Ratio between the NNLO coefficient for the differential distribution in $1-T$ calculated with the new and the original implementation of antenna subtraction. Each plot refers to a single colour factor, indicated on the left. The error bars indicate the Monte Carlo uncertainties.}
    \label{fig:1mT}
\end{figure}

\begin{figure}[h]
    \centering
    \includegraphics[width=\linewidth]{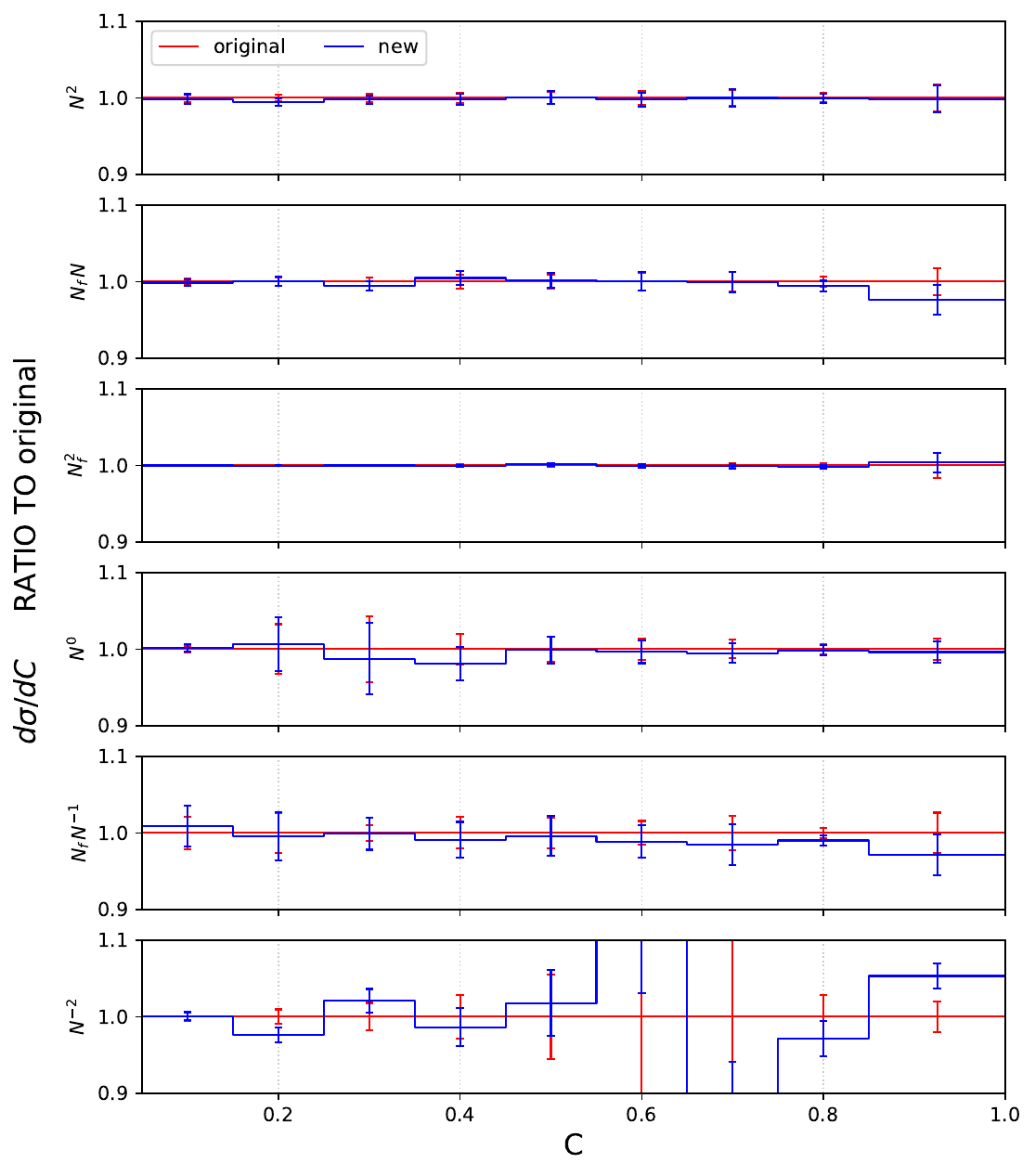}
    \caption{Ratio between the NNLO coefficient for the differential distribution in the C-parameter calculated with the new and the original implementation of antenna subtraction. Each plot refers to a single colour factor, indicated on the left. The error bars indicate the Monte Carlo uncertainties.}
    \label{fig:C}
\end{figure}

\begin{figure}[h]
    \centering
    \includegraphics[width=\linewidth]{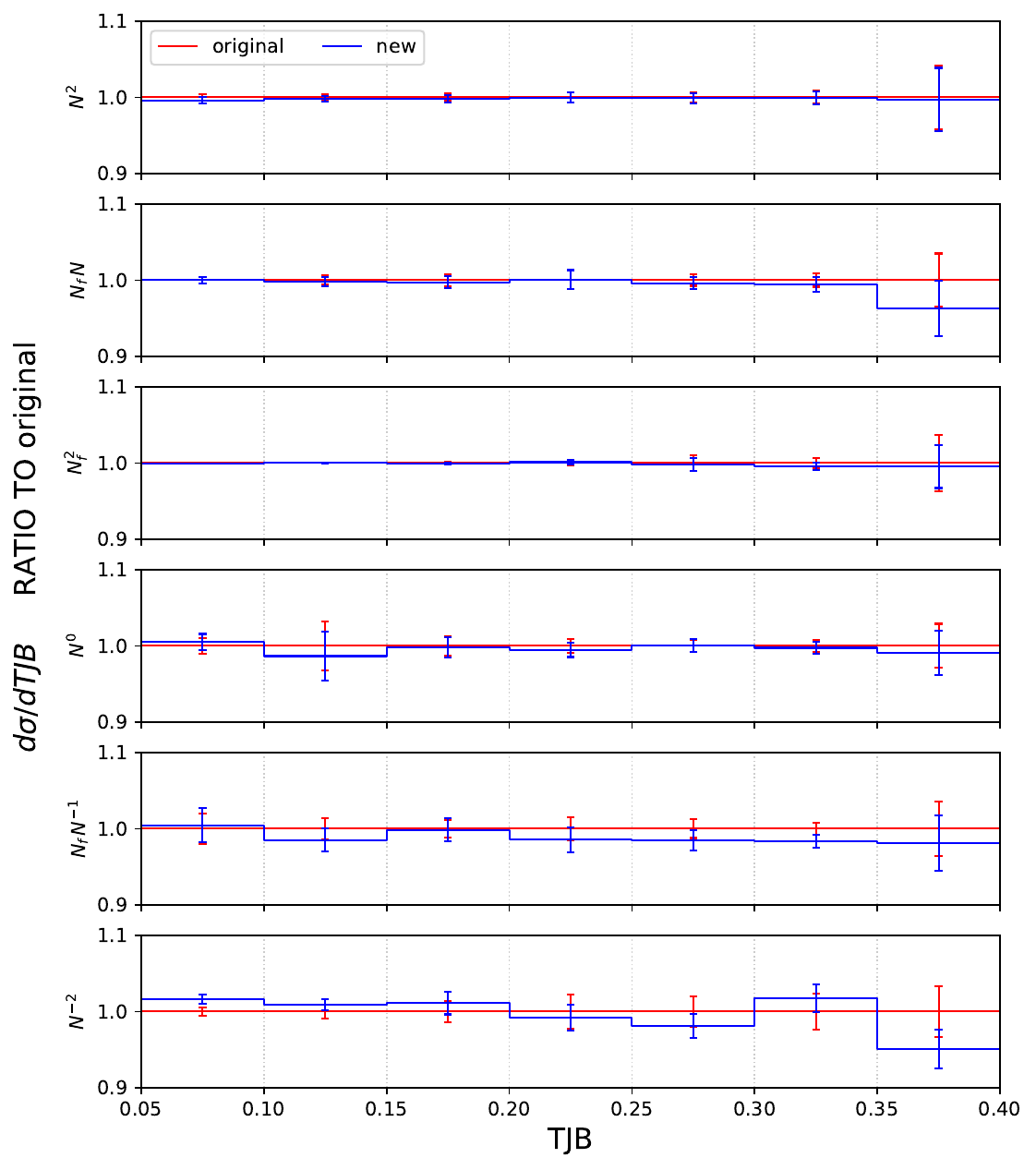}
    \caption{Ratio between the NNLO coefficient for the differential distribution in the total jet broadening calculated with the new and the original implementation of antenna subtraction. Each plot refers to a single colour factor, indicated on the left. The error bars indicate the Monte Carlo uncertainties.}
    \label{fig:TJB}
\end{figure}

\begin{figure}[h]
    \centering
    \includegraphics[width=\linewidth]{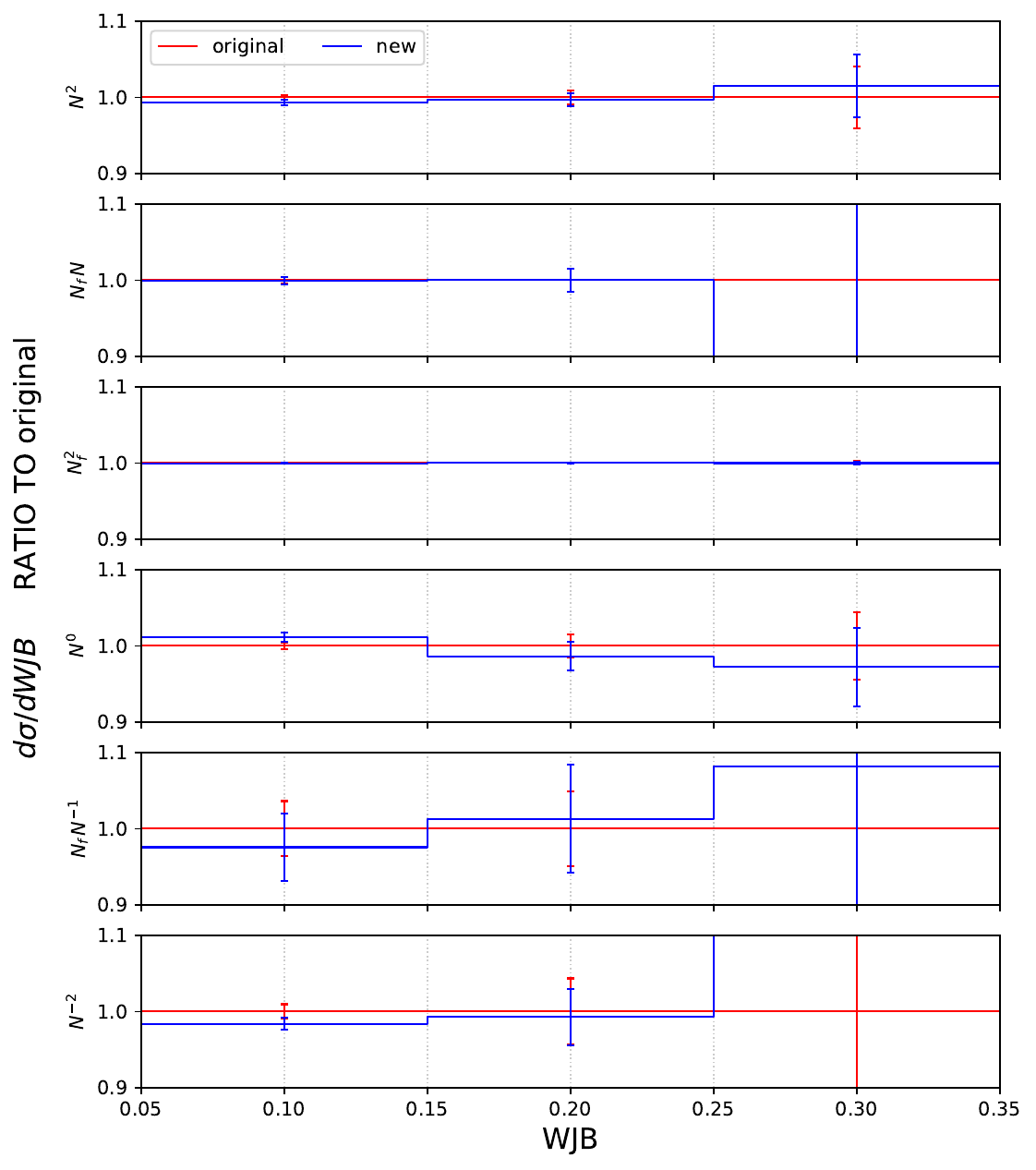}
    \caption{Ratio between the NNLO coefficient for the differential distribution in the wide jet broadening calculated with the new and the original implementation of antenna subtraction. Each plot refers to a single colour factor, indicated on the left. The error bars indicate the Monte Carlo uncertainties.}
    \label{fig:WJB}
\end{figure}

\begin{figure}[h]
    \centering
    \includegraphics[width=\linewidth]{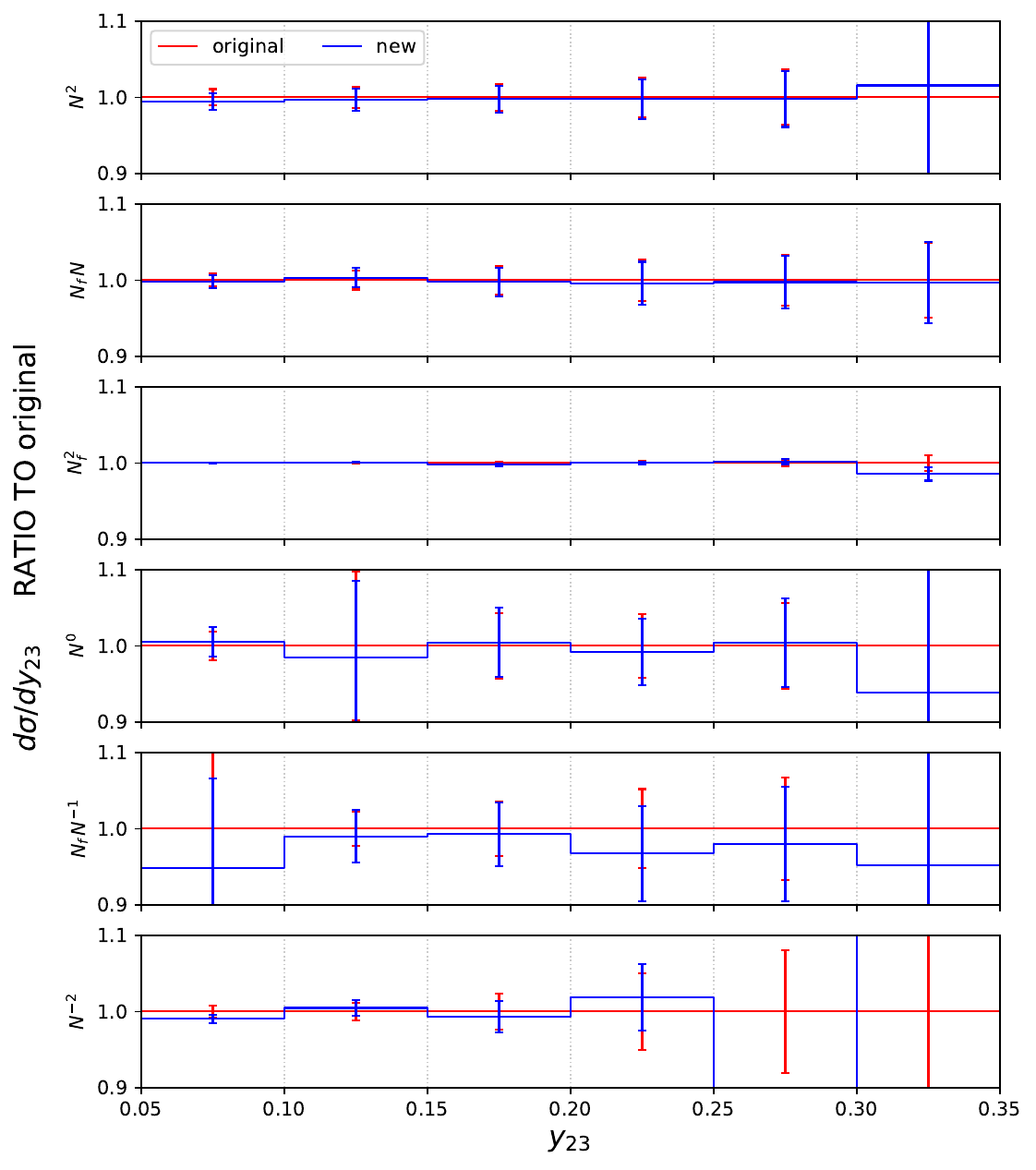}
    \caption{Ratio between the NNLO coefficient for the differential distribution in the jet transition variable $y_{23}$ calculated with the new and the original implementation of antenna subtraction. Each plot refers to a single colour factor, indicated on the left. The error bars indicate the Monte Carlo uncertainties.}
    \label{fig:y23}
\end{figure}

To ensure that the agreement extends beyond the total cross section, in Figures~\ref{fig:1mT}-\ref{fig:y23} we show differential results for the event-shapes listed above. In general, no significant deviations are observed between the new implementation and the traditional one. Large error bars in some bins are due to the numerical value of the differential cross section being very close to zero. 

One of the main motivations of this work is the improvement of the computational performances of the antenna subtraction method. As we mentioned before, the new antenna functions open up the possibility of constructing local subtraction terms for individual colour-ordered matrix elements, summing over colour orderings \textit{a posteriori} by multiplication of the result by a suitable symmetry factor. This is in general not possible with traditional antenna functions, because the almost colour-connected sector requires a sum over colour orderings to achieve local cancellation of infrared singularities. Thanks to this feature, for a generic sub-process with $n$ gluons, one can expect an $n!$ speedup for the point-by-point evaluation of matrix elements and associated subtraction terms. This speedup in general does not directly transfer to the runtime for the full Monte Carlo due to other time-consuming tasks (phase-space generation, observables evaluation, \ldots) and to the presence of several partonic channels. However, this improvement clearly becomes more relevant the higher the multiplicity and the more computational time is required by complicated matrix elements. On top of this, an individual colour-ordered squared matrix element exhibits a much more regular behaviour than the full matrix element, due to the fewer infrared divergences, especially for the double-real correction. This significantly helps the Monte Carlo importance sampling procedure, which we perform through the VEGAS algorithm~\cite{Lepage:1977sw}, with the identification of the relevant phase space region, and hence results in a faster numerical convergence.

For $e^+e^-\to jjj$ at NNLO, the speedup in a specific calculation is affected by many factors, such as the choice of the fiducial phase space, the observables to be computed and the relative size of the double-virtual, real-virtual and double-real contributions. According to the numerical tests we performed, we can in general expect the new implementation to be five to ten times faster than the original one in reaching a given numerical accuracy. For the reasons explained above, the speedup for higher multiplicities will surely be even more significant.


\newpage

\newpage

\newpage

\newpage

\newpage

\newpage
\begin{figure}[h]
    \centering
    \includegraphics[width=\linewidth]{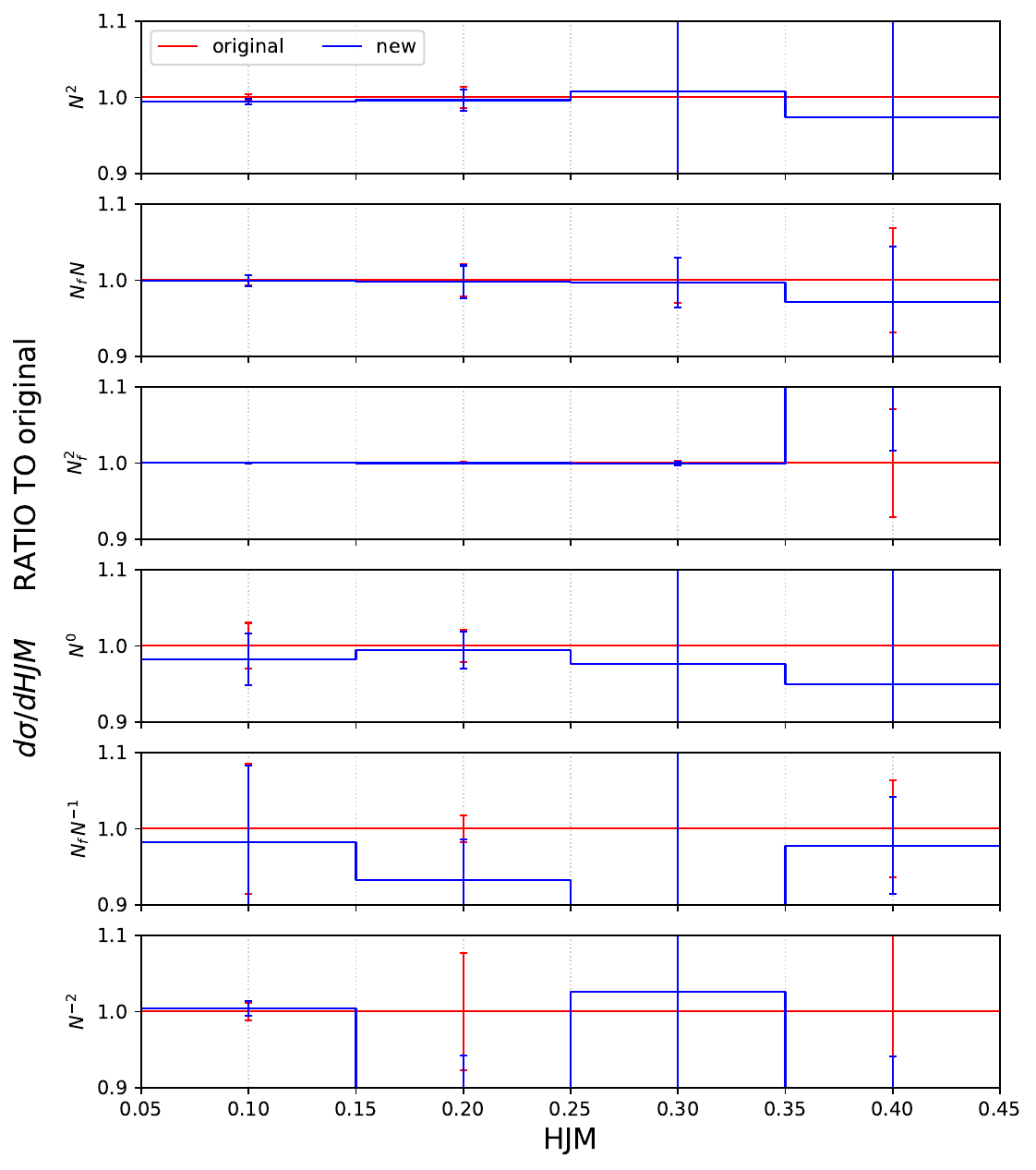}
    \caption{Ratio between the NNLO coefficient for the differential distribution in the heavey jet mass calculated with the new and the original implementation of antenna subtraction. Each plot refers to a single colour factor, indicated on the left. The error bars indicate the Monte Carlo uncertainties.}
    \label{fig:HJM}
\end{figure}
\newpage

\section{Conclusions and outlook}\label{sec:conclusions}

We discussed the definition and implementation of final-state generalised antenna functions for NNLO calculations in QCD, in particular tree-level five-parton three-hard-radiator ($X^0_{5,3}$) and one-loop four-parton three-hard-radiator ( $X^1_{4,3}$) antenna functions. In contrast to traditional antenna functions, generalised ones can be constructed to have more than two hard radiators, relying on the recently proposed idealised antenna algorithm. This feature makes them particularly suitable to remove infrared singularities for almost colour-connected emissions, resulting in a major simplification of the subtraction infrastructure compared to the traditional antenna scheme. One of the main advantages lies in the possibility of assembling local subtraction terms for individual colour-ordered matrix elements, which reduces the computational cost of NNLO calculations. We showed how, through a careful choice of momentum mappings, one can write down an exact factorisation of the phase space leading to a fully analytical integration of the new antenna functions. 

Considering three-jet production at electron-positron colliders as a test case, we demonstrated the complete cancellation of infrared singularities in the new approach. We also performed several numerical tests to assess the correctness of the setup. Most importantly, we were able to reproduce inclusive and differential results obtained with the long-established traditional antenna subtraction method with excellent agreement. We estimate that the calculation with the new scheme is five to ten times faster than the original implementation, depending on the specific computational setup. The speedup increases at higher multiplicities. We remark that the results we obtained not only stand as a thorough validation of generalised antenna functions, but also as a solid check of standard two-hard-radiator antenna functions obtained according to the idealised antenna algorithm, employed here in an NNLO calculation for the first time.

This works paves the way to several future developments. First of all, a natural phenomenological application of the antenna functions computed here is the calculation of the NNLO correction to $e^+e^-\to jjjj$, which would greatly benefit from the improvements and the speedup offered by generalised antenna functions. Secondly, we foresee an extension to hadronic processes through the definition of generalised antenna functions with one or two hard radiators in the initial-state. Most of the concepts and techniques discussed in this paper apply straightforwardly to the initial-state radiation case too, while significant differences lie in the analytical integration of antenna functions and in the bookkeeping of partonic channels. A necessary preliminary step in this direction consists in the continuation of the work in~\cite{Fox:2023bma} with the construction of idealised four-parton initial-final and initial-initial antenna functions. Finally, as a long-term outlook, generalised antenna functions can be assembled for N$^3$LO calculations too. At this perturbative order, the simplifications due to the new antenna functions will be pivotal to achieve a general local subtraction scheme. 

\section*{Acknowledgements}\label{sec:ack}
We thank Oscar Braun-White, Xuan Chen, Aude Gehrmann-De Ridder, Thomas Gehrmann, Alexander Huss,  Petr~Jakub\v{c}{\'i}k, Markus~L\"ochner and Christian Preuss for enlightening discussions and helpful advice during the early stages of this work. 
We also thank Thomas Gehrmann and Christian Preuss for valuable comments on the draft.  
We gratefully acknowledge support from the UK Science and Technology Facilities Council (STFC) under contract ST/X000745/1.
MM is supported by a Royal Society Newton International Fellowship (NIF/R1/232539). 


\appendix
\section{Limits of $X_{5,3}^0$ antenna functions}\label{app:limX53}

We present here the unresolved limits of each $\X$ antenna function. In terms of notation and explicit expressions of projectors and unresolved factors, we follow~\cite{Braun-White:2023sgd}.

\subsection*{$F_{5,3}^{0}(i_g,j_g,k_g,l_g,m_g)$}
\renewcommand{\ant}{F_{5,3}^{0}(i^h,j,k^h,l,m^h)}
\begin{align}
\PDSdown_{jl} \ant &=  S_g(i^h,j,k^h)\, S_g(k^h,l,m^h)\nn \\
\PTCdown_{jkl} \ant &=  P_{gg}^{sub}((j+k)^h,l)
         \, P_{gg}^{sub}(k^h,j)
         \nn \\  & + P_{gg}^{sub}((k+l)^h,j)
         \, P_{gg}^{sub}(k^h,l)\nn \\
\PDCdown_{ij;kl} \ant &=  P_{gg}^{sub}(i^h,j)
          \, P_{gg}^{sub}(k^h,l)\nn \\
\PDCdown_{ij;lm} \ant &=  P_{gg}^{sub}(i^h,j)
          \, P_{gg}^{sub}(m^h,l)\nn \\
\PDCdown_{jk;lm} \ant &=  P_{gg}^{sub}(k^h,j)
          \, P_{gg}^{sub}(m^h,l)\nn \\
\PSdown_{j} \ant &=  S_g(i^h,j,k^h)\, F_3^0(k,l,m)\nn \\
\PSdown_{l} \ant &=  S_g(k^h,l,m^h)\, F_3^0(i,j,k)\nn \\
\PCdown_{ij} \ant &=  P_{gg}^{sub}(i^h,j)\, F_3^0(k,l,m)\nn \\
\PCdown_{jk} \ant &=  P_{gg}^{sub}(k^h,j)\, F_3^0((j+k),l,m)\nn \\
\PCdown_{kl} \ant &=  P_{gg}^{sub}(k^h,l)\, F_3^0(i,j,(k+l))\nn \\
\PCdown_{lm} \ant &=  P_{gg}^{sub}(m^h,l)\, F_3^0(i,j,k) 
\end{align}
\subsection*{$G_{5,3}^{0 (a)}(i_{\Qb},j_Q,k_g,l_g,m_g)$}
\renewcommand{\ant}{G_{5,3}^{0 (a)}(i^h,j,k^h,l,m^h)}
\begin{align}
\PDCdown_{ij;lm} \ant &=  P_{gg}^{sub}(m^h,l)
          \, P_{q\qb}(i^h,j)\nn \\
\PDCdown_{ij;kl} \ant &=  P_{gg}^{sub}(k^h,l)
          \, P_{q\qb}(i^h,j)\nn \\
\PSdown_{l} \ant &=  S_g(k^h,l,m^h)\, G_3^0(k,j,i)\nn \\
\PCdown_{lm} \ant &=  P_{gg}^{sub}(m^h,l)\, G_3^0(k,j,i)\nn \\
\PCdown_{kl} \ant &=  P_{gg}^{sub}(k^h,l)\, G_3^0((k+l),j,i)\nn \\
\PCdown_{ij} \ant &=  P_{q\qb}(i^h,j)\, F_3^0(m,l,k)
\end{align}
\subsection*{$G_{5,3}^{0 (b)}(i_g,j_{\Qb},k_Q,l_g,m_g)$}
\renewcommand{\ant}{G_{5,3}^{0 (b)}(i^h,j,k^h,l,m^h)}
\begin{align}
\PTCdown_{jkl} \ant &=  P_{q\qb}((k+l)^h,j)
         \, P_{qg}(k^h,l)
         \nn \\  & + P_{gg}^{sub}((j+k)^h,l)
         \, P_{q\qb}(k^h,j)\nn \\
\PDCdown_{jk;lm} \ant &=  P_{gg}^{sub}(m^h,l)
          \, P_{q\qb}(k^h,j)\nn \\
\PSdown_{l} \ant &=  S_g(k^h,l,m^h)\, G_3^0(i,j,k)\nn \\
\PCdown_{jk} \ant &=  P_{q\qb}(k^h,j)\, F_3^0(m,l,(j+k))\nn \\
\PCdown_{kl} \ant &=  P_{qg}(k^h,l)\, G_3^0(i,j,(k+l))\nn \\
\PCdown_{lm} \ant &=  P_{gg}^{sub}(m^h,l)\, G_3^0(i,j,k)
\end{align}
\subsection*{$H_{5,3}^{0 (a)}(i_Q,j_{\Qb},k_g,l_{\qb},m_q)$}
\renewcommand{\ant}{H_{5,3}^{0 (a)}(i^h,j,k^h,l,m^h)}
\begin{align}
\PDCdown_{ij;lm} \ant &=  P_{q\qb}(i^h,j)
          \, P_{q\qb}(m^h,l)\nn \\
\PCdown_{ij} \ant &=  P_{q\qb}(i^h,j)\, G_3^0(k,l,m)\nn \\
\PCdown_{lm} \ant &=  P_{q\qb}(m^h,l)\, G_3^0(k,j,i) 
\end{align}
\subsection*{$H_{5,3}^{0 (b)}(i_g,j_{\Qb},k_Q,l_{\qb},m_q)$}
\renewcommand{\ant}{H_{5,3}^{0 (b)}(i^h,j,k^h,l,m^h)}
\begin{align}
\PDCdown_{jk;lm} \ant &=  P_{q\qb}(k^h,j)
          \, P_{q\qb}(m^h,l)\nn \\
\PCdown_{jk} \ant &=  P_{q\qb}(k^h,j)\, G_3^0((j+k),l,m)\nn \\
\PCdown_{lm} \ant &=  P_{q\qb}(m^h,l)\, G_3^0(i,j,k)
\end{align}
\subsection*{$D_{5,3}^{0}(i_q,j_g,k_g,l_g,m_g)$}
\renewcommand{\ant}{D_{5,3}^{0}(i^h,j,k^h,l,m^h)}
\begin{align}
\PDSdown_{jl} \ant &=  S_g(i^h,j,k^h)\, S_g(k^h,l,m^h)\nn \\
\PTCdown_{jkl} \ant &=  P_{gg}^{sub}((j+k)^h,l)
         \, P_{gg}^{sub}(k^h,j)
         \nn \\  & + P_{gg}^{sub}((k+l)^h,j)
         \, P_{gg}^{sub}(k^h,l)\nn \\
\PDCdown_{ij;kl} \ant &=  P_{qg}(i^h,j)
          \, P_{gg}^{sub}(k^h,l)\nn \\
\PDCdown_{ij;lm} \ant &=  P_{qg}(i^h,j)
          \, P_{gg}^{sub}(m^h,l)\nn \\
\PDCdown_{jk;lm} \ant &=  P_{gg}^{sub}(k^h,j)
          \, P_{gg}^{sub}(m^h,l)\nn \\
\PSdown_{j} \ant &=  S_g(i^h,j,k^h)\, F_3^0(k,l,m)\nn \\
\PSdown_{l} \ant &=  S_g(k^h,l,m^h)\, D_3^0(i,j,k)\nn \\
\PCdown_{ij} \ant &=  P_{qg}(i^h,j)\, F_3^0(k,l,m)\nn \\
\PCdown_{jk} \ant &=  P_{gg}^{sub}(k^h,j)\, F_3^0((j+k),l,m)\nn \\
\PCdown_{kl} \ant &=  P_{gg}^{sub}(k^h,l)\, D_3^0(i,j,(k+l))\nn \\
\PCdown_{lm} \ant &=  P_{gg}^{sub}(m^h,l)\, D_3^0(i,j,k)
\end{align}
\subsection*{$E_{5,3}^{0 (a)}(i_q,j_{\Qb},k_Q,l_g,m_g)$}
\renewcommand{\ant}{E_{5,3}^{0 (a)}(i^h,j,k^h,l,m^h)}
\begin{align}
\PTCdown_{jkl} \ant &=  P_{q\qb}((k+l)^h,j)
         \, P_{qg}(k^h,l)
         \nn \\  & + P_{gg}^{sub}((j+k)^h,l)
         \, P_{q\qb}(k^h,j)\nn \\
\PDCdown_{jk;lm} \ant &=  P_{gg}^{sub}(m^h,l)
          \, P_{q\qb}(k^h,j)\nn \\
\PSdown_{l} \ant &=  S_g(k^h,l,m^h)\, E_3^0(i,j,k)\nn \\
\PCdown_{lm} \ant &=  P_{gg}^{sub}(m^h,l)\, E_3^0(i,j,k)\nn \\
\PCdown_{kl} \ant &=  P_{qg}(k^h,l)\, E_3^0(i,j,(k+l))\nn \\
\PCdown_{jk} \ant &=  P_{q\qb}(k^h,j)\, F_3^0(m,l,(j+k))
\end{align}
\subsection*{$E_{5,3}^{0 (b)}(i_q,j_g,k_{\Qb},l_Q,m_g)$}
\renewcommand{\ant}{E_{5,3}^{0 (b)}(i^h,j,k^h,l,m^h)}
\begin{align}
\PTCdown_{jkl} \ant &=  P_{q\qb}((j+k)^h,l)
         \, P_{qg}(k^h,j)
         \nn \\  & + P_{gg}^{sub}((k+l)^h,j)
         \, P_{q\qb}(k^h,l)\nn \\
\PDCdown_{ij;kl} \ant &=  P_{qg}(i^h,j)
          \, P_{q\qb}(k^h,l)\nn \\
\PSdown_{j} \ant &=  S_g(i^h,j,k^h)\, G_3^0(m,l,k)\nn \\
\PCdown_{ij} \ant &=  P_{qg}(i^h,j)\, G_3^0(m,l,k)\nn \\
\PCdown_{jk} \ant &=  P_{qg}(k^h,j)\, G_3^0(m,l,(j+k))\nn \\
\PCdown_{kl} \ant &=  P_{q\qb}(k^h,l)\, D_3^0(i,j,(k+l))
\end{align}
\subsection*{$E_{5,3}^{0 (c)}(i_q,j_g,k_g,l_{\Qb},m_Q)$}
\renewcommand{\ant}{E_{5,3}^{0 (c)}(i^h,j,k^h,l,m^h)}
\begin{align}
\PDCdown_{ij;lm} \ant &=  P_{qg}(i^h,j)
          \, P_{q\qb}(m^h,l)\nn \\
\PDCdown_{jk;lm} \ant &=  P_{gg}^{sub}(k^h,j)
          \, P_{q\qb}(m^h,l)\nn \\
\PSdown_{j} \ant &=  S_g(i^h,j,k^h)\, G_3^0(k,l,m)\nn \\
\PCdown_{ij} \ant &=  P_{qg}(i^h,j)\, G_3^0(k,l,m)\nn \\
\PCdown_{jk} \ant &=  P_{gg}^{sub}(k^h,j)\, G_3^0((j+k),l,m)\nn \\
\PCdown_{lm} \ant &=  P_{q\qb}(m^h,l)\, D_3^0(i,j,k) 
\end{align}
\subsection*{$E_{5,3}^{0 (d)}(i_{\Qb},j_Q,k_{\qb},l_g,m_g)$}
\renewcommand{\ant}{E_{5,3}^{0 (d)}(i^h,j,k^h,l,m^h)}
\begin{align}
\PDCdown_{ij;lm} \ant &=  P_{gg}^{sub}(m^h,l)
          \, P_{q\qb}(i^h,j)\nn \\
\PDCdown_{ij;kl} \ant &=  P_{qg}(k^h,l)
          \, P_{q\qb}(i^h,j)\nn \\
\PSdown_{l} \ant &=  S_g(k^h,l,m^h)\, E_3^0(k,j,i)\nn \\
\PCdown_{lm} \ant &=  P_{gg}^{sub}(m^h,l)\, E_3^0(k,j,i)\nn \\
\PCdown_{kl} \ant &=  P_{qg}(k^h,l)\, E_3^0((k+l),j,i)\nn \\
\PCdown_{ij} \ant &=  P_{q\qb}(i^h,j)\, D_3^0(k,l,m)
\end{align}
\subsection*{$K_{5,3}^{0}(i_q,j_{\Qb},k_Q,l_{\Rb},m_R)$}
\renewcommand{\ant}{K_{5,3}^{0}(i^h,j,k^h,l,m^h)}
\begin{align}
\PDCdown_{jk;lm} \ant &=  P_{q\qb}(k^h,j)
          \, P_{q\qb}(m^h,l)\nn \\
\PCdown_{jk} \ant &=  P_{q\qb}(k^h,j)\, G_3^0((j+k),l,m)\nn \\
\PCdown_{lm} \ant &=  P_{q\qb}(m^h,l)\, E_3^0(i,j,k)
\end{align}
\subsection*{$A_{5,3}^{0}(i_q,j_g,k_g,l_g,m_{\qb})$}
\renewcommand{\ant}{A_{5,3}^{0}(i^h,j,k^h,l,m^h)}
\begin{align}
\PDSdown_{jl} \ant &=  S_g(i^h,j,k^h)\, S_g(k^h,l,m^h)\nn \\
\PTCdown_{jkl} \ant &=  P_{gg}^{sub}((j+k)^h,l)
         \, P_{gg}^{sub}(k^h,j)
         \nn \\  & + P_{gg}^{sub}((k+l)^h,j)
         \, P_{gg}^{sub}(k^h,l)\nn \\
\PDCdown_{ij;kl} \ant &=  P_{qg}(i^h,j)
          \, P_{gg}^{sub}(k^h,l)\nn \\
\PDCdown_{ij;lm} \ant &=  P_{qg}(i^h,j)
          \, P_{qg}(m^h,l)\nn \\
\PDCdown_{jk;lm} \ant &=  P_{gg}^{sub}(k^h,j)
          \, P_{qg}(m^h,l)\nn \\
\PSdown_{j} \ant &=  S_g(i^h,j,k^h)\, D_3^0(m,l,k)\nn \\
\PSdown_{l} \ant &=  S_g(k^h,l,m^h)\, D_3^0(i,j,k)\nn \\
\PCdown_{ij} \ant &=  P_{qg}(i^h,j)\, D_3^0(m,l,k)\nn \\
\PCdown_{jk} \ant &=  P_{gg}^{sub}(k^h,j)\, D_3^0(m,l,(j+k))\nn \\
\PCdown_{kl} \ant &=  P_{gg}^{sub}(k^h,l)\, D_3^0(i,j,(k+l))\nn \\
\PCdown_{lm} \ant &=  P_{qg}(m^h,l)\, D_3^0(i,j,k)
\end{align}
\subsection*{$B_{5,3}^{0}(i_q,j_g,k_{\Qb},l_Q,m_{\qb})$}
\renewcommand{\ant}{B_{5,3}^{0}(i^h,j,k^h,l,m^h)}
\begin{align}
\PTCdown_{jkl} \ant &=  P_{q\qb}((j+k)^h,l)
         \, P_{qg}(k^h,j)
         \nn \\  & + P_{gg}^{sub}((k+l)^h,j)
         \, P_{q\qb}(k^h,l)\nn \\
\PDCdown_{ij;kl} \ant &=  P_{qg}(i^h,j)
          \, P_{q\qb}(k^h,l)\nn \\
\PSdown_{j} \ant &=  S_g(i^h,j,k^h)\, E_3^0(m,l,k)\nn \\
\PCdown_{ij} \ant &=  P_{qg}(i^h,j)\, E_3^0(m,l,k)\nn \\
\PCdown_{jk} \ant &=  P_{qg}(k^h,j)\, E_3^0(m,l,(j+k))\nn \\
\PCdown_{kl} \ant &=  P_{q\qb}(k^h,l)\, D_3^0(i,j,(k+l)) 
\end{align}
\subsection*{$\wt{A}_{5,3}^{0}(i_{\qb},j_{\gamma},k_q,l_g,m_g)$}
\renewcommand{\ant}{\wt{A}_{5,3}^{0}(i^h,j,k^h,l,m^h)}
\begin{align}
\PDSdown_{jl} \ant &=  S_g(i^h,j,k^h)\, S_g(k^h,l,m^h)\nn \\
\PTCdown_{jkl} \ant &=  P_{qg}((j+k)^h,l)
         \, P_{qg}(k^h,j)
         \nn \\  & + P_{qg}((k+l)^h,j)
         \, P_{qg}(k^h,l)
         +R_{q\gamma\gamma}(k^h,j,l)\nn \\
\PDCdown_{ij;kl} \ant &=  P_{qg}(i^h,j)
          \, P_{qg}(k^h,l)\nn \\
\PDCdown_{ij;lm} \ant &=  P_{qg}(i^h,j)
          \, P_{gg}^{sub}(m^h,l)\nn \\
\PDCdown_{jk;lm} \ant &=  P_{qg}(k^h,j)
          \, P_{gg}^{sub}(m^h,l)\nn \\
\PSdown_{j} \ant &=  S_g(i^h,j,k^h)\, D_3^0(k,l,m)\nn \\
\PSdown_{l} \ant &=  S_g(k^h,l,m^h)\, A_3^0(i,j,k)\nn \\
\PCdown_{ij} \ant &=  P_{qg}(i^h,j)\, D_3^0(k,l,m)\nn \\
\PCdown_{jk} \ant &=  P_{qg}(k^h,j)\, D_3^0((j+k),l,m)\nn \\
\PCdown_{kl} \ant &=  P_{qg}(k^h,l)\, A_3^0(i,j,(k+l))\nn \\
\PCdown_{lm} \ant &=  P_{gg}^{sub}(m^h,l)\, A_3^0(i,j,k)
\end{align}
\subsection*{$\wt{B}_{5,3}^{0}(i_{\qb},j_{\gamma},k_q,l_{\Qb},m_Q)$}
\renewcommand{\ant}{\wt{B}_{5,3}^{0}(i^h,j,k^h,l,m^h)}
\begin{align}
\PDCdown_{ij;lm} \ant &=  P_{qg}(i^h,j)
          \, P_{q\qb}(m^h,l)\nn \\
\PDCdown_{jk;lm} \ant &=  P_{qg}(k^h,j)
          \, P_{q\qb}(m^h,l)\nn \\
\PSdown_{j} \ant &=  S_g(i^h,j,k^h)\, E_3^0(k,l,m)\nn \\
\PCdown_{ij} \ant &=  P_{qg}(i^h,j)\, E_3^0(k,l,m)\nn \\
\PCdown_{jk} \ant &=  P_{qg}(k^h,j)\, E_3^0((j+k),l,m)\nn \\
\PCdown_{lm} \ant &=  P_{q\qb}(m^h,l)\, A_3^0(i,j,k)
\end{align}
\subsection*{$\wt{\wt{A}}_{5,3}^{0}(i_{\qb},j_{\gamma},k_q,l_g,m_{\Qb})$}
\renewcommand{\ant}{\wt{\wt{A}}_{5,3}^{0}(i^h,j,k^h,l,m^h)}
\begin{align}
\PDSdown_{jl} \ant &=  S_g(i^h,j,k^h)\, S_g(k^h,l,m^h)\nn \\
\PTCdown_{jkl} \ant &=  P_{qg}((j+k)^h,l)
         \, P_{qg}(k^h,j)
         \nn \\  & + P_{qg}((k+l)^h,j)
         \, P_{qg}(k^h,l)
         +R_{q\gamma\gamma}(k^h,j,l)\nn \\
\PDCdown_{ij;kl} \ant &=  P_{qg}(i^h,j)
          \, P_{qg}(k^h,l)\nn \\
\PDCdown_{ij;lm} \ant &=  P_{qg}(i^h,j)
          \, P_{qg}(m^h,l)\nn \\
\PDCdown_{jk;lm} \ant &=  P_{qg}(k^h,j)
          \, P_{qg}(m^h,l)\nn \\
\PSdown_{j} \ant &=  S_g(i^h,j,k^h)\, A_3^0(k,l,m)\nn \\
\PSdown_{l} \ant &=  S_g(k^h,l,m^h)\, A_3^0(i,j,k)\nn \\
\PCdown_{ij} \ant &=  P_{qg}(i^h,j)\, A_3^0(k,l,m)\nn \\
\PCdown_{jk} \ant &=  P_{qg}(k^h,j)\, A_3^0((j+k),l,m)\nn \\
\PCdown_{kl} \ant &=  P_{qg}(k^h,l)\, A_3^0(i,j,(k+l))\nn \\
\PCdown_{lm} \ant &=  P_{qg}(m^h,l)\, A_3^0(i,j,k) 
\end{align}

\section{Unintegrated $\Y$}\label{app:unintegratedX431}
In this Appendix, we list the three components of the unintegrated $\Y$ antenna functions ($\YL$, $\YM$, $\YR$) listed in Table~\ref{tab:X431}:
\begin{eqnarray}
    A_{4,3;L}^{1}(i_{q},j_{g},k_{\bar{q}},b_{g}) &=& \left(-\frac{1}{\e^2}+\frac{1}{\e}\left(\log\left(\frac{s_{ib}}{\mu^2}\right)-\frac{5}{3}\right)\right)A_{3}^{0}(i,j,k),\\
    A_{4,3;M}^{1}(i_{q},j_{g},k_{\bar{q}},b_{g}) &=& \left(\frac{2}{\e^2}+\frac{1}{\e}\left(\log\left(\frac{s_{ijk}(\mu^2)^2}{s_{ik}s_{\wt{ij}b}s_{\wt{jk}b}}\right)+\frac{10}{3}\right)\right)A_{3}^{0}(i,j,k),\\
    A_{4,3;R}^{1}(i_{q},j_{g},k_{\bar{q}},b_{g}) &=& \left(-\frac{1}{\e^2}+\frac{1}{\e}\left(\log\left(\frac{s_{kb}}{\mu^2}\right)-\frac{5}{3}\right)\right)A_{3}^{0}(i,j,k),\\
    \wh{A}_{4,3;L}^{1}(i_{q},j_{g},k_{\bar{q}},b_{g}) &=& 0 
    ,\\
    \wh{A}_{4,3;M}^{1}(i_{q},j_{g},k_{\bar{q}},b_{g}) &=& 0
    ,\\
    \wh{A}_{4,3;R}^{1}(i_{q},j_{g},k_{\bar{q}},b_{g}) &=& 0
    ,\\
    \wt{A}_{4,3;L}^{1}(i_{q},j_{g},k_{\bar{q}},b_{g}) &=& 0
    ,\\
    \wt{A}_{4,3;M}^{1}(i_{q},j_{g},k_{\bar{q}},b_{g}) &=& 0
    ,\\
    \wt{A}_{4,3;R}^{1}(i_{q},j_{g},k_{\bar{q}},b_{g}) &=& 0
    ,\\
    D_{4,3;L}^{1}(i_{q},j_{g},k_{g},b_{\bar{q}}) &=& 0 ,\\
    D_{4,3;M}^{1}(i_{q},j_{g},k_{g},b_{\bar{q}}) &=& \left(\frac{1}{\e^2}+\frac{1}{\e}\left(\log\left(\frac{s_{ijk}\mu^2}{(s_{ik}+s_{jk})s_{\wt{jk}b}}\right)+\frac{5}{3}\right)\right)D_{3}^{0}(i,j,k),\\
    D_{4,3;R}^{1}(i_{q},j_{g},k_{g},b_{\bar{q}}) &=& \left(-\frac{1}{\e^2}+\frac{1}{\e}\left(\log\left(\frac{s_{kb}}{\mu^2}\right)-\frac{5}{3}\right)\right)D_{3}^{0}(i,j,k),\\
    \wt{D}_{4,3;L}^{1}(i_q,j_g,k_g,b_{\bar{q}}) &=& \left(-\frac{1}{\e^2}+\frac{1}{\e}\left(\log\left(\frac{s_{ib}}{\mu^2}\right)-\frac{3}{2}\right)\right)D_{3}^{0}(i,j,k),\\
    \wt{D}_{4,3;M}^{1}(i_q,j_g,k_g,b_{\bar{q}}) &=& \left(\frac{1}{\e^2}+\frac{1}{\e}\left(\log\left(\frac{(s_{ik}+s_{jk})\mu^2}{s_{ik}s_{\wt{ij}b}}\right)+\frac{3}{2}\right)\right)D_{3}^{0}(i,j,k),\\
    \wt{D}_{4,3;R}^{1}(i_q,j_g,k_g,b_{\bar{q}}) &=& 0,\\
    \wh{D}_{4,3;L}^{1}(i_q,j_g,k_g,b_{\bar{q}}) &=& 0,\\
    \wh{D}_{4,3;M}^{1}(i_q,j_g,k_g,b_{\bar{q}}) &=& -\frac{1}{6\e}D_{3}^{0}(i,j,k),\\
    \wh{D}_{4,3;R}^{1}(i_q,j_g,k_g,b_{\bar{q}}) &=& \frac{1}{6\e}D_{3}^{0}(i,j,k),\\
    E_{4,3;L}^{1}(i_{q},j_{\bar{Q}},k_Q,b_{\bar{q}}) &=& 0,\\
    E_{4,3;M}^{1}(i_{q},j_{\bar{Q}},k_Q,b_{\bar{q}}) &=& \left(\frac{1}{\e^2}+\frac{1}{\e}\left(\log\left(\frac{s_{ijk}\mu^2}{s_{ik}s_{\wt{ij}b}}\right)+\frac{3}{2}\right)\right)E_{3}^{0}(i,j,k),\\
    E_{4,3;R}^{1}(i_{q},j_{\bar{Q}},k_Q,b_{\bar{q}}) &=& \left(-\frac{1}{\e^2}+\frac{1}{\e}\left(\log\left(\frac{s_{kb}}{\mu^2}\right)-\frac{3}{2}\right)\right)E_{3}^{0}(i,j,k),\\
    \wt{E}_{4,3;L}^{1}(i_{q},j_{\bar{Q}},k_Q,b_{\bar{q}}) &=& \left(-\frac{1}{\e^2}+\frac{1}{\e}\left(\log\left(\frac{s_{ib}}{\mu^2}\right)-\frac{3}{2}\right)\right)E_{3}^{0}(i,j,k),\\
    \wt{E}_{4,3;M}^{1}(i_{q},j_{\bar{Q}},k_Q,b_{\bar{q}}) &=& \left(\frac{1}{\e^2}+\frac{1}{\e}\left(\log\left(\frac{\mu^2}{s_{\wt{ij}b}}\right)+\frac{3}{2}\right)\right)E_{3}^{0}(i,j,k),\\
    \wt{E}_{4,3;R}^{1}(i_{q},j_{\bar{Q}},k_Q,b_{\bar{q}}) &=& 0,\\
    F_{4,3;L}^{1}(i_g,j_g,k_g,b_g) &=& 0\\
    F_{4,3;M}^{1}(i_g,j_g,k_g,b_g) &=& \left(\!\frac{1}{\e^2}\!+\!\frac{1}{2\e}\!\left(\!\log\left(\frac{(s_{ij}\!+\!s_{ik})s_{ijk}(\mu^2)^2}{s_{ik}s_{\wt{jk}b}^2(s_{ik}\!+\!s_{jk})}\right)\!+\!\frac{11}{3}\right)\!\right)F_{3}^{0}(i,j,k),\\
    F_{4,3;R}^{1}(i_g,j_g,k_g,b_g) &=& \left(-\frac{1}{\e^2}+\frac{1}{\e}\left(\log\left(\frac{s_{kb}}{\mu^2}\right)-\frac{11}{6}\right)\right)F_{3}^{0}(i,j,k),\\
    \wh{F}_{4,3;L}^{1}(i_g,j_g,k_g,b_g) &=& 
    0,\\
    \wh{F}_{4,3;M}^{1}(i_g,j_g,k_g,b_g) &=& -\frac{1}{3\e}F_3^0(i,j,k)
    ,\\
    \wh{F}_{4,3;R}^{1}(i_g,j_g,k_g,b_g) &=& \frac{1}{3\e}F_3^0(i,j,k)
    ,\\
    G_{4,3;L}^{1}(i_g,j_{\bar{q}},k_{q},b_g) &=& \left(-\frac{1}{\e^2}+\frac{1}{\e}\left(\log\left(\frac{s_{ib}}{\mu^2}\right)-\frac{11}{6}\right)\right)G_{3}^{0}(i,j,k),\\
    G_{4,3;M}^{1}(i_g,j_{\bar{q}},k_{q},b_g) &=& \left(\frac{2}{\e^2}+\frac{1}{\e}\left(\log\left(\frac{s_{ijk}(\mu^2)^2}{s_{ik}s_{\wt{ij}b}s_{\wt{jk}b}}\right)+\frac{21}{6}\right)\right)G_{3}^{0}(i,j,k),\\
    G_{4,3;R}^{1}(i_g,j_{\bar{q}},k_{q},b_g) &=&\left(-\frac{1}{\e^2}+\frac{1}{\e}\left(\log\left(\frac{s_{kb}}{\mu^2}\right)-\frac{5}{3}\right)\right)G_{3}^{0}(i,j,k),\\
    \wh{G}_{4,3;L}^{1}(i_{q},j_{g},k_{\bar{q}},b_{g}) &=& \frac{1}{3\e}G_3^0(i,j,k) 
    ,\\
    \wh{G}_{4,3;M}^{1}(i_{q},j_{g},k_{\bar{q}},b_{g}) &=& -\frac{1}{3\e}G_3^0(i,j,k)
    ,\\
    \wh{G}_{4,3;R}^{1}(i_{q},j_{g},k_{\bar{q}},b_{g}) &=& 0
    ,\\
    \wt{G}_{4,3;L}^{1}(i_{q},j_{g},k_{\bar{q}},b_{g}) &=& 0
    ,\\
    \wt{G}_{4,3;M}^{1}(i_{q},j_{g},k_{\bar{q}},b_{g}) &=& 0
    ,\\
    \wt{G}_{4,3;R}^{1}(i_{q},j_{g},k_{\bar{q}},b_{g}) &=& 0
    .
\end{eqnarray}

\section{Integrated new antenna functions}\label{app:integrated}
\subsection{Integrated $\X$}\label{app:integratedX530}

In this Appendix, we list the integrals of the three components of the $\X$ antenna functions, with each part integrated over the appropriate antenna phase space.  We can write down the integral as, 
\begin{equation}
\label{eq:master}
\calX  (\SAB,\SBC) = \calXM (\SAB,\SBC) + \calXL (\SAB) + \calXR (s\SBC).
\end{equation}
Expressions for $\calXM$, $\calXL$ and $\calXR$ are listed below where 
\begin{equation}
    S_{IK}  = \left( \frac{\SAB}{\mu^2} \right)^{-\e}, \qquad
    S_{KM}  = \left(\frac{\SBC}{\mu^2} \right)^{-\e}.
\end{equation}

\allowdisplaybreaks
\begin{eqnarray}
\label{eq:A530Mint}
\calAM (\SAB,\SBC) &=& S_{IK}S_{KM}\Biggl [
+\frac{1}{\e^4}
+\frac{10}{3\e^3}
+ \frac{1}{\e^2} \left(
\frac{223}{16}
-\frac{4}{3}\pi^2
\right)
\nn\\&&\qquad\qquad
+ \frac{1}{\e} \left(
\frac{673}{12}
-\frac{317}{72}\pi^2
-\frac{68}{3}\zeta_3
\right)
 \nonumber \\
 &&  \qquad\qquad+ \left(
\frac{10799}{48}
-\frac{1789}{96}\pi^2
-\frac{1315}{18}\zeta_3
+\frac{13}{90}\pi^4
\right)
 + \order{\e}\Biggr]
\\
\label{eq:A530Lint}
\calAL (\SAB) &=& S_{IK}^2\Biggl [
-\frac{1}{32\e^2}
+ \frac{1}{\e} \left(
\frac{1487}{1728}
-\frac{1}{48}\pi^2
-\zeta_3
\right)
 \nonumber \\
 && \qquad\qquad + \left(
\frac{82133}{10368}
+\frac{127}{1728}\pi^2
-\frac{25}{8}\zeta_3
-\frac{1}{12}\pi^4
\right)
 + \order{\e}\Biggr]
\\
\label{eq:A530Rint}
\calAR (\SBC) &=& S_{KM}^2\Biggl [
-\frac{1}{32\e^2}
+ \frac{1}{\e} \left(
\frac{1487}{1728}
-\frac{1}{48}\pi^2
-\zeta_3
\right)
 \nonumber \\
 && \qquad\qquad + \left(
\frac{82133}{10368}
+\frac{127}{1728}\pi^2
-\frac{25}{8}\zeta_3
-\frac{1}{12}\pi^4
\right)
 + \order{\e}\Biggr]
\\\label{eq:B530Mint}
\calBM (\SAB,\SBC) &=& S_{IK}S_{KM}\Biggl [
-\frac{1}{3\e^3}
-\frac{4}{3\e^2}
+ \frac{1}{\e} \left(
-\frac{233}{48}
+\frac{29}{72}\pi^2
\right)
\nn\\&&\qquad\qquad
 + \left(
-\frac{1691}{96}
+\frac{16}{9}\pi^2
+\frac{91}{18}\zeta_3
\right)
 + \order{\e}\Biggr]
\\
\label{eq:B530Lint}
\calBL (\SAB) &=& S_{IK}^2\Biggl [
+\frac{1}{72\e^2}
+ \frac{1}{\e} \left(
-\frac{77}{216}
+\frac{1}{24}\pi^2
\right)
\nonumber \\
 &&\qquad\qquad
 + \left(
-\frac{13003}{5184}
-\frac{1}{24}\pi^2
+\frac{9}{4}\zeta_3
\right)
 + \order{\e}\Biggr]
\\
\label{eq:B530Rint}
\calBR (\SBC) &=& S_{KM}^2\Biggl [
-\frac{5}{36\e^2}
-\frac{131}{96\e}
 + \left(
-\frac{21611}{2592}
+\frac{85}{432}\pi^2
\right)
 + \order{\e}\Biggr]
\\\label{eq:A530tMint}
\calAtM (\SAB,\SBC) &=& S_{IK}S_{KM}\Biggl [
+\frac{1}{\e^4}
+\frac{19}{6\e^3}
+ \frac{1}{\e^2} \left(
\frac{637}{48}
-\frac{4}{3}\pi^2
\right)
\nn\\&&\qquad\qquad
+ \frac{1}{\e} \left(
\frac{1919}{36}
-\frac{301}{72}\pi^2
-\frac{68}{3}\zeta_3
\right)
 \nonumber \\
 &&  \qquad+ \left(
\frac{184445}{864}
-\frac{5101}{288}\pi^2
-\frac{1247}{18}\zeta_3
+\frac{13}{90}\pi^4
\right)
 + \order{\e}\Biggr]
\\
\label{eq:A530tLint}
\calAtL (\SAB) &=& S_{IK}^2\Biggl [
-\frac{5}{32\e^2}
+ \frac{1}{\e} \left(
-\frac{43}{64}
-\frac{1}{48}\pi^2
-\frac{1}{2}\zeta_3
\right)
 \nonumber \\
 && \qquad\qquad + \left(
-\frac{435}{128}
+\frac{21}{64}\pi^2
-\frac{17}{8}\zeta_3
-\frac{1}{24}\pi^4
\right)
 + \order{\e}\Biggr]
\\
\label{eq:A530tRint}
\calAtR (\SBC) &=& S_{KM}^2\Biggl [
-\frac{5}{32\e^2}
+ \frac{1}{\e} \left(
-\frac{1057}{1728}
-\frac{1}{48}\pi^2
-\frac{1}{2}\zeta_3
\right)
 \nonumber \\
 &&  \qquad\qquad+ \left(
-\frac{8125}{3456}
+\frac{17}{64}\pi^2
-\frac{17}{8}\zeta_3
-\frac{1}{24}\pi^4
\right)
 + \order{\e}\Biggr]
\\\label{eq:B530tMint}
\calBtM (\SAB,\SBC) &=& S_{IK}S_{KM}\Biggl [
-\frac{1}{3\e^3}
-\frac{5}{4\e^2}
+ \frac{1}{\e} \left(
-\frac{655}{144}
+\frac{29}{72}\pi^2
\right)
\nn\\&&\qquad\qquad
 + \left(
-\frac{14503}{864}
+\frac{239}{144}\pi^2
+\frac{91}{18}\zeta_3
\right)
 + \order{\e}\Biggr]
\\
\label{eq:B530tLint}
\calBtL (\SAB) &=& 0
\\
\label{eq:B530tRint}
\calBtR (\SBC) &=& S_{KM}^2\Biggl [
+ \frac{1}{\e} \left(
-\frac{133}{432}
+\frac{1}{24}\pi^2
\right)
 + \left(
-\frac{517}{144}
+\frac{3}{16}\pi^2
+\frac{9}{4}\zeta_3
\right)
\nonumber \\
 &&\qquad\qquad
 + \order{\e}\Biggr]
\\\label{eq:A530ttMint}
\calAttM (\SAB,\SBC) &=& S_{IK}S_{KM}\Biggl [
+\frac{1}{\e^4}
+\frac{3}{\e^3}
+ \frac{1}{\e^2} \left(
\frac{203}{16}
-\frac{4}{3}\pi^2
\right)
\nn\\&&\qquad\qquad
+ \frac{1}{\e} \left(
\frac{407}{8}
-\frac{95}{24}\pi^2
-\frac{68}{3}\zeta_3
\right)
 \nonumber \\
 && \qquad\qquad + \left(
\frac{1629}{8}
-\frac{1625}{96}\pi^2
-\frac{131}{2}\zeta_3
+\frac{13}{90}\pi^4
\right)
 + \order{\e}\Biggr]
\\
\label{eq:A530ttLint}
\calAttL (\SAB) &=& S_{IK}^2\Biggl [
-\frac{5}{32\e^2}
+ \frac{1}{\e} \left(
-\frac{43}{64}
-\frac{1}{48}\pi^2
-\frac{1}{2}\zeta_3
\right)
 \nonumber \\
 &&\qquad\qquad  + \left(
-\frac{435}{128}
+\frac{21}{64}\pi^2
-\frac{17}{8}\zeta_3
-\frac{1}{24}\pi^4
\right)
 + \order{\e}\Biggr]
\\
\label{eq:A530ttRint}
\calAttR (\SBC) &=& S_{KM}^2\Biggl [
-\frac{5}{32\e^2}
+ \frac{1}{\e} \left(
-\frac{43}{64}
-\frac{1}{48}\pi^2
-\frac{1}{2}\zeta_3
\right)
 \nonumber \\
 && \qquad\qquad + \left(
-\frac{435}{128}
+\frac{21}{64}\pi^2
-\frac{17}{8}\zeta_3
-\frac{1}{24}\pi^4
\right)
 + \order{\e}\Biggr]
\\
\label{eq:D530Mint}
\calDM (\SAB,\SBC) &=& S_{IK}S_{KM}\Biggl [
+\frac{1}{\e^4}
+\frac{7}{2\e^3}
+ \frac{1}{\e^2} \left(
\frac{2095}{144}
-\frac{4}{3}\pi^2
\right)
\nn\\&&\qquad\qquad
+ \frac{1}{\e} \left(
\frac{25277}{432}
-\frac{83}{18}\pi^2
-\frac{68}{3}\zeta_3
\right)
 \nonumber \\
 &&  \qquad\qquad+ \left(
\frac{75949}{324}
-\frac{1867}{96}\pi^2
-76\zeta_3
+\frac{13}{90}\pi^4
\right)
 + \order{\e}\Biggr]
\\
\label{eq:D530Lint}
\calDL (\SAB) &=& S_{IK}^2\Biggl [
-\frac{1}{32\e^2}
+ \frac{1}{\e} \left(
\frac{1487}{1728}
-\frac{1}{48}\pi^2
-\zeta_3
\right)
 \nonumber \\
 && \qquad \qquad+ \left(
\frac{82133}{10368}
+\frac{127}{1728}\pi^2
-\frac{25}{8}\zeta_3
-\frac{1}{12}\pi^4
\right)
 + \order{\e}\Biggr]
\\
\label{eq:D530Rint}
\calDR (\SBC) &=& S_{KM}^2\Biggl [
-\frac{1}{32\e^2}
+ \frac{1}{\e} \left(
\frac{1787}{1728}
-\frac{5}{144}\pi^2
-\zeta_3
\right)
 \nonumber \\
 &&  \qquad\qquad+ \left(
\frac{106825}{10368}
-\frac{89}{1728}\pi^2
-\frac{31}{8}\zeta_3
-\frac{1}{12}\pi^4
\right)
 + \order{\e}\Biggr]
\\\label{eq:E530aMint}
\calEaM (\SAB,\SBC) &=& S_{IK}S_{KM}\Biggl [
-\frac{1}{3\e^3}
-\frac{25}{18\e^2}
+ \frac{1}{\e} \left(
-\frac{523}{108}
+\frac{3}{8}\pi^2
\right)
\nn\\&&\qquad\qquad
 + \left(
-\frac{5339}{324}
+\frac{131}{72}\pi^2
+\frac{61}{18}\zeta_3
\right)
 + \order{\e}\Biggr]
\\
\label{eq:E530aLint}
\calEaL (\SAB) &=& S_{IK}^2\Biggl [
-\frac{5}{36\e^2}
-\frac{131}{96\e}
 + \left(
-\frac{21611}{2592}
+\frac{85}{432}\pi^2
\right)
 + \order{\e}\Biggr]
\\
\label{eq:E530aRint}
\calEaR (\SBC) &=& S_{KM}^2\Biggl [
+\frac{1}{72\e^2}
+ \frac{1}{\e} \left(
-\frac{29}{48}
+\frac{5}{72}\pi^2
\right)
\nonumber \\
 &&\qquad\qquad
 + \left(
-\frac{11803}{1728}
+\frac{17}{72}\pi^2
+\frac{15}{4}\zeta_3
\right)
 + \order{\e}\Biggr]
\\\label{eq:E530bMint}
\calEbM (\SAB,\SBC) &=& S_{IK}S_{KM}\Biggl [
-\frac{1}{3\e^3}
-\frac{4}{3\e^2}
+ \frac{1}{\e} \left(
-\frac{233}{48}
+\frac{29}{72}\pi^2
\right)
\nn\\&&\qquad\qquad
 + \left(
-\frac{1691}{96}
+\frac{16}{9}\pi^2
+\frac{91}{18}\zeta_3
\right)
+ \order{\e}\Biggr]
\\
\label{eq:E530bLint}
\calEbL (\SAB) &=& S_{IK}^2\Biggl [
+\frac{1}{72\e^2}
+ \frac{1}{\e} \left(
-\frac{77}{216}
+\frac{1}{24}\pi^2
\right)
\nn\\&&\qquad\qquad
 + \left(
-\frac{13003}{5184}
-\frac{1}{24}\pi^2
+\frac{9}{4}\zeta_3
\right)
 + \order{\e}\Biggr]
\\
\label{eq:E530bRint}
\calEbR (\SBC) &=& S_{KM}^2\Biggl [
-\frac{5}{36\e^2}
-\frac{131}{96\e}
 + \left(
-\frac{21611}{2592}
+\frac{85}{432}\pi^2
\right)
 + \order{\e}\Biggr]
\\\label{eq:E530cMint}
\calEcM (\SAB,\SBC) &=& S_{IK}S_{KM}\Biggl [
-\frac{1}{3\e^3}
-\frac{47}{36\e^2}
+ \frac{1}{\e} \left(
-\frac{1933}{432}
+\frac{3}{8}\pi^2
\right)
\nn\\&&\qquad\qquad
 + \left(
-\frac{39673}{2592}
+\frac{83}{48}\pi^2
+\frac{61}{18}\zeta_3
\right)
+ \order{\e}\Biggr]
\\
\label{eq:E530cLint}
\calEcL (\SAB) &=& 0
\\
\label{eq:E530cRint}
\calEcR (\SBC) &=& S_{KM}^2\Biggl [
+ \frac{1}{\e} \left(
-\frac{89}{144}
+\frac{5}{72}\pi^2
\right)
 + \left(
-\frac{554}{81}
+\frac{43}{144}\pi^2
+\frac{15}{4}\zeta_3
\right)
\nn\\&&\qquad\qquad
 + \order{\e}\Biggr]
\\\label{eq:E530dMint}
\calEdM (\SAB,\SBC) &=& S_{IK}S_{KM}\Biggl [
-\frac{1}{3\e^3}
-\frac{47}{36\e^2}
+ \frac{1}{\e} \left(
-\frac{689}{144}
+\frac{29}{72}\pi^2
\right)
\nn\\&&\qquad\qquad
 + \left(
-\frac{46085}{2592}
+\frac{751}{432}\pi^2
+\frac{91}{18}\zeta_3
\right)
 + \order{\e}\Biggr]
\\
\label{eq:E530dLint}
\calEdL (\SAB) &=& S_{IK}^2\Biggl [
+ \frac{1}{\e} \left(
-\frac{133}{432}
+\frac{1}{24}\pi^2
\right)
 + \left(
-\frac{517}{144}
+\frac{3}{16}\pi^2
+\frac{9}{4}\zeta_3
\right)
\nn\\&&\qquad\qquad
 + \order{\e}\Biggr]
\\
\label{eq:E530dRint}
\calEdR (\SBC) &=& 0
\\\label{eq:K530Mint}
\calKM (\SAB,\SBC) &=& S_{IK}S_{KM}\Biggl [
+\frac{1}{9\e^2}
+ \frac{1}{\e} \left(
-\frac{197}{216}
+\frac{5}{36}\pi^2
\right)
\nn\\&&\qquad\qquad
 + \left(
-\frac{13685}{1296}
+\frac{1}{18}\pi^2
+\frac{25}{3}\zeta_3
\right)
 + \order{\e}\Biggr]
\\
\label{eq:K530Lint}
\calKL (\SAB) &=& 0
\\
\label{eq:K530Rint}
\calKR (\SBC) &=& S_{KM}^2\Biggl [
+ \frac{1}{\e} \left(
\frac{305}{216}
-\frac{5}{36}\pi^2
\right)
 + \left(
\frac{10823}{648}
-\frac{53}{72}\pi^2
-\frac{15}{2}\zeta_3
\right)
\nn\\&&\qquad\qquad
 + \order{\e}\Biggr]
\\
\label{eq:F530Mint}
\calFM (\SAB,\SBC) &=& S_{IK}S_{KM}\Biggl [
+\frac{1}{\e^4}
+\frac{11}{3\e^3}
+ \frac{1}{\e^2} \left(
\frac{243}{16}
-\frac{4}{3}\pi^2
\right)
\nn\\&& \qquad
+ \frac{1}{\e} \left(
\frac{13187}{216}
-\frac{347}{72}\pi^2
-\frac{68}{3}\zeta_3
\right)
 \nonumber \\
 &&\qquad  + \left(
\frac{316663}{1296}
-\frac{17539}{864}\pi^2
-\frac{1421}{18}\zeta_3
+\frac{13}{90}\pi^4
\right)
 + \order{\e}\Biggr]
\\
\label{eq:F530Lint}
\calFL (\SAB) &=& S_{IK}^2\Biggl [
-\frac{1}{32\e^2}
+ \frac{1}{\e} \left(
\frac{1787}{1728}
-\frac{5}{144}\pi^2
-\zeta_3
\right)
 \nonumber \\
 && \qquad + \left(
\frac{106825}{10368}
-\frac{89}{1728}\pi^2
-\frac{31}{8}\zeta_3
-\frac{1}{12}\pi^4
\right)
 + \order{\e}\Biggr]
\\
\label{eq:F530Rint}
\calFR (\SBC) &=& S_{KM}^2\Biggl [
-\frac{1}{32\e^2}
+ \frac{1}{\e} \left(
\frac{1787}{1728}
-\frac{5}{144}\pi^2
-\zeta_3
\right)
 \nonumber \\
 && \qquad + \left(
\frac{106825}{10368}
-\frac{89}{1728}\pi^2
-\frac{31}{8}\zeta_3
-\frac{1}{12}\pi^4
\right)
 + \order{\e}\Biggr]
\\\label{eq:G530aMint}
\calGaM (\SAB,\SBC) &=& S_{IK}S_{KM}\Biggl [
-\frac{1}{3\e^3}
-\frac{49}{36\e^2}
+ \frac{1}{\e} \left(
-\frac{2035}{432}
+\frac{3}{8}\pi^2
\right)
\nn\\&&\qquad\qquad
 + \left(
-\frac{14083}{864}
+\frac{781}{432}\pi^2
+\frac{61}{18}\zeta_3
\right)
 + \order{\e}\Biggr]
\\
\label{eq:G530aLint}
\calGaL (\SAB) &=& S_{IK}^2\Biggl [
+ \frac{1}{\e} \left(
-\frac{89}{144}
+\frac{5}{72}\pi^2
\right)
 + \left(
-\frac{554}{81}
+\frac{43}{144}\pi^2
+\frac{15}{4}\zeta_3
\right)
\nn\\&&\qquad\qquad
 + \order{\e}\Biggr]
\\
\label{eq:G530aRint}
\calGaR (\SBC) &=& 0
\\\label{eq:G530bMint}
\calGbM (\SAB,\SBC) &=& S_{IK}S_{KM}\Biggl [
-\frac{1}{3\e^3}
-\frac{25}{18\e^2}
+ \frac{1}{\e} \left(
-\frac{523}{108}
+\frac{3}{8}\pi^2
\right)
\nn\\&&\qquad\qquad
 + \left(
-\frac{5339}{324}
+\frac{131}{72}\pi^2
+\frac{61}{18}\zeta_3
\right)
 + \order{\e}\Biggr]
\\
\label{eq:G530bLint}
\calGbL (\SAB) &=& S_{IK}^2\Biggl [
-\frac{5}{36\e^2}
-\frac{131}{96\e}
 + \left(
-\frac{21611}{2592}
+\frac{85}{432}\pi^2
\right)
 + \order{\e}\Biggr]
\\
\label{eq:G530bRint}
\calGbR (\SBC) &=& S_{KM}^2\Biggl [
+\frac{1}{72\e^2}
+ \frac{1}{\e} \left(
-\frac{29}{48}
+\frac{5}{72}\pi^2
\right)
\nn\\&&\qquad\qquad
 + \left(
-\frac{11803}{1728}
+\frac{17}{72}\pi^2
+\frac{15}{4}\zeta_3
\right)
+ \order{\e}\Biggr]
\\\label{eq:H530aMint}
\calHaM (\SAB,\SBC) &=& S_{IK}S_{KM}\Biggl [
+\frac{1}{9\e^2}
+\frac{1}{2\e}
 + \left(
\frac{197}{81}
-\frac{43}{216}\pi^2
\right)
 + \order{\e}\Biggr]
\\
\label{eq:H530aLint}
\calHaL (\SAB) &=& 0
\\
\label{eq:H530aRint}
\calHaR (\SBC) &=& 0
\\\label{eq:H530bMint}
\calHbM (\SAB,\SBC) &=& S_{IK}S_{KM}\Biggl [
+\frac{1}{9\e^2}
+ \frac{1}{\e} \left(
-\frac{197}{216}
+\frac{5}{36}\pi^2
\right)
\nn\\&&\qquad\qquad
 + \left(
-\frac{13685}{1296}
+\frac{1}{18}\pi^2
+\frac{25}{3}\zeta_3
\right)
+ \order{\e}\Biggr]
\\
\label{eq:H530bLint}
\calHbL (\SAB) &=& 0
\\
\label{eq:H530bRint}
\calHbR (\SBC) &=& S_{KM}^2\Biggl [
+ \frac{1}{\e} \left(
\frac{305}{216}
-\frac{5}{36}\pi^2
\right)
 + \left(
\frac{10823}{648}
-\frac{53}{72}\pi^2
-\frac{15}{2}\zeta_3
\right)
\nn\\&& \qquad\qquad
 + \order{\e}\Biggr]
\end{eqnarray}

\subsection{Integrated $\Y$}\label{app:integratedX431}
In this Appendix, we list the integral of the $\Y$ antenna functions over the appropriate antenna phase space.  Because the integration of  each of the $\YL$, $\YM$ and $\YR$ components is proportional to the same scale, 
\begin{equation}
    S_{IK}  = \left( \frac{s_{IK}}{\mu^2} \right)^{-\e}
\end{equation}
we combine the results together as $\calY$.

\begin{eqnarray}
{\cal A}_{4,3}^{1}(s_{IK}) &=& S_{IK}\left(\frac{1}{\e}\left(-\frac{11}{8}+2\zeta_3\right)+\left(-\frac{93}{8}+3\zeta_3+\frac{2\pi^4}{15}\right)+O(\e)\right),\\
{\cal \wh{A}}_{4,3}^{1}(s_{IK}) &=& 0 
,\\
{\cal \wt{A}}_{4,3}^{1}(s_{IK}) &=& 0
,\\
{\cal D}_{4,3}^{1}(s_{IK}) &=& 0,\\
{\cal \wt{D}}_{4,3}^{1}(s_{IK}) &=& S_{IK}\left(\frac{1}{\e}\left(-\frac{179}{108} + 2 \zeta_{3}\right)+\left(-\frac{17461}{1296}+\frac{2\pi^4}{15}+\frac{10\zeta_3}{3}\right)+O(\e)\right)
,\\
{\cal \wh{D}}_{4,3}^{1}(s_{IK}) &=& 0
,\\
{\cal E}_{4,3}^{1}(s_{IK}) &=& S_{IK}\left(\frac{1}{\e}\left(\frac{1}{6}+\frac{\pi^2}{18}\right)+\left(\frac{5}{4}+\frac{\pi^2}{8}+\zeta_{3}\right)+O(\e)\right)
,\\
{\cal \wh{E}}_{4,3}^{1}(s_{IK}) &=& 0
,\\
{\cal \wt{E}}_{4,3}^{1}(s_{IK}) &=& S_{IK}\left(\frac{1}{\e}\left(\frac{59}{216}-\frac{\pi^{2}}{18}\right)+\left(\frac{1303}{648}-\frac{\pi^{2}}{8}-\frac{5\zeta_{3}}{3}\right)+O(\e)\right)
,\\
{\cal F}_{4,3}^{1}(s_{IK}) &=& S_{IK}\left(\frac{1}{\e}\left(-\frac{419}{432}+\zeta_3\right)+\left(-\frac{1241}{162}+\frac{11\zeta_3}{6}+\frac{\pi^4}{15}\right)+O(\e)\right)
,\\
{\cal \wh{F}}_{4,3}^{1}(s_{IK}) &=& 0
,\\
{\cal G}_{4,3}^{1}(s_{IK}) &=& S_{IK}\left(\frac{1}{\e}\frac{95}{216}+\left(\frac{2113}{648} - \frac{2\zeta_3}{3}\right)+O(\e)\right)
,\\
{\cal \wh{G}}_{4,3}^{1}(s_{IK}) &=& 0
,\\
{\cal \wt{G}}_{4,3}^{1}(s_{IK}) &=& 0
.
\end{eqnarray}

\section{Subtraction terms for $e^+e^-\to jjj$ at NNLO}\label{app:sub}

In the following, we give the expressions for the NNLO subtraction terms for $e^+e^-\to jjj$ employing generalised antenna functions. The leading-colour subtraction terms have been considered as examples and thoroughly discussed in Sections~\ref{sec:RRsub}, ~\ref{sec:RVsub} and~\ref{sec:VVsub}. Here we list the subtraction terms for the remaining colour factors. In general, we denote quark momenta with numbers, and gluon momenta with letters.

We follow the notation and the conventions of~\cite{Gehrmann-DeRidder:2007foh}, with a few modifications which we illustrate in the following. The Born-level matrix element is represented by $M_3^0(1_q,i_g,2_{\qb})$. The interference between the one-loop correction and the Born amplitude is split according to its leading-colour, subleading-colour and closed-fermionic-loop components, respectively denoted with $M_3^1(1_q,i_g,2_{\qb})$, $\wt{M}_3^1(1_q,i_g,2_{\qb})$, $\wh{M}_3^1(1_q,i_g,2_{\qb})$. For the single-real emission matrix element, we have to distinguish different partonic channels. The two-quark two-gluon matrix element is split into its leading- and subleading-colour components: $M_{4,qgg\qb}^0(1_q,i_g,j_g,2_{\qb})$ and $\wt{M}_{4,qgg\qb}^0(1_q,i_g,j_g,2_{\qb})$. The four-quark different-flavour matrix element is denoted with $M_{4,q\qb'q'\qb}^0(1_q,4_{\qb'},3_{q'},2_{\qb})$. For the identical-flavour case, we isolate the part of the matrix element which only exists with four same-flavour quarks:
\begin{equation}
    M_{4,q\qb q\qb}^0(1_q,2_{\qb},3_{q},4_{\qb})=\left|A_{4,q\qb q'\qb'}^0(1_q,4_{\qb'},3_{q'},2_{\qb})-A_{4,q\qb q'\qb'}^0(1_q,2_{\qb},3_{q'},4_{\qb'})\right|^2,
\end{equation}
where $A$ is used to denote a scattering amplitude rather than a squared matrix element. An analogous definition is used for the real-virtual and double-real matrix elements~\cite{Gehrmann-DeRidder:2007foh}. Everywhere the dependence of the reduced matrix elements on the electron and positron momenta is understood, as well as the renormalisation scale dependence of loop quantities.

The coloured indices on the left of each line in the subtraction terms follow the scheme described in Sections~\ref{sec:RRsub}, ~\ref{sec:RVsub} and~\ref{sec:VVsub}, to relate different contributions to their (un)integrated counterparts.
\subsection{Double-real subtraction terms}\label{app:RRsub}
We start with the definition of the abelian ($ab.$) double-real subtraction term, which come with the colour factor $(N^2+1)/N^2$:
\begin{flalign}
&{\rm d}\sigma_{NNLO,ab.}^{S} = N_5 \frac{N^2+1}{N^2} d\Phi_5(\{p\}_5;q) \Biggl \{ &\nn \\
\textA{1}&+A_3^{0}(1,i,2)\,\wt{M}_{4,qgg\qb}^0((\wt{1i}),j,k,(\wt{2i}))\,J_3^{(4)}(\lbrace p\rbrace_{4}) &\nn\\
 \textA{2}&+A_3^{0}(1,j,2)\,\wt{M}_{4,qgg\qb}^0((\wt{1j}),i,k,(\wt{2j}))\,J_3^{(4)}(\lbrace p\rbrace_{4}) &\nn\\
 \textA{3}&+A_3^{0}(1,k,2)\,\wt{M}_{4,qgg\qb}^0((\wt{1k}),j,i,(\wt{2k}))\,J_3^{(4)}(\lbrace p\rbrace_{4}) &\nn\\
 &-----------------------&\nn\\
\textB{4}&+\,\wt{A}_4^{0}(1,j,i,2)\,M_3^0((\wt{1ji}),k,(\wt{ji2}))\,J_3^{(3)}(\lbrace p\rbrace_{3}) &\nn\\
 \textB{5}&-\,A_3^{0}(2,i,1)\,A_3^{0}((\wt{1i}),j,(\wt{i2}))\,M_3^0((\wt{(\wt{i1})j}),k,(\wt{j\wt{(i2})}))\,J_3^{(3)}(\lbrace p\rbrace_{3}) &\nn\\
 \textB{6}&-\,A_3^{0}(2,j,1)\,A_3^{0}((\wt{1j}),i,(\wt{j2}))\,M_3^0((\wt{(\wt{j1})i}),k,(\wt{i\wt{(j2})}))\,J_3^{(3)}(\lbrace p\rbrace_{3}) &\nn\\
 \textB{7}&+\,\wt{A}_4^{0}(1,i,k,2)\,M_3^0((\wt{1ik}),j,(\wt{ik2}))\,J_3^{(3)}(\lbrace p\rbrace_{3}) &\nn\\
 \textB{8}&-\,A_3^{0}(2,i,1)\,A_3^{0}((\wt{1i}),k,(\wt{i2}))\,M_3^0((\wt{(\wt{i1})k}),j,(\wt{k\wt{(i2})}))\,J_3^{(3)}(\lbrace p\rbrace_{3}) &\nn\\
 \textB{9}&-\,A_3^{0}(2,k,1)\,A_3^{0}((\wt{1k}),i,(\wt{k2}))\,M_3^0((\wt{(\wt{k1})i}),j,(\wt{i\wt{(k2})}))\,J_3^{(3)}(\lbrace p\rbrace_{3}) &\nn\\
 \textB{10}&+\,\wt{A}_4^{0}(1,j,k,2)\,M_3^0((\wt{1jk}),i,(\wt{jk2}))\,J_3^{(3)}(\lbrace p\rbrace_{3}) &\nn\\
 \textB{11}&-\,A_3^{0}(2,k,1)\,A_3^{0}((\wt{1k}),j,(\wt{k2}))\,M_3^0((\wt{(\wt{k1})j}),i,(\wt{j\wt{(k2})}))\,J_3^{(3)}(\lbrace p\rbrace_{3}) &\nn\\
 \textB{12}&-\,A_3^{0}(2,j,1)\,A_3^{0}((\wt{1j}),k,(\wt{j2}))\,M_3^0((\wt{(\wt{j1})k}),i,(\wt{k\wt{(j2})}))\,J_3^{(3)}(\lbrace p\rbrace_{3}) \Biggr \}.&
\end{flalign}
The subleading-colour double-real subtraction term contributing to the $N^0$ colour factor has three components:
\begin{equation}
    {\rm d}\sigma_{NNLO,N^0}^{S} = {\rm d}\sigma_{NNLO,N^0}^{S,qggg\qb}+{\rm d}\sigma_{NNLO,N^0}^{S,q\qb q\qb g}+\frac{N^2}{N^2+1}{\rm d}\sigma_{NNLO,ab.}^{S},
\end{equation}
with:
\begin{flalign}
&{\rm d}\sigma_{NNLO,N^0}^{S} = -N_5 N^0 d\Phi_5(\{p\}_5;q) \frac{1}{3!} \sum_{(i,j,k) \in P(3,4,5)}  \Biggl \{ &\nn \\
 \textA{1}&+D_3^{0}(1,i,j)\,\wt{M}_{4,qgg\qb}^0((\wt{1i}),(\wt{ij}),k,2)\,J_3^{(4)}(\lbrace p\rbrace_{4}) &\nn\\
 \textA{2}&+D_3^{0}(2,j,i)\,\wt{M}_{4,qgg\qb}^0(1,k,(\wt{ij}),(\wt{2j}))\,J_3^{(4)}(\lbrace p\rbrace_{4}) &\nn\\
 \textA{3}&+A_3^{0}(1,k,2)\,M_{4,qgg\qb}^0((\wt{1k}),i,j,(\wt{2k}))\,J_3^{(4)}(\lbrace p\rbrace_{4}) &\nn\\
 &-----------------------&\nn\\
\textB{4}&+\,A_4^0(1,i,j,2)\,M_3^0((\wt{1ij}),k,(\wt{ij2}))\,J_3^{(3)}(\lbrace p\rbrace_{3}) &\nn\\
 \textB{5}&-\,D_3^{0}(1,i,j)\,A_3^{0}((\wt{1i}),(\wt{ij}),2)\,M_3^0((\wt{(\wt{1i})(\wt{ij})}),k,(\wt{(\wt{ij})2}))\,J_3^{(3)}(\lbrace p\rbrace_{3}) &\nn\\
 \textB{6}&-\,D_3^{0}(2,j,i)\,A_3^{0}(1,(\wt{ij}),(\wt{j2}))\,M_3^0((\wt{1\wt{(ij})}),k,(\wt{(\wt{ij})(\wt{j2})}))\,J_3^{(3)}(\lbrace p\rbrace_{3}) &\nn\\
 &-----------------------&\nn\\
\textC{7}&+\,\wt{A_{5,3}}^{0}(2,k,1,i,j)\,M_3^0(\{2k1ij\},\{1ij\},\{2k1\})\,J_3^{(3)}(\lbrace p\rbrace_{3}) &\nn\\
 \textC{8}&-\,D_3^{0}(1,i,j)\,A_3^{0}(2,k,(\wt{1i}))\,M_3^0([\u{(\wt{1i})}k],(\wt{ji}),[k2])\,J_3^{(3)}(\lbrace p\rbrace_{3}) &\nn\\
 \textC{9}&-\,A_3^{0}(2,k,1)\,D_3^{0}((\wt{k1}),i,j)\,M_3^0([\u{(\wt{1k})}i],[ji],(\wt{k2}))\,J_3^{(3)}(\lbrace p\rbrace_{3}) &\nn\\
\textC{10}&+\,\wt{A_{5,3}}^{0}(1,k,2,j,i)\,M_3^0(\{1k2\},\{2ji\},\{1k2ji\})\,J_3^{(3)}(\lbrace p\rbrace_{3}) &\nn\\
 \textC{11}&-\,D_3^{0}(2,j,i)\,A_3^{0}(1,k,(\wt{2j}))\,M_3^0([1k],(\wt{ij}),[k\u{\wt{(2j})}])\,J_3^{(3)}(\lbrace p\rbrace_{3}) &\nn\\
 \textC{12}&-\,A_3^{0}(1,k,2)\,D_3^{0}((\wt{k2}),j,i)\,M_3^0((\wt{1k}),[ji],[\u{(\wt{k2})}j])\,J_3^{(3)}(\lbrace p\rbrace_{3})\Biggr \},
\end{flalign}
and
\begin{flalign}
&{\rm d}\sigma_{NNLO,N^0}^{S,q\qb q\qb g} = -N_5 N^0 d\Phi_5(\{p\}_5;q) \Biggl \{ &\nn \\
\textA{1}&+A_3^{0}(1,i,3)\,M_{4,q\qb q\qb}^0((\wt{1i}),2,(\wt{i3}),4)\,J_3^{(4)}(\lbrace p\rbrace_{4}) &\nn\\
 \textA{2}&+A_3^{0}(2,i,4)\,M_{4,q\qb q\qb}^0(1,(\wt{2i}),3,(\wt{i4}))\,J_3^{(4)}(\lbrace p\rbrace_{4}) &\nn\\
  &-----------------------&\nn\\
\textB{3}&+2\,C_4^0(1,2,3,4)\,M_3^0((\wt{123}),i,(\wt{234}))\,J_3^{(3)}(\lbrace p\rbrace_{3}) &\nn\\
\textB{4}&+2\,C_4^0(3,2,1,4)\,M_3^0((\wt{321}),i,(\wt{412}))\,J_3^{(3)}(\lbrace p\rbrace_{3}) &\nn\\
\textB{5}&+2\,C_4^0(4,3,2,1)\,M_3^0((\wt{123}),i,(\wt{432}))\,J_3^{(3)}(\lbrace p\rbrace_{3}) &\nn\\
\textB{6}&+2\,C_4^0(2,3,4,1)\,M_3^0((\wt{143}),i,(\wt{234}))\,J_3^{(3)}(\lbrace p\rbrace_{3})\Biggr \}.&
\end{flalign}
The most subleading-colour double-real subtraction term contributing to the $N^{-2}$ colour factor has two components:
\begin{equation}
    {\rm d}\sigma_{NNLO,N^{-2}}^{S} = {\rm d}\sigma_{NNLO,N^{-2}}^{S,q\qb q\qb g}+\frac{1}{N^2+1}{\rm d}\sigma_{NNLO,ab.}^{S},
\end{equation}
with
\begin{flalign}
&{\rm d}\sigma_{NNLO,N^{-2}}^{S,q\qb q\qb g} = N_5 N^{-2} d\Phi_5(\{p\}_5;q) \Biggl \{ &\nn \\
\textA{1}&-A_3^{0}(1,i,4)\,M_{4,q\qb q\qb}^0((\wt{1i}),3,(\wt{i4}),2)\,J_3^{(4)}(\lbrace p\rbrace_{4}) &\nn\\
 \textA{2}&+A_3^{0}(1,i,2)\,M_{4,q\qb q\qb}^0((\wt{1i}),3,4,(\wt{i2}))\,J_3^{(4)}(\lbrace p\rbrace_{4}) &\nn\\
 \textA{3}&+A_3^{0}(1,i,3)\,M_{4,q\qb q\qb}^0((\wt{1i}),(\wt{i3}),4,2)\,J_3^{(4)}(\lbrace p\rbrace_{4}) &\nn\\
 \textA{4}&+A_3^{0}(3,i,4)\,M_{4,q\qb q\qb}^0(1,(\wt{3i}),(\wt{4i}),2)\,J_3^{(4)}(\lbrace p\rbrace_{4}) &\nn\\
 \textA{5}&-A_3^{0}(3,i,2)\,M_{4,q\qb q\qb}^0(1,(\wt{3i}),4,(\wt{2i}))\,J_3^{(4)}(\lbrace p\rbrace_{4}) &\nn\\
 \textA{6}&+A_3^{0}(4,i,2)\,M_{4,q\qb q\qb}^0(1,3,(\wt{4i}),(\wt{2i}))\,J_3^{(4)}(\lbrace p\rbrace_{4}) &\nn\\
  &-----------------------&\nn\\
\textB{7}&+2\,C_4^0(1,3,4,2)\,M_3^0((\wt{134}),i,(\wt{342}))\,J_3^{(3)}(\lbrace p\rbrace_{3}) &\nn\\
\textB{8}&+2\,C_4^0(4,3,1,2)\,M_3^0((\wt{431}),i,(\wt{312}))\,J_3^{(3)}(\lbrace p\rbrace_{3}) &\nn\\
\textB{9}&+2\,C_4^0(2,4,3,1)\,M_3^0((\wt{431}),i,(\wt{243}))\,J_3^{(3)}(\lbrace p\rbrace_{3}) &\nn\\
\textB{10}&+2\,C_4^0(3,1,2,4)\,M_3^0((\wt{124}),i,(\wt{312}))\,J_3^{(3)}(\lbrace p\rbrace_{3}) \Biggr \}.&
\end{flalign}
The leading-$N_f$ double-real subtraction term proportional to $N_f\,N$ after summation over quark flavours reads:
\begin{flalign}
&{\rm d}\sigma_{NNLO,N_fN}^{S} = N_5 N_f N d\Phi_5(\{p\}_5;q)  \Biggl \{ \Bigg[ &\nn \\
 \textA{1}&+\frac{1}{2}\,A_3^{0}(1,i,3)\,M_{4,q\qb'q'\qb}^0((\wt{1i}),(\wt{i3}),4,2)\,J_3^{(4)}(\lbrace p\rbrace_{4}) &\nn\\
 \textA{2}&+\frac{1}{2}\,G_3^{0}(i,3,4)\,M_{4,qgg\qb}^0(1,(\wt{i3}),(\wt{34}),2)\,J_3^{(4)}(\lbrace p\rbrace_{4}) &\nn\\
 \textA{3}&+\frac{1}{2}\,E_3^{0}(2,4,3)\,M_{4,qgg\qb}^0(1,i,(\wt{34}),(\wt{42}))\,J_3^{(4)}(\lbrace p\rbrace_{4}) &\nn\\
 &-----------------------&\nn\\
\textB{4}&+\frac{1}{2}\,\,\overline{E}_4^0(1,i,3,4)\,M_3^0((\wt{1i3}),(\wt{43i}),2)\,J_3^{3}(\lbrace p\rbrace_{3}) &\nn\\
 \textB{5}&-\frac{1}{2}\,\,A_3^{0}(1,i,3)\,E_3^{0}((\wt{1i}),(\wt{3i}),4)\,M_3^0((\wt{(\wt{1i})(\wt{i3})}),(\wt{4 (\wt{3i})}),2)\,J_3^{3}(\lbrace p\rbrace_{3}) &\nn\\
 \textB{6}&-\frac{1}{2}\,\,G_3^{0}(i,3,4)\,D_3^{0}(1,(\wt{i3}),(\wt{34}))\,M_3^0((\wt{1(\wt{i3})}),(\wt{(\wt{i3})(\wt{34})}),2)\,J_3^{3}(\lbrace p\rbrace_{3}) &\nn\\
\textB{7}&+\,E_4^0(2,4,3,i)\,M_3^0(1,(\wt{i34}),(\wt{243}))\,J_3^{3}(\lbrace p\rbrace_{3}) &\nn\\
 \textB{8}&-\frac{1}{2}\,\,E_3^{0}(2,4,3)\,D_3^{0}((\wt{24}),(\wt{43}),i)\,M_3^0(1,(\wt{(\wt{43})i}),(\wt{(\wt{24})(\wt{43})}))\,J_3^{3}(\lbrace p\rbrace_{3}) &\nn\\
 \textB{9}&-\frac{1}{2}\,\,G_3^{0}(i,3,4)\,D_3^{0}(2,(\wt{43}),(\wt{3i}))\,M_3^0(1,(\wt{(\wt{3i})(\wt{43})}),(\wt{(\wt{43}) 2}))\,J_3^{3}(\lbrace p\rbrace_{3}) &\nn\\
 &-----------------------&\nn\\
 \textC{10}&+\frac{1}{2}\,\,B_{5,3}^0(1,i,3,4,2)\,M_3^0(\{1i3\},\{1,i,3,4,2\},\{342\})\,J_3^{3}(\lbrace p\rbrace_{3}) &\nn\\
 \textC{11}&-\frac{1}{2}\,\,A_3^{0}(1,i,3)\,E_3^{0}(2,4,(\wt{3i}))\,M_3^0((\wt{1i}),[4, \u{(\wt{3i})}],(\wt{24}))\,J_3^{3}(\lbrace p\rbrace_{3}) &\nn\\
 \textC{12}&-\frac{1}{2}\,\,E_3^{0}(2,4,3)\,D_3^{0}(1,i,(\wt{34}))\,M_3^0((\wt{1i}),[i, \u{(\wt{34})}],(\wt{24}))\,J_3^{3}(\lbrace p\rbrace_{3}) &\nn\\
 &+(1\leftrightarrow 2,3\leftrightarrow 4) \Bigg]+(1\leftrightarrow 3,2\leftrightarrow 4)\Biggr \}.&
\end{flalign}
The subleading-$N_f$ double-real subtraction term proportional to $N_f\,N^{-1}$ reads:
\begin{flalign}
&{\rm d}\sigma_{NNLO,N_fN^{-1}}^{S} = -N_5 N_f N^{-1} d\Phi_5(\{p\}_5;q)  \Biggl \{ &\nn \\
 \textA{1}&+\frac{1}{2}\,A_3^{0}(3,i,4)\,M_{4,q\qb'q'\qb}^0(1,(\wt{3i}),(\wt{4i}),2)\,J_3^{(4)}(\lbrace p\rbrace_{4}) &\nn\\
 \textA{2}&+\frac{1}{2}\,A_3^{0}(1,i,2)\,M_{4,q\qb'q'\qb}^0((\wt{1i}),3,4,(\wt{2i}))\,J_3^{(4)}(\lbrace p\rbrace_{4}) &\nn\\
 \textA{3}&+\frac{1}{2}\,E_3^{0}(1,3,4)\,\wt{M}_{4,qgg\qb}^0((\wt{13}),(\wt{34}),i,2)\,J_3^{(4)}(\lbrace p\rbrace_{4}) &\nn\\
 \textA{4}&+\frac{1}{2}\,E_3^{0}(2,4,3)\,\wt{M}_{4,qgg\qb}^0(1,i,(\wt{43}),(\wt{24}))\,J_3^{(4)}(\lbrace p\rbrace_{4}) &\nn\\
 &-----------------------&\nn\\
 \textB{5}&+\frac{1}{2}\,\,\wt{E}_4^0(1,3,i,4)\,M_3^0((\wt{13i}),(\wt{3i4}),2)\,J_3^{(3)}(\lbrace p\rbrace_{3}) &\nn\\
 \textB{6}&-\frac{1}{2}\,\,A_3^{0}(3,i,4)\,E_3^{0}(1,(\wt{3i}),(\wt{4i}))\,M_3^0((\wt{1(\wt{3i})}),(\wt{(\wt{3i}) (\wt{4i})}),2)\,J_3^{(3)}(\lbrace p\rbrace_{3}) &\nn\\
 \textB{7}&+\frac{1}{2}\,\,\wt{E}_4^0(2,4,i,3)\,M_3^0(1,(\wt{3i4}),(\wt{i42}))\,J_3^{(3)}(\lbrace p\rbrace_{3}) &\nn\\
 \textB{8}&-\frac{1}{2}\,\,A_3^{0}(3,i,4)\,E_3^{0}(2,(\wt{4i}),(\wt{i3}))\,M_3^0(1,(\wt{(\wt{3i})(\wt{4i})}),(\wt{2 (\wt{4i})}))\,J_3^{(3)}(\lbrace p\rbrace_{3}) &\nn\\
 \textB{9}&+\frac{1}{2}\,\,B_4^0(1,3,4,2)\,M_3^0((\wt{134}),i,(\wt{243}))\,J_3^{(3)}(\lbrace p\rbrace_{3}) &\nn\\
 \textB{10}&-\frac{1}{2}\,\,E_3^{0}(1,3,4)\,A_3^{0}((\wt{13}),(\wt{43}),2)\,M_3^0((\wt{(\wt{13})(\wt{43})}),i,(\wt{2 (\wt{43})}))\,J_3^{(3)}(\lbrace p\rbrace_{3}) &\nn\\
 \textB{11}&+\frac{1}{2}\,\,B_4^0(2,4,3,1)\,M_3^0((\wt{134}),i,(\wt{342}))\,J_3^{(3)}(\lbrace p\rbrace_{3}) &\nn\\
 \textB{12}&-\frac{1}{2}\,\,E_3^{0}(2,4,3)\,A_3^{0}(1,(\wt{34}),(\wt{42}))\,M_3^0((\wt{1(\wt{34})}),i,(\wt{(\wt{34})(\wt{42})}))\,J_3^{(3)}(\lbrace p\rbrace_{3}) &\nn\\
 &-----------------------&\nn\\
 \textC{13}&+\frac{1}{2}\,\,\wt{B}_{5,3}^0(1,i,2,4,3)\,M_3^0((\wt{2i1}),(\wt{342}),[3, 4, 2, i, 1])\,J_3^{(3)}(\lbrace p\rbrace_{3}) &\nn\\
 \textC{14}&-\frac{1}{2}\,\,A_3^{0}(1,i,2)\,E_3^{0}([i,\u{2}],4,3)\,M_3^0([1, i],[3, 4],[4, \u{[i,\u{2}]}])\,J_3^{(3)}(\lbrace p\rbrace_{3}) &\nn\\
 \textC{15}&-\frac{1}{2}\,\,E_3^{0}(2,4,3)\,A_3^{0}((\wt{24}),i,1)\,M_3^0((\wt{1i}),(\wt{34}),(\wt{(\wt{24})i}))\,J_3^{(3)}(\lbrace p\rbrace_{3}) &\nn\\
 \textC{16}&+\frac{1}{2}\,\,\wt{B}_{5,3}^0(2,i,1,3,4)\,M_3^0([2, i, 1, 3, 4],(\wt{134}),(\wt{2i1}))\,J_3^{(3)}(\lbrace p\rbrace_{3}) &\nn\\
 \textC{17}&-\frac{1}{2}\,\,A_3^{0}(2,i,1)\,E_3^{0}([\u{1}, i],3,4)\,M_3^0([\u{[\u{1}, i]}, 3],[3, 4],[i, 2])\,J_3^{(3)}(\lbrace p\rbrace_{3}) &\nn\\
 \textC{18}&-\frac{1}{2}\,\,E_3^{0}(1,3,4)\,A_3^{0}((\wt{13}),i,2)\,M_3^0((\wt{(\wt{13})i}),(\wt{34}),(\wt{i2}))\,J_3^{(3)}(\lbrace p\rbrace_{3}) &\nn\\
 &+(1\leftrightarrow 3,2\leftrightarrow 4)\Biggr \}.&
\end{flalign}
We notice here that the mappings in term \textC{14} are not like any iterated $X_3^0 X_3^0$ mappings we discussed in Section~\ref{sec:RRsub}: we have two dipole mappings in which particle 3 is rescaled both times. The reason for this is that terms \textC{13} and \textC{14} only serve to \textit{reroute} contributions between different layers of the subtraction infrastructure and, when summed, they do not yield additional unresolved behaviour at the double-real level. More precisely, we need the integrated form of \textC{14} at the real-virtual level, so we add it in at the double-real along with term \textC{13} which fully cancels the extra unresolved behaviour and is integrated to the double-virtual level. In order for terms \textC{13} and \textC{14} to cancel locally, the mappings of \textC{14} must match the mappings of \textC{13}, which explains the choice of mappings above. The same holds for terms \textC{16} and \textC{17}. This seemingly unnecessary complication is actually required to preserve the general construction principles and the overall pattern of infrared cancellations common to all the other subtraction terms.
\subsection{Real-virtual subtraction terms}\label{app:RVsub}
The subleading-colour real-virtual subtraction term contributing to the $N^0$ colour factor has three components:
\begin{equation}
    {\rm d}\sigma_{NNLO,N^0}^{S} = {\rm d}\sigma_{NNLO,N^0}^{T,qgg\qb}+{\rm d}\sigma_{NNLO,N^0}^{T,q\qb q\qb},
\end{equation}
with:
\begin{flalign}
&{\rm d}\sigma_{NNLO,N^0}^{T,qgg\qb} = -N_4 N^0 \left(\frac{\alpha_s}{2\pi}\right) d\Phi_4(\{p\}_4;q) \frac{1}{2!} \sum_{(i,j) \in P(3,4)} \Biggl \{ &\nn \\
 \textA{1}&-\bigg[ 
 +\mathcal{D}_{3}^{0}(s_{1j})
 +\mathcal{D}_{3}^{0}(s_{j2})
 +\mathcal{D}_{3}^{0}(s_{1i})
 +\mathcal{D}_{3}^{0}(s_{i2})
 -\mathcal{A}_{3}^{0}(s_{12})\bigg]\,\wt{M}_{4,qgg\qb}^0(1,i,j,2)\,J_3^{(4)}(\lbrace p\rbrace_{4})&\nonumber\\
  \textA{2}&-\mathcal{A}_{3}^{0}(s_{12})\,M_{4,qgg\qb}^0(1,i,j,2)\,J_3^{(4)}(\lbrace p\rbrace_{4})&\nonumber\\
  \textA{3}&-\mathcal{A}_{3}^{0}(s_{12})\,M_{4,qgg\qb}^0(1,j,i,2)\,J_3^{(4)}(\lbrace p\rbrace_{4})&\nonumber\\
  &-----------------------&\nn\\
 \textB{4}&+\bigg[ +\mathcal{D}_{3}^{0}(s_{1i})
 +\mathcal{D}_{3}^{0}(s_{i2})
 -\mathcal{A}_{3}^{0}(s_{12})
\bigg]  A_3^{0}(1,i,2)\,M_3^0((\wt{1i}),j,(\wt{i2}))\,J_3^{(3)}(\lbrace p\rbrace_{3})  &\nonumber\\
 \textB{5}&+\bigg[ +\mathcal{D}_{3}^{0}(s_{1j})
 +\mathcal{D}_{3}^{0}(s_{j2})
 -\mathcal{A}_{3}^{0}(s_{12})
\bigg]  A_3^{0}(1,j,2)\,M_3^0((\wt{1j}),i,(\wt{j2}))\,J_3^{(3)}(\lbrace p\rbrace_{3})  &\nonumber\\
&-----------------------&\nn\\
\textC{6}& +\mathcal{D}_{3}^{0}(s_{1j})
\,A_3^{0}(1,i,2)\,M_3^0([\u{1}, i],j,[i, 2])\,J_3^{(3)}(\lbrace p\rbrace_{3})  &\nonumber\\
 \textC{7}& +\mathcal{D}_{3}^{0}(s_{j2})
\,A_3^{0}(1,i,2)\,M_3^0([1,i],j,[i,\u{2}])\,J_3^{(3)}(\lbrace p\rbrace_{3})  &\nonumber\\
 \textC{8}& +\mathcal{D}_{3}^{0}(s_{1i})
\,A_3^{0}(1,j,2)\,M_3^0([\u{1}, j],i,[j, 2])\,J_3^{(3)}(\lbrace p\rbrace_{3})  &\nonumber\\
 \textC{9}& +\mathcal{D}_{3}^{0}(s_{i2})
\,A_3^{0}(1,j,2)\,M_3^0([1, j],i,[j, \u{2}])\,J_3^{(3)}(\lbrace p\rbrace_{3})  &\nonumber\\
 \textC{10}& +\mathcal{A}_{3}^{0}(s_{12})
\,D_3^{0}(1,i,j)\,M_3^0([\u{1}, i],[i, j],2)\,J_3^{(3)}(\lbrace p\rbrace_{3})  &\nonumber\\
 \textC{11}& +\mathcal{A}_{3}^{0}(s_{12})
\,D_3^{0}(2,j,i)\,M_3^0(1,[ij],[j, \u{2}])\,J_3^{(3)}(\lbrace p\rbrace_{3})  &\nonumber\\
 \textC{12}& +\mathcal{A}_{3}^{0}(s_{12})
\,D_3^{0}(1,j,i)\,M_3^0([\u{1}, j],[j, i],2)\,J_3^{(3)}(\lbrace p\rbrace_{3})  &\nonumber\\
 \textC{13}& +\mathcal{A}_{3}^{0}(s_{12})
\,D_3^{0}(2,i,j)\,M_3^0(1,[j, i],[i, \u{2}])\,J_3^{(3)}(\lbrace p\rbrace_{3})  &\nonumber\\
&-----------------------&\nn\\
  \textD{14}&+\,A_{3}^{1}(1,i,2)\,M_3^0((\wt{1i}),j,(\wt{i2}))\,J_3^{(3)}(\lbrace p\rbrace_{3})&\nonumber\\
  \textD{15}&+\,A_{3}^{1}(1,j,2)\,M_3^0((\wt{1j}),i,(\wt{j2}))\,J_3^{(3)}(\lbrace p\rbrace_{3})&\nonumber\\
  \textD{16}&+\,A_3^{0}(1,i,2)\,M_3^1((\wt{1i}),j,(\wt{i2}))\,J_3^{(3)}(\lbrace p\rbrace_{3})&\nonumber\\
  \textD{17}&+\,A_3^{0}(1,j,2)\,M_3^1((\wt{1j}),i,(\wt{j2}))\,J_3^{(3)}(\lbrace p\rbrace_{3})&\nonumber\\
  \textD{18}&+\,\wt{D}_{3}^{1}(1,i,j)\,M_3^0((\wt{1i}),(\wt{ij}),2)\,J_3^{(3)}(\lbrace p\rbrace_{3})&\nonumber\\
  \textD{19}&+\,\wt{D}_{3}^{1}(2,j,i)\,M_3^0(1,(\wt{ij}),(\wt{j2}))\,J_3^{(3)}(\lbrace p\rbrace_{3})&\nonumber\\
  \textD{20}&+\,\wt{D}_{3}^{1}(1,j,i)\,M_3^0((\wt{1j}),(\wt{ij}),2)\,J_3^{(3)}(\lbrace p\rbrace_{3})&\nonumber\\
  \textD{21}&+\,\wt{D}_{3}^{1}(2,i,j)\,M_3^0(1,(\wt{ij}),(\wt{i2}))\,J_3^{(3)}(\lbrace p\rbrace_{3}),&\nonumber\\
  \textD{22}&+\,D_3^{0}(1,i,j)\,\wt{M}_3^1((\wt{1i}),(\wt{ij}),2)\,J_3^{(3)}(\lbrace p\rbrace_{3}),&\nonumber\\
  \textD{23}&+\,D_3^{0}(2,j,i)\,\wt{M}_3^1(1,(\wt{ij}),(\wt{j2}))\,J_3^{(3)}(\lbrace p\rbrace_{3}),&\nonumber\\
  \textD{24}&+\,D_3^{0}(1,j,i)\,\wt{M}_3^1((\wt{1j}),(\wt{ij}),2)\,J_3^{(3)}(\lbrace p\rbrace_{3}),&\nonumber\\
  \textD{25}&+\,D_3^{0}(2,i,j)\,\wt{M}_3^1(1,(\wt{ij}),(\wt{i2}))\,J_3^{(3)}(\lbrace p\rbrace_{3}),&\nonumber\\
&-----------------------&\nn\\
  \textE{26}&+\,\wt{A}_{4,3}^{1}(1,i,2;j)\,M_3^0(\{1i\},j,\{i2\})\,J_3^{(3)}(\lbrace p\rbrace_{3})&\nonumber\\
  \textE{27}&+\,\wt{A}_{4,3}^{1}(1,j,2;i)\,M_3^0(\{1j\},i,\{j2\})\,J_3^{(3)}(\lbrace p\rbrace_{3})&\nonumber\\
  \textE{28}&+\,\wt{D}_{4,3}^{1}(1,i,j;2)\,M_3^0(\{1i\},\{ij\},2)\,J_3^{(3)}(\lbrace p\rbrace_{3})&\nonumber\\
  \textE{29}&+\,\wt{D}_{4,3}^{1}(2,j,i;1)\,M_3^0(1,\{ij\},\{j2\})\,J_3^{(3)}(\lbrace p\rbrace_{3})&\nonumber\\
  \textE{30}&+\,\wt{D}_{4,3}^{1}(1,j,i;2)\,M_3^0(\{1j\},\{ij\},2)\,J_3^{(3)}(\lbrace p\rbrace_{3})&\nonumber\\
  \textE{31}&+\,\wt{D}_{4,3}^{1}(2,i,j;1)\,M_3^0(1,\{ij\},\{i2\})\,J_3^{(3)}(\lbrace p\rbrace_{3}) \Biggr \},&
\end{flalign}
and
\begin{flalign}
&{\rm d}\sigma_{NNLO,N^0}^{T,q\qb q\qb} = N_4 N^0 \left(\frac{\alpha_s}{2\pi}\right) d\Phi_4(\{p\}_4;q) \Biggl \{ &\nn \\
 \textA{1}&+\bigg[ 
 +\mathcal{A}_{3}^{0}(s_{13})
 +\mathcal{A}_{3}^{0}(s_{24})\bigg]\,M_{4,q\qb q\qb}^0(1,2,3,4)\,J_3^{(4)}(\lbrace p\rbrace_{4})\Biggr \}.&
\end{flalign}
The most subleading-colour double-real subtraction term contributing to the $N^{-2}$ colour factor has two components:
\begin{equation}
    {\rm d}\sigma_{NNLO,N^{-2}}^{T} = {\rm d}\sigma_{NNLO,N^{-2}}^{T,qgg\qb}+{\rm d}\sigma_{NNLO,N^{-2}}^{T,q\qb q\qb},
\end{equation}
with
\begin{flalign}
&{\rm d}\sigma_{NNLO,N^{-2}}^{T,qgg\qb} = N_4 N^{-2} \left(\frac{\alpha_s}{2\pi}\right) d\Phi_4(\{p\}_4;q) \Biggl \{ &\nn \\
  \textA{1}&-\mathcal{A}_{3}^{0}(s_{12})\,\wt{M}_{4,qgg\qb}^0(1,i,j,2)\,J_3^{(3)}(\lbrace p\rbrace_{4})&\nonumber\\
 &-----------------------&\nn\\
 \textB{2}& +\mathcal{A}_{3}^{0}(s_{12})
\,A_3^{0}(1,i,2)\,M_3^0((\wt{1i}),j,(\wt{i2}))\,J_3^{(3)}(\lbrace p\rbrace_{3})  &\nonumber\\
 \textB{3}& +\mathcal{A}_{3}^{0}(s_{12})
\,A_3^{0}(1,j,2)\,M_3^0((\wt{1j}),i,(\wt{j2}))\,J_3^{(3)}(\lbrace p\rbrace_{3})  &\nonumber\\
&-----------------------&\nn\\
  \textD{4}&+\,A_3^{0}(1,i,2)\,\wt{M}_3^1((\wt{1i}),j,(\wt{i2}))\,J_3^{(3)}(\lbrace p\rbrace_{3})&\nonumber\\
  \textD{5}&+\,A_3^{0}(1,j,2)\,\wt{M}_3^1((\wt{1j}),i,(\wt{j2}))\,J_3^{(3)}(\lbrace p\rbrace_{3})&\nonumber\\
  \textD{6}&+\,\wt{A}_{3}^{1}(1,i,2)\,M_3^0((\wt{1i}),j,(\wt{i2}))\,J_3^{(3)}(\lbrace p\rbrace_{3})&\nonumber\\
  \textD{7}&+\,\wt{A}_{3}^{1}(1,j,2)\,M_3^0((\wt{1j}),i,(\wt{j2}))\,J_3^{(3)}(\lbrace p\rbrace_{3}) \Biggr \},&
\end{flalign}
and
\begin{flalign}
&{\rm d}\sigma_{NNLO,N^{-2}}^{T,q\qb q\qb} = N_4 N^{-2} \left(\frac{\alpha_s}{2\pi}\right) d\Phi_4(\{p\}_4;q) \Biggl \{ &\nn \\
 \textA{1}&+\bigg[ 
 +\mathcal{A}_{3}^{0}(s_{13})
 -\mathcal{A}_{3}^{0}(s_{14})
 -\mathcal{A}_{3}^{0}(s_{12})\bigg]\,M_{4,q\qb q\qb}^0(1,2,3,4)\,J_3^{(4)}(\lbrace p\rbrace_{4})&\nonumber\\
 \textA{2}&+\bigg[ 
 -\mathcal{A}_{3}^{0}(s_{23})
 +\mathcal{A}_{3}^{0}(s_{24})
 -\mathcal{A}_{3}^{0}(s_{34})\bigg]\,M_{4,q\qb q\qb}^0(1,2,3,4)\,J_3^{(4)}(\lbrace p\rbrace_{4}) \Biggr \}.&
\end{flalign}
The leading-$N_f$ real-virtual subtraction term contributing to the $N_f N$ colour factor has two components:
\begin{equation}
    {\rm d}\sigma_{NNLO,N_fN}^{T} = {\rm d}\sigma_{NNLO,N_fN}^{T,qgg\qb}+{\rm d}\sigma_{NNLO,N_fN}^{T,q\qb' q'\qb},
\end{equation}
with
\begin{flalign}
&{\rm d}\sigma_{NNLO,N_fN}^{T,qgg\qb} = N_4 N_f \,N \left(\frac{\alpha_s}{2\pi}\right) d\Phi_4(\{p\}_4;q) \Biggl \{ &\nn \\
  \textA{1}&-\mathcal{G}_{3}^{0}(s_{ij})\,M_{4,qgg\qb}^0(1,i,j,2)\,J_3^{(4)}(\lbrace p\rbrace_{4})&\nonumber\\
  \textA{2}&-\mathcal{E}_{3}^{0}(s_{1i})\,M_{4,qgg\qb}^0(1,i,j,2)\,J_3^{(4)}(\lbrace p\rbrace_{4})&\nonumber\\
  \textA{3}&-\mathcal{E}_{3}^{0}(s_{2j})\,M_{4,qgg\qb}^0(1,i,j,2)\,J_3^{(4)}(\lbrace p\rbrace_{4})&\nonumber\\
&-----------------------&\nn\\
 \textB{4}& +\mathcal{G}_{3}^{0}(s_{ij})
\,D_3^{0}(1,i,j)\,M_3^0((\wt{1i}),(\wt{ij}),2)\,J_3^{(3)}(\lbrace p\rbrace_{3})  &\nonumber\\
 \textB{5}& +\mathcal{G}_{3}^{0}(s_{ij})
\,D_3^{0}(2,j,i)\,M_3^0(1,(\wt{ij}),(\wt{j2}))\,J_3^{(3)}(\lbrace p\rbrace_{3})  &\nonumber\\
 \textB{6}& +\mathcal{E}_{3}^{0}(s_{1i})
\,D_3^{0}(1,i,j)\,M_3^0((\wt{1i}),(\wt{ij}),2)\,J_3^{(3)}(\lbrace p\rbrace_{3})  &\nonumber\\
 \textB{7}& +\mathcal{E}_{3}^{0}(s_{2j})
\,D_3^{0}(2,j,i)\,M_3^0(1,(\wt{ij}),(\wt{j2}))\,J_3^{(3)}(\lbrace p\rbrace_{3})  &\nonumber\\
&-----------------------&\nn\\
\textC{8}& +\mathcal{E}_{3}^{0}(s_{1i})
\,D_3^{0}(2,j,i)\,M_3^0(1,[\u{i}j],[j2])\,J_3^{(3)}(\lbrace p\rbrace_{3})  &\nonumber\\
\textC{9}& +\mathcal{E}_{3}^{0}(s_{2j})
\,D_3^{0}(1,i,j)\,M_3^0([1i],[i\u{j}],2)\,J_3^{(3)}(\lbrace p\rbrace_{3})  &\nonumber\\
&-----------------------&\nn\\  
 \textD{10}&+\,D_3^{0}(1,i,j)\,\wh{M}_3^1((\wt{1i}),(\wt{ij}),2)\,J_3^{(3)}(\lbrace p\rbrace_{3})&\nonumber\\
  \textD{11}&+\,\hat{D}_{3}^{1}(1,i,j)\,M_3^0((\wt{1i}),(\wt{ij}),2)\,J_3^{(3)}(\lbrace p\rbrace_{3})&\nonumber\\
  \textD{12}&+\,D_3^{0}(2,j,i)\,\wh{M}_3^1(1,(\wt{ij}),(\wt{2j}))\,J_3^{(3)}(\lbrace p\rbrace_{3})&\nonumber\\
  \textD{13}&+\,\hat{D}_{3}^{1}(2,j,i)\,M_3^0(1,(\wt{ij}),(\wt{2j}))\,J_3^{(3)}(\lbrace p\rbrace_{3})&\nonumber\\
&-----------------------&\nn\\
  \textE{14}&+\,\widehat{D}_{4,3}^1(1,i,j,2)\,M_3^0(\{1i\},\{ij\},2)\,J_3^{(3)}(\lbrace p\rbrace_{3})&\nonumber\\
  \textE{15}&+\,\widehat{D}_{4,3}^1(2,j,i,1)\,M_3^0(1,\{ij\},\{2j\})\,J_3^{(3)}(\lbrace p\rbrace_{3}) \Biggr \},&
\end{flalign}
and

\begin{flalign}
&{\rm d}\sigma_{NNLO,N_fN}^{T,q\qb' q'\qb} = N_4 N_f \,N \left(\frac{\alpha_s}{2\pi}\right) d\Phi_4(\{p\}_4;q) \Biggl \{ &\nn \\
 \textA{1}&-\bigg[
 +\mathcal{A}_{3}^{0}(s_{14})
 +\mathcal{A}_{3}^{0}(s_{23})\bigg]\,M_{4,q\qb'q'\qb}^0(1,4,3,2)\,J_3^{(4)}(\lbrace p\rbrace_{4})&\nn\\
 &-----------------------&\nn\\
 \textB{2}& +\mathcal{A}_{3}^{0}(s_{14})
\,E_3^{0}(1,4,3)\,M_3^0((\wt{14}),(\wt{43}),2)\,J_3^{(3)}(\lbrace p\rbrace_{3})  &\nn\\
 \textB{3}& +\mathcal{A}_{3}^{0}(s_{14})
\,E_3^{0}(4,1,2)\,M_3^0(3,(\wt{21}),(\wt{14}))\,J_3^{(3)}(\lbrace p\rbrace_{3})  &\nn\\
 \textB{4}& +\mathcal{A}_{3}^{0}(s_{23})
\,E_3^{0}(3,2,1)\,M_3^0((\wt{32}),(\wt{21}),4)\,J_3^{(3)}(\lbrace p\rbrace_{3})  &\nn\\
 \textB{5}& +\mathcal{A}_{3}^{0}(s_{23})
\,E_3^{0}(2,3,4)\,M_3^0(1,(\wt{43}),(\wt{32}))\,J_3^{(3)}(\lbrace p\rbrace_{3})  &\nn\\
  &-----------------------&\nn\\
 \textC{6}& +\mathcal{A}_{3}^{0}(s_{14})
\,E_3^{0}(3,2,1)\,M_3^0([3, 2],[2, \u{1}],4)\,J_3^{(3)}(\lbrace p\rbrace_{3})  &\nn\\
 \textC{7}& +\mathcal{A}_{3}^{0}(s_{14})
\,E_3^{0}(2,3,4)\,M_3^0(1,[\u{4}, 3],[3, 2])\,J_3^{(3)}(\lbrace p\rbrace_{3})  &\nn\\
 \textC{8}& +\mathcal{A}_{3}^{0}(s_{23})
\,E_3^{0}(1,4,3)\,M_3^0([1, 4],[4, \u{3}],2)\,J_3^{(3)}(\lbrace p\rbrace_{3})  &\nn\\
 \textC{9}& +\mathcal{A}_{3}^{0}(s_{23})
\,E_3^{0}(4,1,2)\,M_3^0(3,[\u{2}, 1],[1, 4])\,J_3^{(3)}(\lbrace p\rbrace_{3})  &\nn\\
   &-----------------------&\nn\\
  \textD{10}&+\frac{1}{2}\,\,E_3^{0}(1,4,3)\,M_3^1((\wt{14}),(\wt{43}),2)\,J_3^{(3)}(\lbrace p\rbrace_{3})&\nn\\
  \textD{11}&+\frac{1}{2}\,\,E_{3}^{1}(1,4,3)\,M_3^0((\wt{14}),(\wt{43}),2)\,J_3^{(3)}(\lbrace p\rbrace_{3})&\nn\\
  \textD{12}&+\frac{1}{2}\,\,E_3^{0}(2,3,4)\,M_3^1(1,(\wt{43}),(\wt{32}))\,J_3^{(3)}(\lbrace p\rbrace_{3})&\nn\\
  \textD{13}&+\frac{1}{2}\,\,E_{3}^{1}(2,3,4)\,M_3^0(1,(\wt{43}),(\wt{32}))\,J_3^{(3)}(\lbrace p\rbrace_{3})&\nn\\
  \textD{14}&+\frac{1}{2}\,\,E_3^{0}(3,2,1)\,M_3^1((\wt{32}),(\wt{21}),4)\,J_3^{(3)}(\lbrace p\rbrace_{3})&\nn\\
  \textD{15}&+\frac{1}{2}\,\,E_{3}^{1}(3,2,1)\,M_3^0((\wt{32}),(\wt{21}),4)\,J_3^{(3)}(\lbrace p\rbrace_{3})&\nn\\
  \textD{16}&+\frac{1}{2}\,\,E_3^{0}(4,1,2)\,M_3^1(3,(\wt{21}),(\wt{14}))\,J_3^{(3)}(\lbrace p\rbrace_{3})&\nn\\
  \textD{17}&+\frac{1}{2}\,\,E_{3}^{1}(4,1,2)\,M_3^0(3,(\wt{21}),(\wt{14}))\,J_3^{(3)}(\lbrace p\rbrace_{3})&\nn\\
     &-----------------------&\nn\\
   \textE{18}&+\frac{1}{2}\,\,E_{4,3}^1(1,4,3,2)\,M_3^0(\{14\},\{43\},2)\,J_3^{(3)}(\lbrace p\rbrace_{3})&\nn\\
  \textE{19}&+\frac{1}{2}\,\,E_{4,3}^1(2,3,4,1)\,M_3^0(1,\{43\},\{32\})\,J_3^{(3)}(\lbrace p\rbrace_{3})&\nn\\
  \textE{20}&+\frac{1}{2}\,\,E_{4,3}^1(3,2,1,4)\,M_3^0(\{32\},\{21\},4)\,J_3^{(3)}(\lbrace p\rbrace_{3})&\nn\\
  \textE{21}&+\frac{1}{2}\,\,E_{4,3}^1(4,1,2,3)\,M_3^0(3,\{21\},\{14\})\,J_3^{(3)}(\lbrace p\rbrace_{3}) \Biggr \}.&
\end{flalign}
The subleading-$N_f$ real-virtual subtraction term contributing to the $N_f N^{-1}$ colour factor has two components:
\begin{equation}
    {\rm d}\sigma_{NNLO,N_f N^{-1}}^{T} = {\rm d}\sigma_{NNLO,N_f N^{-1}}^{T,qgg\qb}+{\rm d}\sigma_{NNLO,N_f N^{-1}}^{T,q\qb' q'\qb},
\end{equation}
with
\begin{flalign}
&{\rm d}\sigma_{NNLO,N_f N^{-1}}^{T,qgg\qb} = -N_4 N_f \,N^{-1} \left(\frac{\alpha_s}{2\pi}\right) d\Phi_4(\{p\}_4;q) \Biggl \{ &\nn \\
 \textA{1}&-\bigg[
 +\mathcal{E}_{3}^{0}(s_{1i})
 +\mathcal{E}_{3}^{0}(s_{2i})\bigg]\,\wt{M}_{4,qgg\qb}^0(1,i,j,2)\,J_3^{(4)}(\lbrace p\rbrace_{4})&\nn\\
 \textA{2}&-\bigg[ 
 +\mathcal{E}_{3}^{0}(s_{1j})
 +\mathcal{E}_{3}^{0}(s_{2j})\bigg]\,\wt{M}_{4,qgg\qb}^0(1,i,j,2)\,J_3^{(4)}(\lbrace p\rbrace_{4})&\nn\\
 &-----------------------&\nn\\
 \textB{3}&+\bigg[ +\mathcal{E}_{3}^{0}(s_{1j})
 +\mathcal{E}_{3}^{0}(s_{2j})
\bigg]  A_3^{0}(1,j,2)\,M_3^0((\wt{1j}),i,(\wt{j2}))\,J_3^{(3)}(\lbrace p\rbrace_{3})  &\nn\\
 \textB{4}&+\bigg[ +\mathcal{E}_{3}^{0}(s_{1i})
 +\mathcal{E}_{3}^{0}(s_{2i})
\bigg]  A_3^{0}(1,i,2)\,M_3^0((\wt{1i}),j,(\wt{i2}))\,J_3^{(3)}(\lbrace p\rbrace_{3})  &\nn\\
 &-----------------------&\nn\\
 \textC{5}&+\bigg[ +\mathcal{E}_{3}^{0}(s_{1i})
 +\mathcal{E}_{3}^{0}(s_{2i})
\bigg]  A_3^{0}(1,j,2)\,M_3^0((\wt{1j}),i,(\wt{j2}))\,J_3^{(3)}(\lbrace p\rbrace_{3})  &\nn\\
 \textC{6}&+\bigg[ +\mathcal{E}_{3}^{0}(s_{1j})
 +\mathcal{E}_{3}^{0}(s_{2j})
\bigg]  A_3^{0}(1,i,2)\,M_3^0((\wt{1i}),j,(\wt{i2}))\,J_3^{(3)}(\lbrace p\rbrace_{3})  &\nn\\
  &-----------------------&\nn\\
  \textD{7}&+\,A_3^{0}(1,i,2)\,\wh{M}_3^1((\wt{1i}),j,(\wt{i2}))\,J_3^{(3)}(\lbrace p\rbrace_{3})&\nn\\
  \textD{8}&+\,A_3^{0}(1,j,2)\,\wh{M}_3^1((\wt{1j}),i,(\wt{j2}))\,J_3^{(3)}(\lbrace p\rbrace_{3})&\nn\\
  \textD{9}&+\,\hat{A}_{3}^{1}(1,i,2)\,M_3^0((\wt{1i}),j,(\wt{i2}))\,J_3^{(3)}(\lbrace p\rbrace_{3})&\nn\\
  \textD{10}&+\,\hat{A}_{3}^{1}(1,j,2)\,M_3^0((\wt{1j}),i,(\wt{j2}))\,J_3^{(3)}(\lbrace p\rbrace_{3}) \Biggr \},&
\end{flalign}
and
\begin{flalign}
&{\rm d}\sigma_{NNLO,N_fN^{-1}}^{T,q\qb' q'\qb} = N_4 N_f \,N^{-1} \left(\frac{\alpha_s}{2\pi}\right) d\Phi_4(\{p\}_4;q) \Biggl \{ &\nn \\
 \textA{1}&+\bigg[
 +\mathcal{A}_{3}^{0}(s_{12})
 +\mathcal{A}_{3}^{0}(s_{43})
 \bigg]\,M_{4,q\qb'q'\qb}^0(1,4,3,2)\,J_3^{(4)}(\lbrace p\rbrace_{4})&\nn\\
   &-----------------------&\nn\\ 
 \textB{2}& -\mathcal{A}_{3}^{0}(s_{43})
\,E_3^{0}(1,4,3)\,M_3^0((\wt{14}),(\wt{43}),2)\,J_3^{(3)}(\lbrace p\rbrace_{3})  &\nn\\
 \textB{3}& -\mathcal{A}_{3}^{0}(s_{12})
\,E_3^{0}(4,1,2)\,M_3^0(3,(\wt{12}),(\wt{41}))\,J_3^{(3)}(\lbrace p\rbrace_{3})  &\nn\\
  &-----------------------&\nn\\
\textC{4}& -\mathcal{A}_{3}^{0}(s_{12})
\,E_3^{0}(1,4,3)\,M_3^0([\u{1}, 4],[4, 3],2)\,J_3^{(3)}(\lbrace p\rbrace_{3})  &\nn\\
 \textC{5}& -\mathcal{A}_{3}^{0}(s_{43})
\,E_3^{0}(4,1,2)\,M_3^0(3,[1, 2],[\u{4}, 1])\,J_3^{(3)}(\lbrace p\rbrace_{3})  &\nn\\
    &-----------------------&\nn\\
  \textD{6}&-\,E_3^{0}(1,4,3)\,\wt{M}_{3,f_1}^{1}((\wt{14}),(\wt{43}),2)\,J_3^{(3)}(\lbrace p\rbrace_{3})&\nn\\
  \textD{7}&-\,\wt{E}_{3}^{1}(1,4,3)\,M_3^0((\wt{14}),(\wt{43}),2)\,J_3^{(3)}(\lbrace p\rbrace_{3})&\nn\\
  \textD{8}&-\,E_3^{0}(4,1,2)\,\wt{M}_{3,f_2}^{1}(3,(\wt{12}),(\wt{41}))\,J_3^{(3)}(\lbrace p\rbrace_{3})&\nn\\
  \textD{9}&-\,\wt{E}_{3}^{1}(4,1,2)\,M_3^0(3,(\wt{12}),(\wt{41}))\,J_3^{(3)}(\lbrace p\rbrace_{3})&\nn\\
    &-----------------------&\nn\\
  \textE{10}&-\,\wt{E}_{4,3}^{1}(1,4,3,2)\,M_3^0(\{14\},\{43\},2)\,J_3^{(3)}(\lbrace p\rbrace_{3})&\nn\\
  \textE{11}&-\,\wt{E}_{4,3}^{1}(4,1,2,3)\,M_3^0(3,\{12\},\{41\})\,J_3^{(3)}(\lbrace p\rbrace_{3}) \Biggr \},&
\end{flalign} 
where the $f_{i}$ indicates the flavour of the quarks attached to the $Z$/$\gamma$.

The $N_f^2$ real-virtual subtraction term is given by:
\begin{flalign}
&{\rm d}\sigma_{NNLO,N_f^2}^{T,q\qb^{'} q^{'}\qb} = N_4 N_f^{2} \left(\frac{\alpha_s}{2\pi}\right) d\Phi_4(\{p\}_4;q) \Biggl \{ &\nn \\
  \textD{1}&+\frac{1}{2}\,\,E_3^{0}(1,4,3)\,\wh{M}_{3,f_1}^{1}(2,(\wt{43}),(\wt{14}))\,J_3^{(3)}(\lbrace p\rbrace_{3})&\nn\\
  \textD{2}&+\frac{1}{2}\,\,E_3^{0}(2,4,3)\,\wh{M}_{3,f_1}^{1}((\wt{24}),(\wt{43}),1)\,J_3^{(3)}(\lbrace p\rbrace_{3})&\nn\\
  \textD{3}&+\frac{1}{2}\,\,\hat{E}_{3}^{1}(1,4,3)\,M_3^0(2,(\wt{43}),(\wt{14}))\,J_3^{(3)}(\lbrace p\rbrace_{3})&\nn\\
  \textD{4}&+\frac{1}{2}\,\,\hat{E}_{3}^{1}(2,4,3)\,M_3^0((\wt{24}),(\wt{43}),1)\,J_3^{(3)}(\lbrace p\rbrace_{3})&\nn\\
  \textD{5}&+\frac{1}{2}\,\,E_3^{0}(4,2,1)\,\wh{M}_{3,f_2}^{1}(3,(\wt{21}),(\wt{24}))\,J_3^{(3)}(\lbrace p\rbrace_{3})&\nn\\
  \textD{6}&+\frac{1}{2}\,\,E_3^{0}(3,2,1)\,\wh{M}_{3,f_2}^{1}((\wt{32}),(\wt{21}),4)\,J_3^{(3)}(\lbrace p\rbrace_{3})&\nn\\
  \textD{7}&+\frac{1}{2}\,\,\hat{E}_{3}^{1}(4,2,1)\,M_3^0(3,(\wt{21}),(\wt{24}))\,J_3^{(3)}(\lbrace p\rbrace_{3})&\nn\\
  \textD{8}&+\frac{1}{2}\,\,\hat{E}_{3}^{1}(3,2,1)\,M_3^0((\wt{32}),(\wt{21}),4)\,J_3^{(3)}(\lbrace p\rbrace_{3}) \Biggr \},&
\end{flalign}
where the $f_{i}$ indicates the flavour of the quarks attached to the $Z$/$\gamma$.
\subsection{Double-virtual subtraction terms}\label{app:VVsub}
The subleading-colour double-virtual subtraction term contributing to the $N^0$ reads:
\begin{flalign}
&{\rm d}\sigma_{NNLO,N^0}^{U} = N_3 N^0 \left(\frac{\alpha_s}{2\pi}\right)^2d\Phi_3(\{p\}_3;q)\, J_3^{(3)}(\lbrace p\rbrace_{3}) \Biggl \{ &\nn \\
 \textB{1}&-\left(2\mathcal{C}_{4}^{0}(s_{12})-\frac{1}{2}\mathcal{\wt{A}}_{4}^{0}(s_{12})+\mathcal{A}_{4}^{0}(s_{12})\right)M_3^0(1,i,2)
 &\nn\\
 \textC{2}&-\left(\mathcal{\wt{A}}_{5,3}^{0}(s_{12},s_{i2})+\mathcal{\wt{A}}_{5,3}^{0}(s_{12},s_{1i})\right)M_3^0(1,i,2)
 &\nn\\
     &-----------------------&\nn\\
 \textD{3}&-\left(\mathcal{A}_{3}^{1}(s_{12})+\frac{b_{0}}{\e}\left(\left(\frac{s_{12}}{\mu^2}\right)^{-\e}-1\right)\mathcal{A}_{3}^{0}(s_{12})\right)M_3^0(1,i,2)
 &\nn\\
  \textD{4}&+\left(\mathcal{\wt{D}}_{3}^{1}(s_{1i})+\mathcal{\wt{D}}_{3}^{1}(s_{i2})\right)M_3^0(1,i,2) &\nn \\
  \textD{5}&-\mathcal{A}_{3}^{0}(s_{12})M_3^1(1,i,2)
 &\nn\\
 \textD{6}&-\left(\mathcal{D}_{3}^{0}(s_{1i})+\mathcal{D}_{3}^{0}(s_{i2})\right)\wt{M}_3^{1}(1,i,2)
 &\nn\\
 \textE{7}&-\left(\mathcal{\wt{A}}_{4,3}^{1}(s_{12})+\mathcal{\wt{D}}_{4,3}^{1}(s_{1i})+\mathcal{\wt{D}}_{4,3}^{1}(s_{i2})\right)M_3^{0}(1,i,2) \Biggr \}.&
\end{flalign}
The most subleading-colour double-virtual subtraction term contributing to the $N^{-2}$ colour factor reads:
\begin{flalign}
&{\rm d}\sigma_{NNLO,N^{-2}}^{U} = N_3 N^{-2} \left(\frac{\alpha_s}{2\pi}\right)^2d\Phi_3(\{p\}_3;q) \,J_3^{(3)}(\lbrace p\rbrace_{3}) \Biggl \{ &\nn \\
 \textB{1}&-\bigg[ 
 \frac{1}{2}\,\,\tilde{\cal A}_{4}^{0}(s_{12}) 
 +2\,{\cal C}_{4}^{0}(s_{12}) 
 \bigg] 
 M_3^0(1,i,2)
 &\nonumber\\
&-----------------------&\nn\\
 \textD{2}& 
 -\,{\cal A}_{3}^{0}(s_{12}) 
 \wt{M}_3^1(1,i,2)
 &\nonumber\\
 \textD{3}&  
 -\,\tilde{\cal{A}}_{3}^{1}(s_{12})  
 M_3^0(1,i,2) \Biggr \}.&
\end{flalign}
The leading-$N_f$ double-virtual subtraction term proportional to $N_f\,N$ reads:
\begin{flalign}
&{\rm d}\sigma_{NNLO,N_fN}^{U} = N_3 N_f\,N \left(\frac{\alpha_s}{2\pi}\right)^2d\Phi_3(\{p\}_3;q) \, J_3^{(3)}(\lbrace p\rbrace_{3})\Biggl \{ &\nn \\
  \textB{1}&-\left(\mathcal{E}_{4}^{0}(s_{1i})+\mathcal{E}_{4}^{0}(s_{i2})+\frac{1}{2}\mathcal{\bar{E}}_{4}^{0}(s_{1i})+\frac{1}{2}\mathcal{\bar{E}}_{4}^{0}(s_{i2})\right)
 M_3^0(1,i,2)
 &\nn\\
 \textC{2}&-\left(\frac{1}{2}\mathcal{B}_{5,3}^{0}(s_{1i},s_{i2})+\frac{1}{2}\mathcal{B}_{5,3}^{0}(s_{i2},s_{1i})\right)
 M_3^0(1,i,2)
 &\nn\\
 &-----------------------&\nn\\
 \textD{3}&-\left(\mathcal{D}_{3}^{0}(s_{1i})+\mathcal{D}_{3}^{0}(s_{i2})\right)\wh{M}_3^1(1,i,2)
 &\nn\\
 \textD{4}&-\frac{1}{2}\left(\mathcal{E}_{3}^{0}(s_{1i})+\mathcal{E}_{3}^{0}(s_{i2})\right)M_3^1(1,i,2)
 &\nn\\
 \textD{5}&-\left(\mathcal{\hat{D}}_{3}^{1}(s_{1i})+\frac{b_{0,F}}{\e}\left(\left(\frac{s_{1i}}{\mu^2}\right)^{-\e}-1\right)\mathcal{D}_{3}^{0}(s_{1i})\right)
 M_3^0(1,i,2)
 &\nn\\
 \textD{6}&-\left(\mathcal{\hat{D}}_{3}^{1}(s_{i2})+\frac{b_{0,F}}{\e}\left(\left(\frac{s_{i2}}{\mu^2}\right)^{-\e}-1\right)\mathcal{D}_{3}^{0}(s_{i2})\right)
 M_3^0(1,i,2)
 &\nn\\
 \textD{7}&-\frac{1}{2}\left(\mathcal{E}_{3}^{1}(s_{1i})+\frac{b_0}{\e}\left(\left(\frac{s_{1i}}{\mu^2}\right)^{-\e}-1\right)\mathcal{E}_{3}^{0}(s_{1i})\right)
 M_3^0(1,i,2)
 &\nn\\
 \textD{8}&-\frac{1}{2}\left(\mathcal{E}_{3}^{1}(s_{i2})+\frac{b_0}{\e}\left(\left(\frac{s_{i2}}{\mu^2}\right)^{-\e}-1\right)\mathcal{E}_{3}^{0}(s_{i2})\right)
 M_3^0(1,i,2)
 &\nn\\
 \textE{9}&-\frac{1}{2}\left(\mathcal{E}_{4,3}^{1}(s_{1i})+\mathcal{E}_{4,3}^{1}(s_{i2})\right)
 M_3^0(1,i,2) \Biggr \}.&
\end{flalign}
The subleading-$N_f$ double-virtual subtraction term proportional to $N_f\,N^{-1}$ reads:
\begin{flalign}
&{\rm d}\sigma_{NNLO,N_fN^{-1}}^{U} = N_3 N_f\,N^{-1} \left(\frac{\alpha_s}{2\pi}\right)^2d\Phi_3(\{p\}_3;q)\, J_3^{(3)}(\lbrace p\rbrace_{3}) \Biggl \{ &\nn \\
\textB{1}&+\frac{1}{2}\left(2\mathcal{E}_{4}^{0}(s_{12})+\mathcal{\wt{E}}_{4}^{0}(s_{1i})+\mathcal{\wt{E}}_{4}^{0}(s_{i2})\right)
 M_3^0(1,i,2)
 &\nn\\
 \textC{2}&+\frac{1}{2}\left(\mathcal{\wt{B}}_{5,3}^{0}(s_{12},s_{i2})+\mathcal{\wt{B}}_{5,3}^{0}(s_{12},s_{1i})\right)
 M_3^0(1,i,2)
 &\nn\\
 &-----------------------&\nn\\
 \textD{3}&+\mathcal{A}_{3}^{0}(s_{12})\wh{M}_3^1(1,i,2)&\nn\\
 \textD{4}&+\frac{1}{2}\left(\mathcal{E}_{3}^{0}(s_{1i})+\mathcal{E}_{3}^{0}(s_{i2})\right)
 \wt{M}_3^{1}(1,i,2)
 &\nn\\
 \textD{5}&+\frac{1}{2}\left(\mathcal{\wt{E}}_{3}^{1}(s_{1i})+\mathcal{\wt{E}}_{3}^{1}(s_{i2})\right)
 M_3^0(1,i,2)
 &\nn\\
 \textD{6}&+\left(\mathcal{\hat{A}}_3^1(s_{12})+\frac{b_{0,F}}{\e}\left(\bigg( \frac{s_{12}}{\mu^2}\bigg)^{-\epsilon}-1\right)\mathcal{A}_3^0(s_{12})\right)
 M_3^0(1,i,2)
 &\nn\\
 \textE{7}&+\frac{1}{2}\left(\mathcal{\wt{E}}_{4,3}^{1}(s_{1i})+\mathcal{\wt{E}}_{4,3}^{1}(s_{i2})\right)
 M_3^0(1,i,2) \Biggr \}.&
\end{flalign}
The double-virtual subtraction term proportional to $N_f^2$ reads:
\begin{flalign}
&{\rm d}\sigma_{NNLO,N_f^{2}}^{U} = N_3 N_f^2 \left(\frac{\alpha_s}{2\pi}\right)^2d\Phi_3(\{p\}_3;q)\, J_3^{(3)}(\lbrace p\rbrace_{3}) \Biggl \{ &\nn \\
 \textD{1}&-\frac{1}{2}\bigg[
\,\,{\cal E}_{3}^{0}(s_{1i}) +{\cal E}_{3}^{0}(s_{i2})
 \bigg] \wh{M}_3^1(1,i,2)&\nn\\
\textD{2}&-\frac{1}{2}\bigg[ 
 \hat{\cal{E}}_{3}^{1}(s_{1i}) +\frac{b_{0,F}}{\epsilon}\left(\bigg( \frac{s_{1i}}{\mu^2}\bigg)^{-\epsilon}-1\right){\cal{E}}_{3}^{0}(s_{1i})\bigg] M_3^0(1,i,2)
 &\nn\\
  \textD{3}&-\frac{1}{2}\bigg[ 
 \hat{\cal{E}}_{3}^{1}(s_{i2}) +\frac{b_{0,F}}{\epsilon}\left(\bigg( \frac{s_{i2}}{\mu^2}\bigg)^{-\epsilon}-1\right){\cal{E}}_{3}^{0}(s_{i2})\bigg] M_3^0(1,i,2)\Biggr \}.&
\end{flalign}

\bibliographystyle{jhep}
\bibliography{bib2}{}
\end{document}